\documentclass[journal]{IEEEtran}

\usepackage{ucs} 
\usepackage{graphicx} 
\usepackage{xcolor}

\usepackage{amsmath}
\usepackage{amssymb}
\usepackage{xr}
\usepackage{xfrac}
\usepackage{nicefrac}
\usepackage{hyperref}
\usepackage{enumerate}
\usepackage{cite}
\usepackage{url}
\usepackage{flushend}

\newcommand\tran{s} 
\newcommand\dec{i}  
\newcommand\recv{r} 
\newcommand\tp{t}    
\newcommand\cm{ } 
\newcommand\ep{e} 

\begin{document}
\urlstyle{sf}
\title{Quantum CDMA Communication Systems
}

\author{Mohammad~Rezai and~Jawad~A.~Salehi,~\IEEEmembership{Fellow,~IEEE},
  \thanks{Mohammad~Rezai is with Sharif Quantum Center and Electrical Engineering Department, Sharif University of Technology, Tehran, Iran (e-mail:~\url{m.rezai@ee.sharif.edu}).}
  \thanks{Jawad~A.~Salehi is with Sharif Quantum Center and Electrical Engineering Department, Sharif University of Technology, Tehran, Iran (e-mail:~\url{jasalehi@sharif.edu}).}}

\IEEEpubid{0018-9448 \text{\textcopyright} 2021 IEEE. Personal use is permitted, but republication/redistribution requires IEEE permission. }
\maketitle

\begin{abstract}
        Barcoding photons, atoms, and any quantum states can provide a host of functionalities that could benefit future quantum communication systems and networks beyond today's imagination. As a significant application of barcoding photons, we introduce code division multiple-access (CDMA) communication systems for various applications.
        In this context, we introduce and discuss the fundamental principles of a novel quantum CDMA (QCDMA) technique based on spectrally encoding and decoding of continuous-mode quantum light pulses.
        In particular, we present the mathematical models of various QCDMA modules that are fundamental in describing an ideal and typical QCDMA system, such as quantum signal sources, quantum spectral encoding phase operators, M$\times$M quantum broadcasting star-coupler, quantum spectral phase decoding operators, and the quantum receivers.
Following the above discussions, we then elaborate on a QCDMA system with M users. In describing a QCDMA system, this paper considers a unified approach where the input continuous-mode quantum light pulses can take on any form of pure states such as quantum coherent (Glauber) states and quantum number (Fock) states.
        The mathematical presentation is independent of the form of the input pure quantum states.
        We show that the spectrally encoded quantum states of the light at the quantum star-coupler output are not, in general, factorized states, except for input Glauber states. 
        For input number states, one can observe features like entanglement
and quantum interference.
        Moreover and interestingly, as a consequence of Heisenberg's uncertainty principle, the quantum signals sent by photon number states obtain complete phase uncertainty at the time of measurement. Therefore, at the receiver output, the multiaccess inter-signal interference vanishes. 
        Due to Heisenberg's uncertainty principle, the received signal intensity at the photodetector's output, right at the time of measurement, changes from coherent detection scheme for input Glauber states to incoherent detection scheme for input number states. 
         We also would like to highlight that the mathematical models and tools developed in this paper, in the context of QCDMA, become
         very useful for developing and analyzing other quantum multiple-access techniques based on wavelength, space and time domain.
         Furthermore, our mathematical model is valuable in signal design and data modulations of point-to-point quantum communications, quantum pulse shaping, and quantum radar signals and systems where the inputs are continuous mode quantum signals.
        QCDMA may open up novel possibilities for various quantum technologies such as data privacy, distributed quantum computation, quantum internet, and anti-jamming quantum communication systems.        
\end{abstract}

\IEEEpeerreviewmaketitle
\section{Introduction}
\begin{figure}[!t]
          \centering
          \includegraphics[width=\columnwidth]{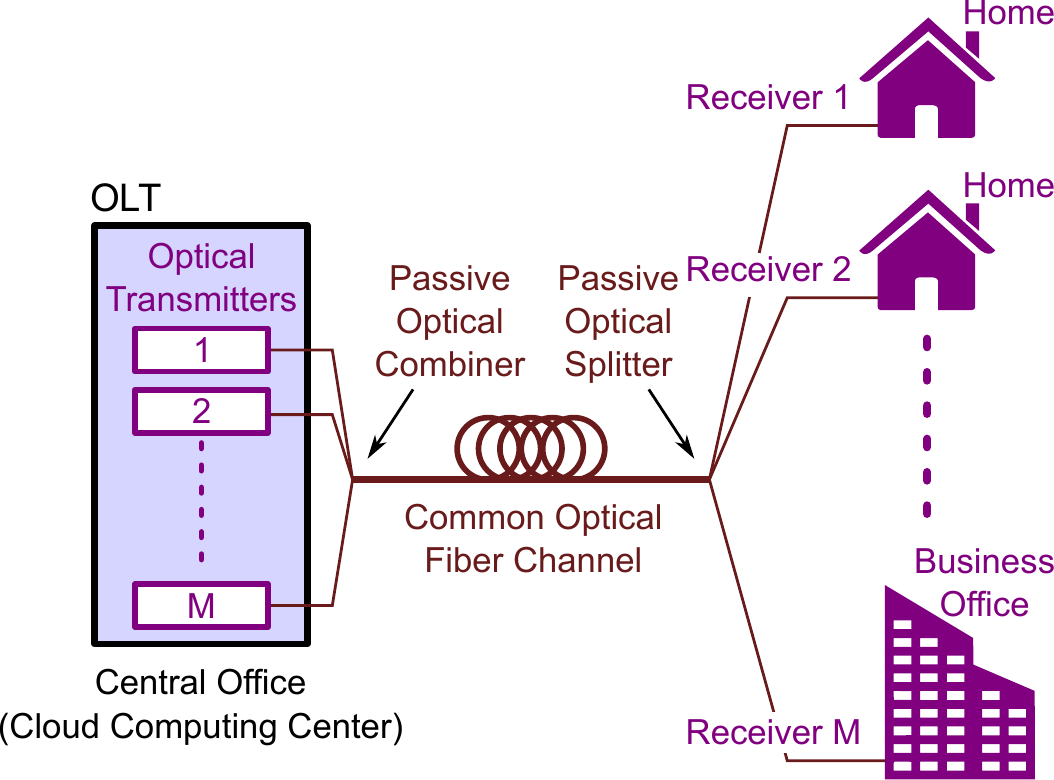}
          \caption{\textbf{A typical and simplified last-mile technology.} It is based on a passive optical network (PON) for fiber-to-the-home (business) architecture.
            The combiner, fiber channel, and splitter cascade can be
 mathematically modeled as star-coupler, see Fig.~\ref{fig:cdma}.
          }\label{fig:lastmile}
        \end{figure}

        \IEEEPARstart{A}{mong} many advances in disruptive quantum information technologies, quantum communication systems and networks enjoy special attention; due to many crucial and interesting quantum features in quantum signals, such as superposition, quantum interference and entanglement~\cite{helstrom_1976,cariolaro_2015,wilde_2013,razavi_2018,kimble_n_2008,wehner_s_2018,yard_tit_2011,zhang_sr_2013,escartin_2015_jstonqe,sharma_oqe_2020,giovannetti_prl_2004,brent_pra_2005,guha_pra_2007,shapiro_jstqe_2009,guha_itp_2011,wilde_ita_2012,wilde_prl_2012,xu_qip_2013}.
        On the other hand, due to optical fiber's superior channel characterization, optical communications is a dominant means of transporting information bits across the globe~\cite{papen_blahut_2019}.
        \IEEEpubidadjcol
        One of the essential features of optical fiber telecommunications infrastructure is its ability to deliver information bits from the backbone through a relatively short distance to desired users or customers at home or businesses~\cite{razavi_2018}.
        This so-called last-mile technology infrastructure requires a multiple-access scheme that would enable it to deliver the information bits distinctively to various users employing a typical optical fiber communication channel.
        Using the above techniques, we can push the intelligence of the network's switching and routing to its peripheries, such as the user's unit modules, making the network's central control processing as simple as possible and simultaneously reducing the cost of a communications network infrastructure.
        In optical fiber-based communications, fiber-to-the-home (FTTH) or business (FTTB) based on passive optical network (PON) is a preferred last-mile technology for future ultra high speed, high bandwidth services to individual home subscribers due to its simplicity and reduced cost~\cite{green_b_2005}.    
        Figure~\ref{fig:lastmile} shows a simplified FTTH architecture where a central node (optical line terminators (OLT)) such as a central switching node or cloud computing center, connected via a shared optical fiber channel to neighborhood's vicinity.
        Services provided via information bits to various users using a shared optical fiber channel require advanced multiple-access technologies.
        Most multiple-access techniques are accomplished either in frequency (wavelength), time, or code domain.        
        This paper focuses on a multiple-access technique in the code domain, also known as code division multiple-access (CDMA).
        In a CDMA technique, communication between a pair of users (for example, information delivery from sender~$\tran$ to receiver~$\recv$) requires that the $\tran$th sender encodes or imprints upon its output information bits a pseudorandom binary sequence (barcode) that corresponds to the $\recv$th user's signature sequence.
        Various CDMA techniques have found many wireless~\cite{aburgheff_2007,viterbi_1995} and optical networks~\cite{salehi_tran_1989} application due to their specific advantages such as high spectral efficiency, distributivity, data privacy, and asynchronous transmission amongst the users in the network.
        In a typical CDMA system, many transmitters send their encoded information bits across a shared channel to receivers.
        An intended receiver must recover its information amongst many other interfering, users’ coded signals known as multiaccess interfering signals and using the common optical channel.        
        Therefore,
        a CDMA network's successful design would depend on the signature sequences' or barcodes' proper design for various network users.
        The most popular and successful signature sequences in most CDMA techniques are binary phase-shifting generated by maximum length shift registers~\cite{golomb_b_1967}.
        These signature sequences are also known as pseudorandom sequences. In this paper, we also use the same class of binary sequences. For the sake of mathematical simplicity in system analysis, assume the sequences are pure random binary sequences~\cite{salehi_jlwt_1990}.

        \begin{figure*}[htp!]
	\includegraphics[width=\textwidth]{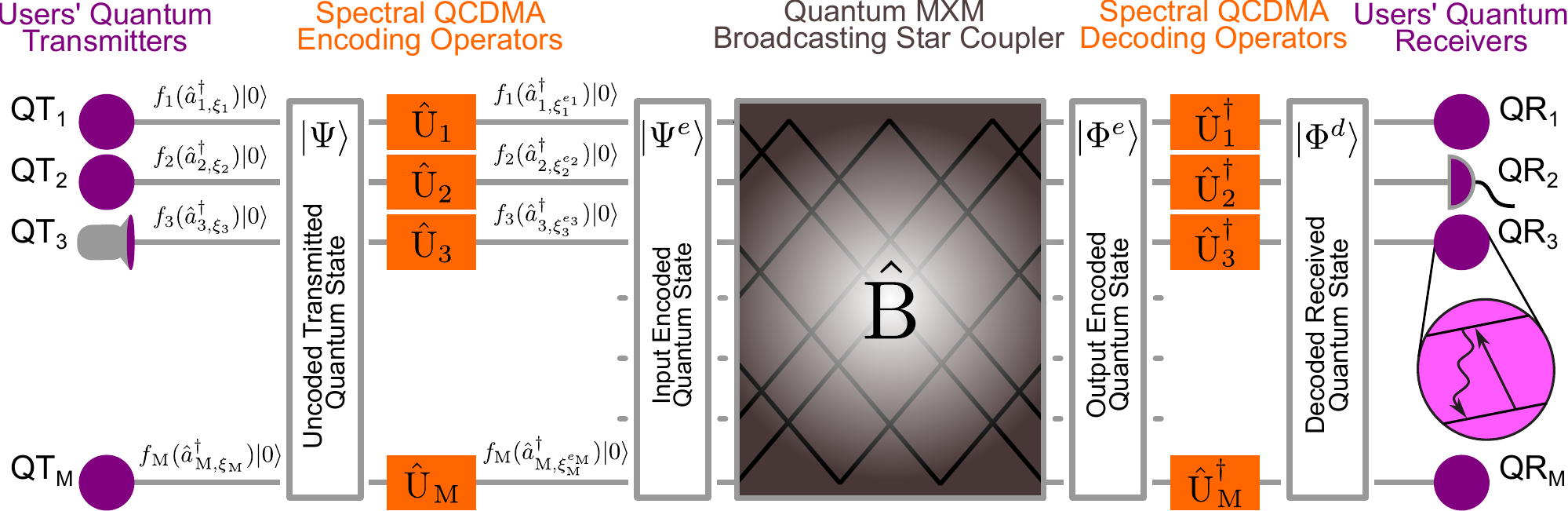}
	\caption{\textbf{A typical quantum optical code-division multiple-access communications system (QCDMA) with spectral encoder/decoder.}
          The network is composed of M distinct pure state emitters with corresponding photon-wavepacket $\xi_{\tran}, \ \tran \in{1,\ldots, \text{M}}$.
          The wavepackets can be different, but in many applications, they are taken to be the same. 
          Each quantum transmitter or sender (QT$_{\tran}$) encodes its quantum signal using a spectral phase-shifting operator ($\hat{\text{U}}_{\tran}$) with a distinct encoding mask.
          A broadcasting star-coupler can be modeled as a sequence of beamsplitters that broadcasts all quantum transmitters' signals to all receivers.
          Each quantum receiver (QR$_\recv$) decodes its corresponding signal using the conjugate mask.
          Consequently,
          quantum receiver QR$_{\recv}$ decodes and receives only the signals from node QT$_{\tran}$ (the schematic takes the assumption of $\recv=\tran$).
        }\label{fig:cdma}
      \end{figure*}
      
        \section{Quantum CDMA}
        Figure~\ref{fig:cdma} depicts a simplified and typical QCDMA communication system with M users.
        It is composed of M quantum transmitters (QT) and M quantum receivers (QR).
        In this paper, we assume the transmitters emit pure states such as a single-photon~$\lvert 1 \rangle $ or a number state~$\lvert n \rangle $ (Fock states) or coherent light  $\lvert \alpha \rangle$ (Glauber state), where $\alpha$ denotes the complex amplitude of the optical-electrical field.
        A quantum receiver can be modeled as an ideal photo-detector or other quantum structures such as a single atom~\cite{boozer_prl_2007} or a spin system~\cite{yang_np_2016}.
        Each user sends its quantum signal (pulse) into a QCDMA spectral encoding operator unit ($\hat{\text{U}} $) prior to entering into an M$\times$M broadcasting star-coupler ($\hat{\text{B}} $) .
        The star-coupler broadcasts each user's spectrally encoded quantum signal to all the users' quantum receivers.
        The receivers with the spectral encoding information ($\hat{\text{U}}$) can decode the signal into its original form (via $\hat{\text{U}}^{-1} = \hat{\text{U}}^{\dagger}$) and interface it with their corresponding quantum receivers.
        Note that the superscript dagger ($\dagger$) implies complex conjugate transpose of operator~$\hat{\text{U}}$.

        To encode the spectrum of quantum signals, we apply an encoding mechanism (see Fig.~\ref{fig:modem}a) that can generally be conceived in three steps.
        The first step is to represent a quantum signal in its corresponding frequency domain.
        Mathematically, the frequency domain representation can be achieved by the inverse Fourier transformation (IFT) of the quantum signal from the time domain into its spectral domain.
        Once we realize the signal in the frequency domain, in the second step, each frequency component~$\omega$ of the quantum signal is multiplied by its corresponding and assigned encoding operator $\hat{\text{U}}(\omega)$.
        For example, the operator $\hat{\text{U}}$ can randomly phase-shift the input quantum light pulse's spectral components. 
        Finally, in the third step, the signal's encoded spectral components are reassembled by an inverse operation performed in the first step.
        Mathematically the inverse operation is equivalent to the Fourier transformation (FT) of the spectrally encoded quantum signal into its time domain.

        This general concept of spectral encoding/decoding can be realized experimentally by a variety of methods.       
        For example, arrayed waveguide gratings (AWGs) and fiber Bragg grating (FBG) can provide this functionality on an integrated chip~\cite{okamoto_b_2000}. 
        Figure~\ref{fig:modem}b shows an alternative and simple experimental realization of such a spectral encoder/decoder~\cite{weiner_ol_1988}. 
        It is composed of a pair of diffraction gratings and lenses. 
        The first set of grating and lens decomposes a quantum light pulse's spectral components into $N_c$ distinct and non-overlapping frequency bands and place them in a one-to-one mapping operation into different spatial modes.
        This mapping enables us to apply distinct phase-shifts to each band of the spectral components~\cite{weiner_ol_1988}.
        This scheme can employ a spatially patterned phase mask, as depicted in Fig.~\ref{fig:modem}b, to each spatially and spectrally distinct band.
        The second set of grating and lens reassembles the $N_c$ spatially separated and independently encoded spectral bands of the quantum light pulse back into a single spatial mode.
        If the mask shifts the phase of each spectral component by an independent and identically distributed random value, the encoded quantum light pulse's spectral components interfere, and the original quantum pulse spreads in the time domain, see Fig.~\ref{fig:modem}c.
        However, the spectrally phase-encoded quantum pulse would be reconstructed to its original shape if one applies the encoding mask's conjugate to the encoded pulse, implementing the same mechanism as the encoding process.
        If the decoding mask is not the encoding mask's conjugate, the signal would not reconstruct its original pulse shape and remain as a low-intensity time spread quantum signal.
        The improperly decoded signals act as a multiaccess interfering signal at the front-end of the desired user's quantum receiver, thereby degrading the desired or the intended quantum receiver's performance.
        For more application of this technique in quantum information processing and its experimental demonstration, we refer the interested readers to Ref.~\cite{kues_np_2019}.

 \section{Quantum Light Pulses}              
        In the context of spectrally encoding/decoding, the quantum signals generated by a QCDMA user's transmitter come in the form of quantum light pulses with a spectrum covering a range of frequencies with spectral amplitude $\xi(\omega)$, the so-called photon-wavepacket.
        It is shown~\cite{loudon_2000}, the corresponding photon-wavepacket creation operator is as follows         
        \begin{equation}
          \begin{split}
            \hat a^ \dagger _{\xi}
            &=\int  d\omega \,  \xi  (\omega) \hat a ^\dagger (\omega) =  \int  d\tp \,  \xi  (\tp) \hat a ^\dagger (\tp)\, ,\\
          \end{split}
          \label{eqn:adximain}
        \end{equation}
        where, $\xi  (\tp)$ represents the temporal amplitude of the photon-wavepacket.
        $  \hat a ^\dagger (\omega)$ and  $ \hat a^{\dagger} (\tp) $ represent continuous mode creation operators in the frequency and time domains, respectively.
        Since $ \hat a (\tp) $ is the Fourier transform of $ \hat a(\omega)$,       
        from Eq.~\eqref{eqn:adximain}, one can deduce that the temporal amplitude of the photon-wavepacket denoted by $ \xi (\tp)$ is also the Fourier transform function of $\xi(\omega)$.
        The duality of time and frequency formalism enables us to choose either frequency or time to expand quantum light pulse or photon-wavepacket.
        Let us assume that the frequency domain wavepacket of the user's quantum light pulse has a Gaussian pulse shape with the central frequency of $\omega_0$; therefore, its spectral amplitude can be expressed as follows 
        \begin{equation}
          \xi (\omega )=\frac{1}{\sqrt[4]{2 \pi  \Delta ^2}} e^{-\frac{(\omega -\omega_0)^2}{4 \Delta ^2}} e^{i \omega \tp_0  }  e^{ -\frac{i \omega_0 \tp_0}{2}}\, ,
          \label{eqn:xiGomega}
        \end{equation}
        where $2 \Delta$ and $\tp_0$ correspond respectively to the effective spectral bandwidth and the central time where the pulse is peaked.

    For a Gaussian spectral wavepacket (Eq.~\eqref{eqn:xiGomega}), the temporal wavepacket is as follows
        \begin{equation}
          \begin{split}
            \xi (t )&=\sqrt[4]{\frac{2 }{\pi \tau_p ^2}} e^{-\frac{ (\tp -\tp_0)^2}{\tau_p^2}} e^{-i \omega_0 \tp  }  e^{ \frac{i \omega_0 \tp_0}{2}}\, .
          \end{split}
          \label{eqn:xit}
        \end{equation}

        Equation~\eqref{eqn:xit} shows that the temporal wavepacket is also Gaussian but with an effective pulse duration $\tau_p = 2\sigma_p= \nicefrac{1}{\Delta}$\, , where $\sigma_p$ is the standard deviation of the temporal Gaussian pulse.
        \par
        One should differentiate quantum optical Gaussian states, states with Gaussian Wigner quasiprobability distribution in phase space, from quantum states with Gaussian line-shape (photon-wavepacket) considered in this paper.        
        For instance, number states are not Gaussian states but can have a Gaussian-shaped spectrum. On the other hand, Glauber (Coherent) states are Gaussian states even though their spectrum may not be Gaussian.
        Furthermore, let us highlight that in this paper, we use the Gaussian wavepacket only as an example to demonstrate the QCDMA system.
        The mathematical model is equally valid for almost any photon-wavepacket.        
        For instance, as a more practical case, a time-bin qubit can be represented by a single-photon with photon-wavepacket $\eta(\tp)= b_1 \xi_1(\tp)+ b_2\xi_2(\tp)$, where $\xi_1(\tp)$ and $\xi_2(\tp)$ are respectively Gaussian-wavepackets with central times~$\tp_1$ and~$\tp_2$
       (that is, in Eq.~\eqref{eqn:xit} for the temporal shape of $\xi_1(\tp)$ ($\xi_2(\tp)$), $\tp_0$ is substituted with $\tp_1$ ($\tp_2$)).
        Coefficients~$b_1$ and $b_2$ are the probability amplitudes, $|b_1|^2+|b_2|^2=1$.
        From Eq.~\eqref{eqn:adximain}, one can show that $ \hat a^ \dagger _{\eta}= b_1\hat a^ \dagger _{\xi_1}+ b_2 \hat a^ \dagger _{\xi_2}$; therefore, a single-photon with photon-wavepacket $\eta$ corresponds to a time-bin qubit~$\lvert \psi \rangle =  \hat a^ \dagger _{\eta}  \lvert 0 \rangle= ( b_1\hat a^ \dagger _{\xi_1}+ b_2 \hat a^ \dagger _{\xi_2})\lvert 0 \rangle = b_1 \lvert 1_{\xi_1}\rangle + b_2\lvert 1_{\xi_2} \rangle=\lvert 1_{\eta} \rangle$.
        A similar argument applies for frequency-bin qubits where the central frequency of $\xi_1$ is $\omega_0 = \omega_1$, and the central frequency of $\xi_2$ is $\omega_0=\omega_2$.

        {
          \begin{figure}[!t]
            \centering
	\includegraphics[width=\columnwidth]{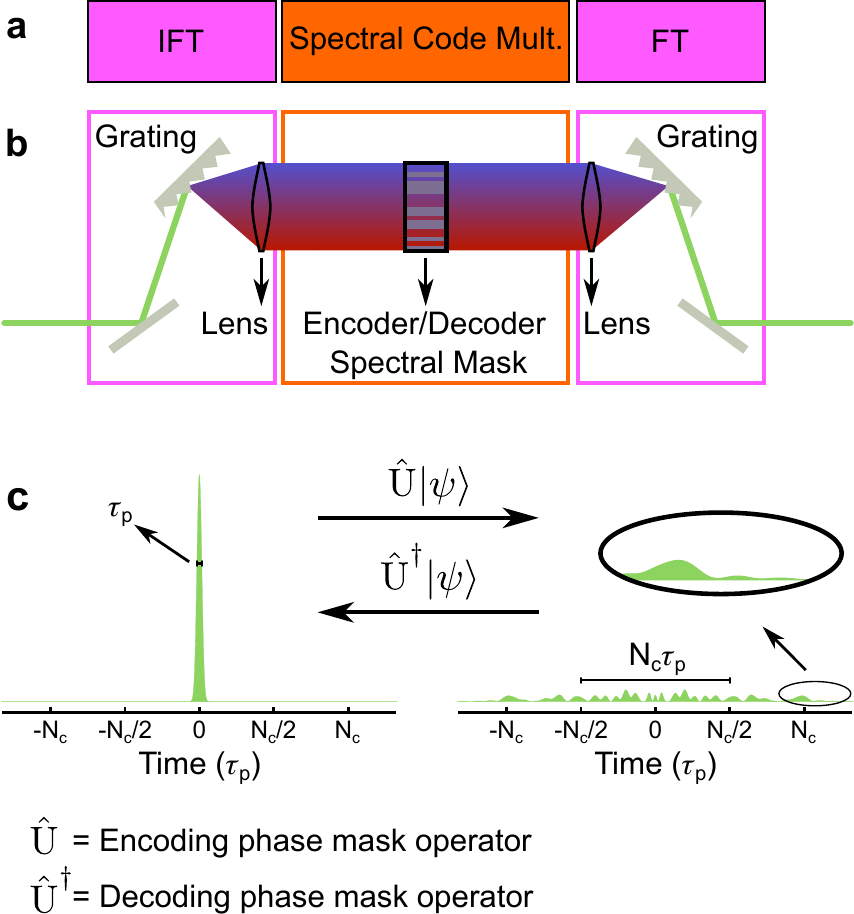}
	\caption{\textbf{Quantum pseudorandom spectral phase-shifting operator.}
          \textbf{a} The general structure of spectral encoding/decoding devices.
          The Inverse Fourier transform (IFT) section, which comprises a grating and a lens in b, decomposes a quantum signal's constituent frequencies. 
          The spectral code multiplier applies the code's corresponding phase-shift to each quantum signal frequency component.
          The Fourier transform (FT), which also comprises a grating and a lens in b, reconstruct the signal.
          \textbf{b} A demonstration of the random spectral phase-shifter.         
          The left grating and lens disperse the spectrum.
          The mask phase-shifts different sections of the spectrum by a randomly chosen value.
          The right grating and lens reassemble the light to a single beam.
          \textbf{c} Spreading and despreading a sharp quantum pulse in the time domain.
          A random spectral phase-shifter spreads a quantum light pulse with effective pulse duration $\tau_p$ in the time domain to approximately~$N_c \tau_p$. In this image $N_c = 31$.
          The quantum light pulse can be reconstructed when the conjugated mask is used in the decoder's spectral phase-shifter.
          Both masks together shift the phase of all frequencies by the same value, and consequently, the quantum pulse reappears.         
          }\label{fig:modem}
        \end{figure}

        \begin{figure}[!t]
          \centering
          \includegraphics[width=\columnwidth]{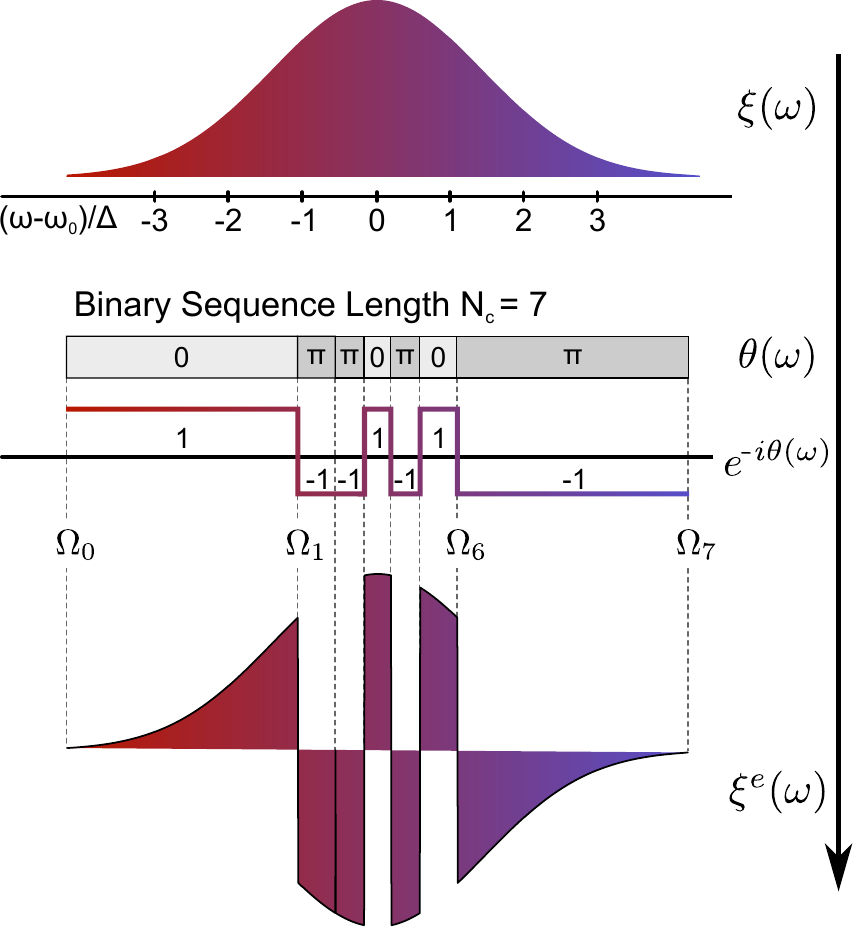}
          \caption{\textbf{Barcoding Photons.}
            Binary spectral encoding.
            \textbf{a} is a Gaussian photon-wavepacket. 
            Its spectrum is divided into seven chips of equivalent area.
            Each chip forms a domain to encode a phase on it.
            \textbf{b} shows a mask which phase-shifts each chip by a value of $0$ or $\pi$, and consequently, the corresponding wavepacket amplitudes are multiplied by $+1$ or $-1$, respectively.
            \textbf{c} depicts this amplitude's sign alternation.
            \textbf{d} shows the transformed spectrum of the Gaussian wavepacket due to the spectrally encoded phase mask.}\label{fig:coding}
        \end{figure}
        }
\section{Quantum Transmitter's Signal}

        This paper assumes that each user's quantum source transmits a pure quantum light pulse state. Furthermore, the photons emitted by various user's quantum sources have identical spectral wavepackets $\xi(\omega)\,$.
        Therefore, one can write the transmitted pure quantum light pulse state as a superposition of the photon's number states $\lvert n_{\xi} \rangle$ with wavepacket $\xi$, that is        
        \begin{equation}
          \begin{split}
            \lvert \psi \rangle &= \sum_n c_n \lvert n_{\xi} \rangle =f(\hat a^ \dagger _{\xi}) \lvert 0 \rangle \, ,
          \end{split}
          \label{eqn:si=f(adag)}
        \end{equation}
        where function $f(\hat a^ \dagger _{\xi})$ is an analytic, infinitely differentiable function, $c_n$s correspond to the coefficients of the Taylor series of function $f(\hat a^ \dagger _{\xi})$, and $ \lvert 0 \rangle$ denotes the vacuum state~\cite{louisell_1990}.
        
        This formalism (Eq.~\eqref{eqn:si=f(adag)}) represents a general class of pure quantum light states. 
        For example, in the case of a single-photon emitter, function $f(\hat a^ \dagger _{\xi})$ is the first order power function
        , that is, $f(\hat a^ \dagger _{\xi})=\hat a^ \dagger _{\xi}$.
         However, it takes an exponential form for a coherent (Glauber) state, that is $f(\hat a^ \dagger _{\xi})=\exp( -|\alpha|^ 2/2 + \alpha \hat a^ \dagger _{\xi})$. 

        Intensity is a quantity measured in most relevant experiments.
        For example, in QCDMA, users' quantum receivers may measure the light intensity emerging from their quantum spectral decoders.
        Therefore, in what follows, we calculate the intensity and its dependence on time for quantum states represented by Eq.~\eqref{eqn:si=f(adag)}.
        The positive part of the electric field operator is proportional to the annihilation operator as $\hat E^{+}(\omega) = C_{\omega} \hat a (\omega)$, where proportionality factor~$C_{\omega}$ in free space depends on the vacuum permittivity~$\epsilon_0$, the cross-section of the electromagnetic field~$A$, the Planck constant $\hbar$, and the speed of light~$c$ as $C_{\omega}=\sqrt{\frac{\hbar \omega}{4 \pi \epsilon_0 c A}}$.
        \par
        For simplicity, we apply the narrowband approximation, that is, the spectral width of photon-wavepacket $\xi(\omega)$ is small compared to its central frequency~\cite{loudon_2000}, that is $\Delta \ll \omega_0$.
        Consequently, factor~$C_{\omega}$ remains almost invariant for the whole range of photon-wavepacket and can be replaced with the constant $C_{\omega_0}=\sqrt{\frac{\hbar \omega_0}{4 \pi \epsilon_0 c A}}$.
        In order to simplify our expressions, we choose the value of $C_{\omega_0}$ as the unit of electric field $\hat E^{+}(\omega)$ (note that the ladder operators $\hat a$ and~$\hat a^{\dagger}$ are unitless), implying $\hat E^{+}(\omega) \equiv \hat a (\omega)$ and consequently $\hat E^{+}(\tp) \equiv \hat a (\tp)$.        
        Therefore, the instantaneous-intensity measurement reads
        \begin{equation}
          \begin{split}
            I(\tp) = \langle \psi \rvert  \hat a ^ \dagger (\tp) \hat a(\tp) \lvert \psi \rangle=\lvert  \hat a(\tp) \lvert \psi \rangle \rvert^{2}=\lvert  \hat a(\tp) f(\hat a^ \dagger _{\xi}) \lvert 0 \rangle  \rvert^{2} \, .
          \end{split}
          \label{eqn:I(t)}
        \end{equation}
        This equation states that the instantaneous-intensity at time~$\tp$  is the state's square amplitude (probability) after a photon at time~$\tp$ is eliminated (by a photodetector, for example)\cite{rezai_prx_2018}.
        Since the commutation $[\hat a(\tp), \hat a_ \xi ^ \dagger] =\xi(\tp)$ holds, one can deduce the following commutation relation
        \begin{equation}
          \begin{split}
            [\hat a(\tp),f(\hat a_ \xi ^ \dagger)] = \xi (\tp) f'(\hat a_ \xi ^ {\dagger}) \, ,
          \end{split}
          \label{eqn:[f,a]}
        \end{equation}
        where the superscript $'$ (prime) indicates the derivative of the function $f(\hat a_ \xi ^ \dagger)$ with respect to~$\hat a_ \xi ^ \dagger$~\cite{louisell_1990}.
        Equation~\eqref{eqn:[f,a]} indicates that $\hat a(\tp) \lvert \psi \rangle=  \xi (\tp)f'(\hat a^ \dagger _{\xi}) \lvert 0\rangle$\,.
        Therefore for a non-vacuum state ($f(\hat a^ \dagger _{\xi}) \neq 1$), the instantaneous-intensity at time~$\tp$ is proportional to the amplitude square of the quantum light pulse wavepacket at time~$\tp$, that is       
        \begin{equation}
          \begin{split}
            I(\tp) = \bar I  \lvert \xi (\tp) \rvert^2 \, ,
          \end{split}
          \label{eqn:Ixi}
        \end{equation}
        where $\bar I$ is the mean intensity (for more details, see appendix~\ref{app:I(t)}).
        Thus Eq.~\eqref{eqn:Ixi} shows that the temporal shape of the instantaneous-intensity is independent of function~$f(\hat a^ \dagger_{\xi})\, $, and purely depends on the temporal wavepacket shape ($\xi(\tp)$) of the quantum light pulse.
        For instance, the temporal shape of instantaneous-intensity measurement would be the same for quantum number states (Fock states), coherent states (Glauber states) if their photons have the same wavepacket.
        
          \section{Quantum Spectral CDMA Encoding Operator}
        In a QCDMA, one can describe the effect of the encoder (decoder) module (see Fig.~\ref{fig:modem}) by a spectral phase-shifting operator denoted by $\hat{\text{U}}$ ($\hat{\text{U}}^{-1}=\hat{\text{U}}^{\dagger}$ for decoder) and expressed mathematically as 
        \begin{equation}
          \begin{split}
            \hat{\text{U}} =  e^{-i  \sum_{\omega} \, \theta(\omega) \hat a^{\dagger} (\omega) \hat a (\omega) } = \prod_{\omega} e^{-i \, \theta(\omega) \hat a^{\dagger} (\omega) \hat a (\omega) } = \prod _{\omega} \hat{\text{U}} (\omega)\, ,
          \end{split}
          \label{eqn:phs}
        \end{equation}
        where $\theta(\omega)$ is the phase change applied at frequency $\omega $ upon the quantum light pulse's spectral wavepacket $\xi(\omega)$, and symbol~$\prod$ denotes the tensor-products (some authors write it by symbol~$\bigotimes$).
        Considering the Heisenberg picture, the unitary operator $\hat{\text{U}}$ transforms the field operator $\hat a^\dagger (\omega)$ with relation $\hat{\text{U}} \hat a^\dagger (\omega) \hat{\text{U}^ \dagger} = \hat{\text{U}}(\omega) \hat a^\dagger (\omega) \hat{\text{U}^ \dagger}(\omega) = \hat a^\dagger (\omega) e^{-i \theta(\omega)}$.
        We use this transformation to express how the encoder enables the light pulse's quantum state to evolve into a coded state.
        The encoder's evolution (known as the Schr{\"o}dinger picture) on input quantum light pulse state $\lvert \psi \rangle$ gives state $\lvert \psi^{\ep} \rangle$, where superscript~$\ep$ implies encoded state.
        The encoded state is expressible as
        \begin{equation}
          \begin{split}
            \lvert \psi^{\ep} \rangle=\hat{\text{U}}  \lvert \psi \rangle = f( \hat{\text{U}} \,  \hat a_ \xi ^ \dagger  \, \hat{\text{U}}^ \dagger ) \lvert 0 \rangle = f( \hat a^ \dagger _{\xi^{\ep}}) \lvert 0 \rangle  \, .
          \end{split}
          \label{eqn:Utheta psi}
        \end{equation}
        where $\xi^{\ep} (\omega)= \xi (\omega)e^{-i \theta(\omega)}$. 
        Therefore, one can conclude that the spectral encoding operator in a typical QCDMA changes only the spectral wavepacket ($\xi(\omega)$) of the input quantum light pulse.
        This conclusion is not limited to the pure state Eq.~\eqref{eqn:si=f(adag)}, but it is also valid for a quantum mixed state.
        Specifically, the spectral encoding operator transforms a mix of states~$\lvert \psi_i \rangle = f_i(\hat a_ {\xi_i} ^ \dagger) \lvert 0 \rangle$ with corresponding statistical weights~$p_i$, denoted by density matrix~$\rho=\sum_{i} p_i\lvert \psi_i \rangle  \langle \psi_i \rvert$, to another encoded mixed state with encoded density matrix~$\rho^{\ep}=\hat{\text{U}} \rho \hat{\text{U}}^{\dagger}=\sum_{i} p_i\lvert \psi^{\ep}_i \rangle  \langle \psi^{\ep}_i \rvert$, where $\lvert \psi^{\ep}_i \rangle = \hat{\text{U}}  \lvert \psi_i \rangle= f_i(\hat a_ {\xi^{\ep}_i} ^ \dagger) \lvert 0 \rangle$. 
              
        Suppose we choose binary values of $0$ and $\pi$ for the amount of the phase-shift. In that case, the operator can implement a pseudorandom binary sequence (barcode) onto the spectrum of the incoming quantum light pulse.
        Let $N_c$ denote the length of the code (pseudorandom sequence) used by the encoding operator.
        In the context of QCDMA application, we divide the wavepacket spectrum into $N_c$ spectral chips with boundaries ($\Omega_0, \Omega_1, ...,\Omega_{N_c}$) such that the mean absolute square of the spectral wavepacket $\xi (\omega)$ is the same for each spectral chip ($\int_{\Omega_k}^{\Omega_{k+1}} \lvert \xi (\omega)\rvert^2 d \omega =\frac{1}{N_c}$).
        Each spectral chip is used for encoding one phase-shift.
        \par        
        Figure~\ref{fig:coding} shows how a binary code of length 7 transforms the quantum light pulse's spectral wavepacket $ \xi(\omega)$.
        The Fourier transform of the spectrally encoded wavepacket $\xi^{\ep}(\omega)$ displays the shape of the quantum light pulse's wavepacket in the time domain, that is $\text{FT} (\xi^{\ep}(\omega))=\xi^{\ep}(\tp)$.
        Subsequently, the temporal structure of the instantaneous-intensity, as indicated in Eq.~\eqref{eqn:Ixi}, is $I(\tp) = \bar I \lvert \xi^{\ep} (\tp) \rvert^2$ and is depicted in Fig.~\ref{fig:modem}c.
        Figure~\ref{fig:modem}c illustrates that a random binary phase-shift spreads the energy of a sharply peaked quantum light pulse, with duration~$\tau_p$, in the time domain by an amount of approximately equal to $N_c \tau_p$.
        This spreading of energy can, other than QCDMA, have many critical applications in secure communication systems such as information privacy, anti-jamming, or hiding signals by creating signals with a low probability of intercept~\cite{torrieri_b_2018}.

        \section{M$\times$M Quantum CDMA Broadcasting Star-Coupler}
        In a typical QCDMA network, the encoded quantum signal of each user is broadcasted, in a quantum sense (see appendix~\ref{app:sc}), by a passive medium, which can be modeled by a star-coupler, an M$\times$M beamsplitter~\cite{salehi_jlwt_1990}, or equivalently by a mesh of 2$\times$2 beamsplitters~\cite{reck_prl_1994,clements_o_2016} as depicted in Fig.~\ref{fig:cdma}. 
        It is worth noting that even though we are using beamsplitters for quantum communications, a sequence of beamsplitters and phase-shifters can also be used to perform universal quantum computation~\cite{knill_n_2001}.
        Here, M is the number of active users in a QCDMA system.
        Each of these M users is composed of two polarization modes.
        However, in this paper, for simplicity, we assume that all the transmitted quantum signals have the same polarization; also, the optical channel does not alter this polarization.
        Therefore, we ignore the polarization mode in our calculation.
        One can integrate this degree of freedom if needed~\cite{rezai_qst_2019}.
        In this paper, we consider a lossless network.        
        Then, according to the conservation of energy, an M$\times$M star-coupler can be represented by a unitary matrix $\mathbf{\underline B }$, which transforms the field operators of the coupler's input ports $ \hat a_{\tran}$ to output ports $ \hat a'_{\recv}$
        with path connection amplitude $B_{\recv \tran}$ between output port $\recv$ and input port $\tran$,
        as follows
        \begin{equation}
          \hat a'_{\recv}(\omega) =\sum_{\tran=1}^{\text{M}} B_{\recv \tran} \hat a_{\tran}(\omega) \, , \qquad \recv \in{1,\ldots,\text{M}} \, .
          \label{eqn:heis}
        \end{equation}
        The matrix $\mathbf{\underline  B}$ is assumed to be independent of frequency $\omega$, implying the network's response is independent of the input electromagnetic field's frequency, at least for the frequency range covered by the photon-wavepackets of the transmitted quantum light pulses.        
        We also assume that transmitting quantum sources are independent of each other.
        Specifically, the all-inclusive transmitted quantum state~$\lvert \Psi \rangle$ is a factorized state of the senders' quantum states, that is $    \lvert \Psi\rangle  =\prod_{\tran=1}^{\text{M}} \lvert \psi_{\tran}\rangle =  \prod_{\tran=1}^{\text{M}} f_{\tran}(\hat a^ \dagger _{\tran,\xi_{\tran} }) \lvert 0 \rangle $,
        where  $\lvert \psi_{\tran} \rangle$  corresponds to the transmitted signal of the $\tran$th sender~(QT$_\tran$).        
        Therefore, the input state to the QCDMA M$\times$M star-coupler, $\lvert \Psi^{\ep} \rangle$, is also a factorized state and takes the form as follows
        \begin{equation}
          \begin{split}
            \lvert \Psi^{\ep} \rangle & =\prod_{\tran=1}^{\text{M}} \lvert \psi_{\tran}^{\ep} \rangle =  \prod_{\tran=1}^{\text{M}} f_{\tran}(\hat a^ \dagger _{\tran,\xi^{\ep_{\tran}}_{\tran} }) \lvert 0 \rangle \, .
          \end{split}
          \label{eqn:Psi^c}
        \end{equation}
        where $\lvert \psi_{\tran}^{\ep} \rangle$ denotes the $\tran$th sender's encoded quantum signal, and $\hat a^ \dagger _{\tran,\xi^{\ep_{\tran}}_{\tran}}$ creates a photon with an encoded wavepacket shape of $\xi^{\ep_{\tran}}_{\tran}$ (spectral modes) on behalf of the $\tran$th quantum transmitter (the $\tran$th input coupler port or path mode). 
        The superscript $\ep_{\tran}$ indicates photons' encoding by the spectral phase-shifting operator corresponding to the barcode of sender~$\tran$ (see Fig.~\ref{fig:cdma}).
        \par
        Equation~\eqref{eqn:heis} illustrates the system's evolution in the Heisenberg picture, where quantum states remain unchanged, but the operators evolve.
        Alternatively, one can adopt the Schr{\"o}dinger picture and associate the evolution of the system with quantum state ($\lvert \Psi^{\ep} \rangle$) transformation as $\hat{\text{B}} ^{\dagger} \lvert \Psi^{\ep} \rangle$, where  $\hat{\text{B}} ^{\dagger}$ denotes star-coupler's transformation operator on input quantum states.
        The equivalency of these two pictures indicates that $\hat{\mathbf{a}} ' = \mathbf{\underline B} \,  \hat{\mathbf{a}} := \hat{\text{B}} \,  \hat{\mathbf{a}} \, \hat{\text{B}}^\dagger $, where $\hat{\mathbf{a}}$  and $\hat{\mathbf{a}}'$ are the column vector representation of the field operators for the input and output ports of the network's  M$\times$M star-coupler, respectively~\cite{leonhardt_rpp_2003,legero_aamo_2006}.
        That is $\hat{\mathbf{a}}= (\hat a_1(\omega),  \hat a_2(\omega),  \hdots, \hat a_{\tran}(\omega), \hdots,  \hat a_{\text{M}}(\omega))^{\top} \,$, where $\top$ denotes the transpose operation.
        In appendix~\ref{app:sc}, we show that the light's quantum state at the star-coupler's output, $\lvert \Phi^{\ep} \rangle$, can be expressed as follows~\cite{leonhardt_book_1997,furusawa_2015}
        \begin{equation}
          \begin{split}
            \lvert \Phi^{\ep} \rangle =\hat{\text{B}}^{\dagger} \lvert \Psi^{\ep} \rangle =\prod_{\tran=1}^{\text{M}} f_{\tran}(\sum_{\recv=1}^{\text{M}} B_{\recv \tran} \hat a_{\recv,\xi^{\ep_{\tran}}_{\tran}}^{ \dagger}) \lvert 0 \rangle \, ,
          \end{split}
          \label{eqn:outsc}
        \end{equation}
        where $\hat a_{\recv,\xi^{\ep_{\tran}}_{\tran}}^{ \dagger}$ corresponds to the field creation operator at the $\recv$th output port with the $\tran$th input encoded photon-wavepacket.
        It is worth noting that Eq.~\eqref{eqn:outsc} represents the broadcasting characteristic of a star-coupler, and we refer to it as the state broadcasting equation.
        Unlike the input quantum state~$ \lvert \Psi^{\ep} \rangle $ of the M$\times$M star-coupler (Eq.~\eqref{eqn:Psi^c}), the output pure quantum state~$\lvert \Phi^{\ep} \rangle$ is not, in general, factorizable.
        Therefore, it can exhibit quantum features such as entanglement~\cite{shih_prl_1988,rezai_o_2019} and quantum interference~\cite{hong_prl_1987}.
        \par
        Let us highlight that since we have assumed the input signals are pure quantum states and the optical elements' operations in the QCDMA scheme are expressed by unitary operators, the quantum state remains pure during the whole process in QCDMA communications.
        Accordingly, one can easily observe entanglement in these pure states, such as in the star-coupler's output, Eq.~\eqref{eqn:outsc}, when they are not factorizable.
        If the optical elements' operation were not unitary, i.e., the optical elements cause photon loss or decoherence of the quantum state;
        the signal's quantum state would appear as a density matrix (similar to when the input quantum signal is a mixed state), which obliges more careful analysis in studying quantum effects such as entanglement~\cite{guo_sr_2014}.      

        \section{Decoding of Quantum  CDMA Signals}
        Let us assume that each M transmitting users encode their quantum signals by distinct pseudorandom binary, $\{0, \pi \}$, phase sequence of length~$N_c$~\cite{salehi_jlwt_1990,weiner_ol_1988}.
        Encoding via a pseudorandom binary phase-shift spreads the original photon-wavepacket pulse in the time domain by a factor of approximately $N_c$, consequently reduces the central peak on average by a factor of $N_c$ compared to the original pulse.
        That is $\mathbb E \{ \lvert \xi^{\ep}(\tp_0) \rvert^2 \} =  ( \nicefrac{1}{N_c})\lvert \xi(\tp_0) \rvert^2$ , where $\mathbb E$ denotes statistical expectation operation.
        The spreading and amplitude reduction is due to destructive interference between different spectral components.        
        Figure~\ref{fig:modem} shows that a spectral phase-shifting operator with a conjugate phase mask can reconstruct the encoded time spread quantum signal back to its original despread quantum pulse shape.
        Let us assume the output port~$\recv_0  \in{1,\ldots,\text{M}}$ of the star-coupler going to the receiver~$\recv_0$ of the network is attached to a conjugate spectral phase-shifting operator of sender~$\tran_0 \in{1,\ldots,\text{M}}$, which is $\hat{\text{U}} ^{\dagger}_{\tran_0}$.
        Considering Eq.~\eqref{eqn:Utheta psi}, this conjugate operator reshapes the photons' wavepackets existing in port $\recv_0$ of Eq.~\eqref{eqn:outsc}.
        As a result, this transformation reconstructs the photon with the wavepacket of $\xi_{\tran_0}$, i.e., $\hat a^{\dagger}_{\recv_0,\, \xi^{\ep_{\tran_0}}_{\tran_0}} \rightarrow \hat a^{\dagger}_{\recv_0, \, \xi_{\tran_0}} $.
        However, the spectral phase-shifts of photons that are not conjugate to phase mask $\tran_0$ remain random ($\hat a^{\dagger}_{\recv_0, \, \xi^{\ep_{\tran}}_{\tran}} \xrightarrow{\tran \neq \tran_0} \hat a^{\dagger}_{\recv_0, \, \xi^{\ep_{\tran} d_{\tran_0}}_{\tran}}$), where $\xi^{\ep_{\tran} d_{\tran_0}}_{\tran}$ indicates an improperly decoded quantum signal of sender~$\tran \neq \tran_0$.
        Its superscript, $\ep_{\tran} d_{\tran_0}$, shows the flow of the spectral phase-shifting process.
        Explicitly, the  $\tran$th encoding barcode, $\ep_{\tran}$, is followed by $\tran_0$th decoding barcode, $d_{\tran_0}$.
        The randomness of the final spectral phase-shift is because two consecutive and independent random phase-shifting masks amount to another random phase-shifting mask.
        However, if a random phase-shifting mask follows its conjugate mask results in a fixed phase-shift to all spectrum, $ \ep_{\tran_0} d_{\tran_0} =1 \ \  \forall \ \omega$, hence $\xi^{\ep_{\tran_0} d_{\tran_0}}_{\tran_0} = \xi_{\tran_0}$.
        
        In an ideal QCDMA network, the channel is symmetric, meaning all of the input and output ports are equivalent, and also, the light intensity of all the emitted signals are the same.
        In such a network, the correctly decoded wavepacket quantum light pulse, $\xi^{\ep_{\tran_0} d_{\tran_0}}_{\tran_0} = \xi_{\tran_0}$, sticks out above the rest and interacts with a following intended quantum device, or it can be detected by an ideal photodetector if needed.        
        Even a single-photon can interact with a quantum device as experimentally demonstrated for atoms~\cite{ritter_n_2012} and molecules~\cite{rezus_prl_2012}, provided that the wavepacket of the received photon overlaps with the spectral lineshape of the electronic transition of the quantum device.
        Therefore, the quantum device interacts with the properly decoded photon while it does not interact with improperly decoded photons.
        Likewise, since the properly decoded photon transforms back to a sharp pulse, it can be detected with an appropriate photodetector. On the other hand, other improperly decoded quantum signals appear as low-level background noise at the photodetectors' front-end.

\section{Example 1: QCDMA via Continuous Mode Coherent (Glauber) States}
        \begin{figure}[!t]
          \centering
          \includegraphics[width=\columnwidth]{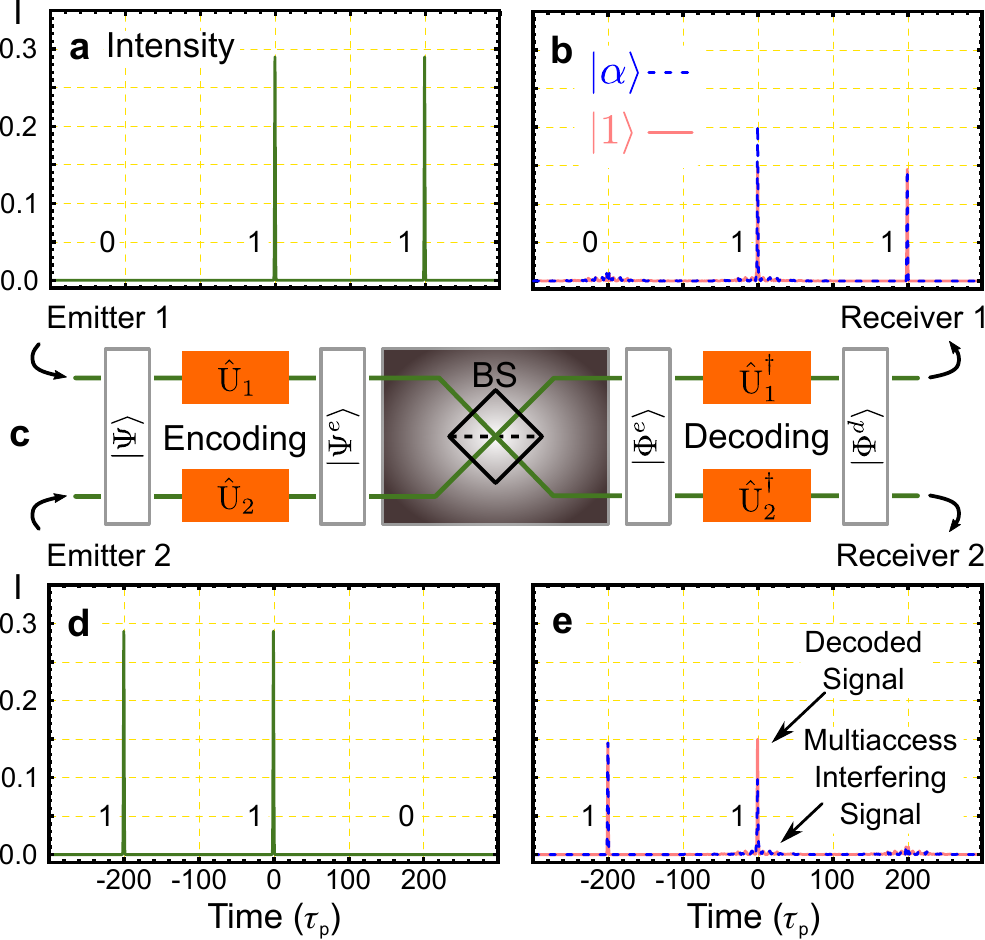}
         \caption{\textbf{A short-pulsed quantum CDMA network simulation.} \textbf{c} shows the schematic of the network.
           \textbf{a} and \textbf{d} show the sequence of pulses sent by transmitter~1 and~2, respectively.
         \textbf{b} and \textbf{e} are respectively, the detected signals by receiver~1 and receiver~2.
         The dashed blue lines are for the Glauber states inputs, and the solid red lines are for the Fock number states inputs.
       }\label{fig:cdma_coherent}
     \end{figure}
     
        Let us demonstrate the QCDMA model by a simulation example.
        We consider a network composed of users transmitting pulsed coherent (Glauber) light, $\lvert \psi \rangle =e^{-\frac{\lvert \alpha \rvert^2 }{2}} e^{\alpha  \hat a_{\xi}^{ \dagger}} \lvert 0 \rangle$, where the pulse intensity is $|\alpha|^2$.
        For binary on-off keying (OOK) signaling, binary one is represented by sending a Glauber light pulse with $\alpha =1$, which corresponds to single-photon average intensity, and binary zero with  $\alpha=0$, which corresponds to a vacuum state  $\lvert 0 \rangle$.
        Let us assume that each of the M transmitted quantum beams are randomly phase-shifted by a binary code
        and enter into a star-coupler.
        Equation~\eqref{eqn:outsc} expresses the output of the star-coupler.
        For input Glauber states, the pure quantum state of Eq.~\eqref{eqn:outsc}, reduces to a factorized state $\lvert \Phi^{\ep}  \rangle=\prod_{\recv =1}^{\text{M}}\lvert \phi^\ep_{\recv}  \rangle$, where $\lvert \phi^\ep_{\recv}  \rangle = \lvert \sum_{\tran =1}^{\text{M}}  B_{\recv  \tran} \alpha_{\tran} \xi^{\ep_{\tran}}_{\tran}\rangle$ is the coded Glauber state going to decoder and receiver~$\recv, \  \recv \in{1,\ldots,\text{M}}  $.
        The star-coupler output quantum state $\lvert \Phi^{\ep}  \rangle$ for input Glauber states does not show any entanglements between its constituent receivers signals~$\lvert \phi^\ep_{\recv}  \rangle$.
        The lack of entanglement at the M$\times$M quantum star-coupler's output makes the input Glauber states good candidates for studying quantum spectral encoding/decoding without tackling or addressing other effects of quantum mechanics such as entanglement and quantum interference~\cite{rezai_qst_2019}.
        \par
        To highlight and comprehend the main aspects of quantum signal processing embedded at the output of a QCDMA communication system, let us, for instance, study the light's output quantum state at quantum receive~$\recv_0$ (QR$_{\recv_0}$). First, note that the received quantum state at receiver~$\recv_0$ is a pure-state: $\rho^\ep_{\recv_0}=\text{Tr}_{\recv \neq \recv_0 } \lvert \Phi^{\ep}\rangle \langle \Phi^{\ep} \rvert= \lvert \phi^\ep_{\recv_0}\rangle \langle \phi^\ep_{\recv_0} \rvert$.
         Quantum receiver ${\recv_0}$ (QR$_{\recv_0}$), to decode the quantum signal of sender~$\tran_0$ (QT$_{\tran_0}$), spectrally phase-shifts incoming quantum light by the corresponding conjugated mask ($ \hat{\text{U}}^{\dagger}_{\tran_0}$), and therefore the quantum state at the output of decoder~$\hat{\text{U}}^{\dagger}_{\tran_0}$ becomes
        \begin{equation}
          \begin{split}
            \lvert \phi^{d}_{\recv_0}  \rangle&= \hat{\text{U}}^{\dagger}_{\tran_0}\lvert \phi^\ep_{\recv_0}  \rangle\\
            &= \Big \lvert B_{\recv_0  \tran_0} \, \alpha_{\tran_0} \xi_{\tran_0} + \sum^{\text{M}}_{\tran \neq \tran_0}  B_{\recv_0 \tran} \,\alpha_{\tran} \xi^{\ep_{\tran}  d_{\tran_0}}_{\tran}\Big \rangle \, .
          \end{split}
          \label{eqn:glaubephi_r0}
        \end{equation}
        The amplitude of the above Glauber state is composed of two terms.
        The first term, $ B_{\recv_0 \, \tran_0} \alpha_{\tran_0} \xi_{\tran_0}$, is proportional to the successfully decoded quantum pulse of the desired sender~$\tran_0$ (QT$_{\tran_0}$). 
        The second term, $\sum^{\text{M}}_{\tran \neq \tran_0}   B_{\recv_0 \, \tran} \alpha_{\tran} \xi^{\ep_{\tran} d_{\tran_0}}_{\tran}$, is the multiaccess interfering signal, corresponding to other users' transmitted quantum signals (QT$_{\tran \neq \tran_0}$) that are not properly decoded and behaves as the background noise at the front-end of the intended or the desired user's quantum receiver, i.e., QR$_{\recv_0}$.
        The strength of this multiaccess background noise depends strongly upon the number of active users M and the code length $N_c$ used in an operational QCDMA communication system.
        For example, on the one hand, increasing the number of users, M, would increase the interfering background multiaccess noise and degrade the system's performance. On the other hand, increasing the code length~$N_c$ weakens the multiaccess signal, thereby enhancing the QCDMA system's performance.
        An in-depth study of such a QCDMA system's performance is beyond the present paper's scope and will be discussed in a follow-up paper.
        \par
        Figure~\ref{fig:cdma_coherent} shows a two users QCDMA system with two quantum transmitters and two quantum receivers.
        The intensity measurements are at the quantum receivers' photodetectors end.
        We consider three scenarios, user one transmitting binary one when user two transmits binary zero.
        In scenario two, user one transmits binary zero, and user two sends binary one.
        In the third scenario, both users send binary one; this scenario is a worst-case scenario.
        Each user encodes its Glauber pulse by a distinct binary sequence of length $N_c=63$.
        The encoded quantum signals enter into the 2$\times$2 star-coupler and are distributed between two users' receivers.
        According to the state broadcasting equation~\eqref{eqn:outsc}, a 50/50 balanced star-coupler 
        with transform matrix
        \begin{equation}
          \mathbf{\underline B}=
            \begin{pmatrix}
              B_{11} & B_{12}\\
              B_{21} & B_{22} 
            \end{pmatrix}=
            \frac{1}{\sqrt 2 }
            \begin{pmatrix}
              1 & 1\\
              -1 & 1 
            \end{pmatrix}\,\,
            \label{eqn:2x2B}
          \end{equation}
        performs equal signal broadcasting.
        We assume receiver~1 and receiver~2 have the decoding mask~1 and~2, respectively, and with that, they apply the spectral phase-shift operation on their respective received quantum signal.
        The dashed blue lines in Fig.~\ref{fig:cdma_coherent}b and Fig.~\ref{fig:cdma_coherent}e show the intensity measurement of the decoded signals by receiver~1 and receiver~2, respectively.
        These figures demonstrate that receiver~1 receives OOK's sequence signal of transmitter~1 shown in Fig.~\ref{fig:cdma_coherent}a, and receiver~2 receives the transmitted signals by user~2 shown in Fig.~\ref{fig:cdma_coherent}d.
        Let us study the challenging scenario when both senders send binary one, $\alpha_\tran=\alpha_1 = \alpha_2 =1$.
        According to Eq.~\eqref{eqn:glaubephi_r0} for $\recv_0=\tran_0=1$, and as is shown in appendix~\ref{sec:qcdmaglauber}, the temporal shape of the intensity measured by receiver~1 would be
         \begin{equation}
          \begin{split}
            I_1(\tp)  
             &=\Big \lvert B_{1 1}  \xi_{1}(\tp) + \sum_{\tran=2}^{\text{M}}  B_{1 \tran}  \xi^{\ep_{\tran}  d_{1}}_{\tran}(\tp)\Big\rvert^2\\
            &=\frac{1}{2}\Big \lvert \xi_{1}(\tp)+ \xi^{\ep_2 d_1}_{2}(\tp)\Big \rvert^2\, ,            
          \end{split}
          \label{eqn:glauber_ir0=1}
        \end{equation}
        where $\xi^{\ep_2 d_1}_{2}(\tp)$ indicates that the encoded photon-wavepacket of transmitter~2 ($\xi^{\ep_2}_{2}(\tp)$) is spectrally phase-shifted by the decoder of transmitter~1 ($\hat{\text{U}}^{\dagger}_1$).
         The receiver structure in Eq.~\eqref{eqn:glauber_ir0=1} is, in optical communications~\cite{gagliardi_book_1995}, well recognized as a coherent detection scheme where both signals, first, add-in their complex amplitude (signals with amplitude and phase) and then photo-detected by absolute square. A well-known example is Michelson's interference experiment.
         In QCDMA, if we sample the light instantaneous-intensity~\eqref{eqn:glauber_ir0=1} at the photon-wavepackets' central time~$\tp_0$ (i.e., $ I_1(\tp_0)$); since $\xi_{1}(\tp_0)=1$ for a peak-normalized pulse, and $\xi^{\ep_2 d_1}_{2}(\tp_0)$ is a Gaussian random variable 
         with mean zero and the variance of approximately~$1/N_c$; the average instantaneous-intensity of receiver~1 at the sampling time~$\tp_0$ becomes $ \mathbb E \{I_{1}(\tp_0)\}=\frac{1}{2} \mathbb E \{\lvert \xi_{1}(\tp_0)+ \xi^{\ep_2 d_1}_{2}(\tp_0) \rvert^2 \} =\frac{1}{2}(1+\frac{1}{N_c}) $. 
         \par
          For the intensity of the second receiver, $I_2(\tp)$, the same argument as above is also valid, however the signals add in with a minus sign, $I_2(\tp)=\frac{1}{2}\lvert \xi_{2}(\tp)- \xi^{\ep_1 d _{2} }_{1} (\tp) \rvert^2$.
         Consequently, the average intensity at the central time~$\tp_0$ is
         $ \mathbb E \{I_{2}(\tp_0)\}=\frac{1}{2} \mathbb E \{\lvert \xi_{2}(\tp_0)- \xi^{\ep_1 d_2}_{1}(\tp_0) \rvert^2 \} =\frac{1}{2}(1+\frac{1}{N_c}) $.

\section{Example 2: QCDMA via Continuous Mode Photon Number (Fock) States}

        Let us present a QCDMA network in which the quantum transmitters send their corresponding quantum signals via a continuous mode number state, specifically transmitter~$\tran$ sends $\lvert \psi_{\tran} \rangle= \lvert n_{\xi_\tran}\rangle =\nicefrac{1}{\sqrt{n_{\tran}!}} \, (\hat a_{\xi_{\tran}}^{\dagger})^{n_{\tran}}\lvert 0 \rangle$.
        The transmitters spectrally encode their number state light pulses, which transforms their photon-wavepacket from $\xi_{\tran}$ to $\xi^{\ep_{\tran}}_{\tran}$; therefore, the encoded quantum state of transmitter~$\tran$ is $\lvert \psi^{\ep}_{\tran}\rangle = \lvert n_{\xi^{\ep_\tran}_{\tran}}\rangle = \nicefrac{1}{\sqrt{n_{\tran}!}} \, (\hat a_{\xi^{\ep_\tran}_{\tran}}^{\dagger})^{n_{\tran}} \lvert 0\rangle$.
        For a network of M quantum transmitters and M quantum receivers, the encoded signals enter into an M$\times$M star-coupler.
        Quantum state vector~$\lvert \Phi^{\ep}\rangle$ of the signals at the star-coupler's output (Eq.~\eqref{eqn:outsc}) and before entering the spectral QCDMA decoding operators is
        \begin{equation}
          \begin{split}
            \lvert \Phi^{\ep}  \rangle = \prod_{\tran=1}^{\text{M}}  \frac{1}{\sqrt{n_{\tran}!}} \left(\sum^{\text{M}}_{\recv=1} B_{\recv \tran} \hat a_{\recv,\xi^{\ep_{\tran}}_{\tran}}^{\dagger} \right )^{n_{\tran}}  \lvert 0\rangle \, .
          \end{split}
          \label{eqn:Phifockmain}
        \end{equation}        
        As an example, let us consider a $2\times2$ network, where transmitters transmit a continuous mode single-photon $\lvert 1_{\xi_{\tran}}\rangle= \hat a_{\xi_{\tran}}^{\dagger} \lvert 0\rangle$ pulse.
        The output state (Eq.~\eqref{eqn:Phifockmain}) for a 50/50 balanced star-coupler
        with the above transformation matrix $ \mathbf{\underline B}$ (i.e., Eq.~\eqref{eqn:2x2B}) reduces to
        \begin{equation}
          \begin{split}
            \lvert \Phi^{\ep} \rangle            
            &=            
            B_{1 1}  B_{1 2} \lvert (1_{\xi^{\ep_1}_{1}},1_{\xi^{\ep_2}_{2}})\rangle \lvert 0\rangle
            +B_{1 1} B_{2 2} \lvert 1_{\xi^{\ep_1}_{1}}\rangle\lvert 1_{\xi^{\ep_2}_{2}}\rangle \\
           & \ \  +B_{2 1}  B_{1 2} \lvert  1_{\xi^{\ep_2}_{2}} \rangle \lvert 1_{\xi^{\ep_1}_{1}}\rangle
           +B_{2 1} B_{2 2}  \lvert 0 \rangle \lvert (1_{\xi^{\ep_1}_{1}}, 1_{\xi^{\ep_2}_{2}}) \rangle\\
            &= \frac{1}{2} \Big( \lvert (1_{\xi^{\ep_1}_{1}},1_{\xi^{\ep_2}_{2}})\rangle \lvert 0\rangle
            + \lvert 1_{\xi^{\ep_1}_{1}}\rangle\lvert 1_{\xi^{\ep_2}_{2}}\rangle \\
           &\ \  - \lvert  1_{\xi^{\ep_2}_{2}} \rangle \lvert 1_{\xi^{\ep_1}_{1}}\rangle
           -  \lvert 0 \rangle \lvert (1_{\xi^{\ep_1}_{1}}, 1_{\xi^{\ep_2}_{2}}) \rangle \Big)\, ,
         \end{split}
       \end{equation}
       where the first term, $\lvert (1_{\xi^{\ep_1}_{1}},1_{\xi^{\ep_2}_{2}})\rangle \lvert 0\rangle$, indicates that both transmitter's signals end up in the first receiver and the second receiver receives a vacuum state~$\lvert 0\rangle$.
       The second term, $\lvert 1_{\xi^{\ep_1}_{1}}\rangle\lvert 1_{\xi^{\ep_2}_{2}}\rangle$, is when the transmitter~1(2) signal goes to receiver~1(2).
       The third term, $ \lvert  1_{\xi^{\ep_2}_{2}} \rangle \lvert 1_{\xi^{\ep_1}_{1}}\rangle$, is the opposite of the second term and shows when the signal of transmitter~1(2) goes to receiver~2(1).
       For the last term, $ \lvert 0 \rangle \lvert (1_{\xi^{\ep_1}_{1}}, 1_{\xi^{\ep_2}_{2}}) \rangle$, as opposed to the first term, both signals end up in receiver~2, and receiver~1 receives a vacuum state.
       It is worth noting that if the wavepackets of the input photons are identical, $\xi^{\ep_1}_{1}=\xi^{\ep_2}_{2}=\xi$, the Hong-Ou-Mandel~\cite{hong_prl_1987} destructive interference between the second and the third terms transforms the output state of the $2\times2$ star-coupler into $ \lvert \Phi^{\ep} \rangle = \frac{1}{\sqrt 2} \Big( \lvert 2_{\xi}\rangle \lvert 0\rangle  -  \lvert 0 \rangle \lvert 2_{\xi} \rangle \Big)$, where $ \lvert 2_{\xi}\rangle= \frac{1}{\sqrt{2}} \hat a ^{\dagger 2}_{\xi}\lvert 0\rangle$ .
       \par
       Let us assume, in an M$\times$M network, receivers~$1,2,\hdots,\text{M}$ intend to receive the quantum signals sent by transmitters~$1,2,\hdots,\text{M}$, respectively (as shown in Fig.~\ref{fig:cdma}, for more general case see appendix~\ref{sec:qcdmafock}.).
       Therefore, the receivers spectrally phase-shift their received quantum light by the conjugate of the corresponding transmitter's encoder.       
       This decoding reconstructs the photon-wavepacket of the desired signal back into a sharp pulse and transforms quantum state~$\lvert \Phi^{\ep} \rangle$ (Eq.~\eqref{eqn:Phifockmain}) into the following state
       \begin{equation}
          \begin{split}
            \lvert \Phi^{d} \rangle &=\hat{\textbf{U}}^ \dagger \, \ \lvert \Phi^{\ep}  \rangle\\
             &=   \prod_{\tran=1}^{\text{M}}  \frac{1}{\sqrt{n_{\tran}!}} \left(  B_{\tran \tran} \hat a_{\tran ,\xi_{\tran}}^{\dagger} + \sum^{\text{M}}_{\recv \neq \tran} B_{\recv \tran} \hat a_{\recv,\xi^{\ep_{\tran} d_{\recv}}_{\tran}}^{\dagger} \right )^{n_{\tran}}  \lvert 0\rangle 
           \end{split}
           \label{eqn:PhidFockmain}
        \end{equation}               
        And for example, if receiver~1 measures its decoded light intensity, it would give
        \begin{equation}
          I_{1}(\tp) =\frac{1}{\text{M}}\left(n_1 \lvert\xi_{1}(\tp)\rvert^2 + \sum_{\tran=2}^{\text{M}}  n_{\tran} \lvert\xi^{\ep_{\tran} d_{1}}_{\tran} (\tp)\rvert^2 \right)\, ,
          \label{eqn:fock_ir0=1}
        \end{equation}
        for a balanced star-coupler, $\lvert B_{\recv \tran} \rvert^2= 1/\text{M}$.
        For a $2\times2$ network with single-photon inputs ($n_1=n_2=1$), Eq.~\eqref{eqn:fock_ir0=1} reduces to $I_1(\tp) = \frac{1}{2} \left(|\xi_{1}(\tp)|^2+ | \xi^{\ep_2 d_1}_{2}(\tp) |^2 \right)$.               
        Interestingly, the received intensity of Eq.~\eqref{eqn:fock_ir0=1} is fundamentally different from the received intensity of Eq.~\eqref{eqn:glauber_ir0=1}, where the input to the receiver is based on Glauber states.
        The received intensity in  Eq.~\eqref{eqn:fock_ir0=1}  is known as an incoherent detection scheme as opposed to Eq.~\eqref{eqn:glauber_ir0=1}, where the received intensity is based on a coherent detection scheme.
        In the incoherent detection scheme, the output signal intensity $I_1(\tp)$ is based on the sum of the individual input signals' intensities into the photodetector.
        In incoherent detection, there is no inter-signal interference amongst the input signals into the photodetector as if the signals were initially incoherent, which is not the case at all.
        As a matter of fact, the number states generated by the quantum transmitters are assumed to be entirely coherent, and that is why we can apply encoding/decoding phase operators to alter the phase of their frequency components and thereby changing their temporal shape from a peaked pulse into a random-looking time spread signal and vice versa.
        This fundamental difference in the detection scheme between the number states and Glauber states is a consequence of Heisenberg's uncertainty principle.       
        Specifically, since number state electromagnetic wave~$\lvert n \rangle$  has definite amplitude, its corresponding phase angle at the time of measurement is equally likely to have any random value between~$0$ and~$2\pi$.
        These arising pure randomness in quantum phase angles of number states at the time of measurement are the root cause in changing the Glauber states' coherent detection into incoherent detection of number states even though we are, in both examples, using the same setup for detecting their corresponding input signals.
        Therefore, we can conclude that due to Heisenberg's uncertainty principle, the quantum phase fluctuation at the time of measurement in number states alters the behavior or the functionality of the receiver structure from a coherent detection scheme for Glauber states into the incoherent detection scheme for number states.

        Consider the simulation of QCDMA via number states~$\lvert n_{\xi_{\tran}} \rangle $ for the corresponding OOK's sequences shown in Fig.~\ref{fig:cdma_coherent}a and Fig.~\ref{fig:cdma_coherent}d.
        For binary one, transmitters send a single-photon state, $n_{\tran}=1$, and for binary zero, a vacuum state, $n_{\tran}=0$, is transmitted.
        Like QCDMA via continuous mode Glauber states, the intensity measurement of the decoded signals by receiver~1 and receiver~2 is shown in Fig.~\ref{fig:cdma_coherent}b and Fig.~\ref{fig:cdma_coherent}e, respectively, but with solid red lines.
        The difference between the solid red lines for number states and the dashed blue lines for the Glauber states is due to the inter-signal interference $\text{Re} (| \xi_{1}(\tp)  \xi^{\ep_2 d_1 \star}_{2}(\tp) |)$ that appears in Eq.~\eqref{eqn:glauber_ir0=1}, $ I_1(\tp)= \frac{1}{2} (| \xi_{1}(\tp) |^2+ |  \xi^{\ep_2 d_1}_{2}(\tp) |^2 )  +\text{Re}( \xi_{1}(\tp)  \xi^{\ep_2 d_1 \star}_{2}(\tp))$ for Glauber states but, because of the complete phase uncertainty, vanishes for number states in Eq.~\eqref{eqn:fock_ir0=1}, $I_1(\tp) = \frac{1}{2} (|\xi_{1}(\tp)|^2+ | \xi^{\ep_2 d_1}_{2}(\tp) |^2 )$.       
The inter-signal interference effect on the light intensity is opposites for receiver~1 and receiver~2. 
        This interference effect can be reduced by increasing code length~$N_c$ and choosing orthogonal codes. 
        
\section{Conclusion}
        This paper introduces novel quantum code division multiple-access (QCDMA) networks based on spectrally encoding/decoding input quantum light pulses.
        Our study and mathematical model can treat any input pure quantum state of light, such as Glauber state, number state, and squeezed state.
        In other words, our mathematical development uses a unified approach applicable for any input pure quantum state with an arbitrary spectral wavepacket.
        \par
        Two limiting cases of quantum signals, namely particle-like number (Fock) states and wave-like coherent (Glauber) states, are detailed as examples for the quantum transmitters' signals.
        Quantum features like entanglement, and quantum interference appears amongst the quantum signals propagating to different quantum receivers for number states inputs.
        However, for coherent state inputs, the quantum state vector of receivers is factorized.
        Furthermore, a quantum receiver's received-intensities would be different in the QCDMA via number states and the corresponding QCDMA via Glauber states, by an amount called inter-signal interference.
        We discussed that the above difference in the received intensities at the time of measurement is due to Heisenberg's uncertainty principle.
        In other words, because of the complete quantum phase uncertainty of particle-like single-photons, there is no inter-signal interference; however, the inter-signal interference emerges for wave-like Glauber states.        
        \par
        It is worth noting that in this paper, the encoding and decoding operations apply to the spectrum of photons; however, the mathematical model is easily extendable to temporal encoding and decoding of quantum light states, namely direct sequence QCDMA.

        \appendices
  \renewcommand{\theequation}{\thesection.\arabic{equation}}
  \section{\textbf{Continuous Mode State of a Quantum Light Pulse}}
\label{app:I(t)}
         We define quantum light pulses as quantum states composed of identical pulsed-shaped single-photons, where their quantum field creation operator is
        \begin{subequations}
          \begin{align}
            \hat a^ \dagger _{\xi}
            &=\int  d\omega \,  \xi  (\omega) \hat a ^\dagger (\omega) =  \int  d\tp \,  \xi  (\tp) \hat a ^\dagger (\tp)\, ,\\         
        \intertext{which is known as the photon-wavepacket creation operator~\cite{loudon_2000}, and then the corresponding photon-wavepacket annihilation operator reads}
            \hat a _{\xi}
            &=\int  d\omega \,  \xi^{\star}  (\omega) \hat a (\omega) =  \int  d\tp \,  \xi^{\star}  (\tp) \hat a (\tp)\, ,
          \end{align}
          \label{eqn:axi}
        \end{subequations}
        where photon-wavepacket $\xi(\omega)$ indicates the spectral amplitude, and
        $\xi  (\tp)$ represents the temporal amplitude of the photon-wavepacket.
        $  \hat a ^\dagger (\omega)$ and $\hat a (\omega)$ denote continuous mode creation and destruction operators in the frequency domain, respectively, and $ \hat a^{\dagger} (\tp) $ and $ \hat a(\tp)$ represent the corresponding creation and destruction operators in the time domain.
        Since $ \hat a (\tp) $ is the Fourier transform of $ \hat a(\omega)$
        \begin{equation}
          \begin{split}
            \hat a  (\tp)& =\frac{1}{\sqrt{2 \pi}} \int d \omega \,  \hat a (\omega) e^{-i \omega \tp} \, ,\\
          \end{split}
          \label{eqn:fta}
        \end{equation}
        from Eq.~\eqref{eqn:axi}, one can deduce that the temporal amplitude of the photon-wavepacket denoted by $ \xi (\tp)$ is also the Fourier transform of $\xi(\omega)$, that is
        \begin{equation}
          \begin{split}
            \xi (\tp)& =\frac{1}{\sqrt{2 \pi}} \int d \omega \,  \xi (\omega) e^{-i \omega \tp} \, .
          \end{split}
          \label{eqn:ftxi}
        \end{equation}

         Now, we proceed further and consider a pure quantum state of light composed of photons with the same photon-wavepacket that can generally be expressed by a function of creation operator ($f(\hat a_ \xi ^ \dagger)$):
        \begin{subequations}
          \begin{align}             \lvert \psi \rangle &= \sum_n c_n \lvert n_{\xi} \rangle =f(\hat a^ \dagger _{\xi}) \lvert 0 \rangle \\
            \intertext{and}
            \langle \psi \rvert &= \sum_n \langle n_{\xi} \rvert c^{\star}_n  = \langle 0 \rvert f^{\star}(\hat a _{\xi}) \, .
          \end{align}
          \label{eqn:si=f(adag)app}
        \end{subequations}
        Function~$f(\hat a_ \xi ^ \dagger)$ is an analytic, arbitrary differentiable function, then representable by the Taylor series:
        \begin{equation}
          f(\hat a_ \xi ^ \dagger)= f(0) +f' (0) \hat a_ \xi ^ \dagger + \frac {f'' (0)}{2 !} \hat a_ \xi ^{\dagger^2} + \frac{f'''(0)}{3!}\hat a_ \xi ^ {\dagger^3} \dots \, ,
          \label{eqn:exoff}
        \end{equation}
        comparing with Eq.~\eqref{eqn:si=f(adag)app}, one can write
        \begin{equation}
          c_n=  \frac {f^{(n)} (0)}{\sqrt{n !}} \, ,
        \end{equation}
        where $f^{(n)} (0)$ is the $n$th derivative of function $f(x)$ with respect to~$x$ at~$x=0$\,.

        As Eq.~\eqref{eqn:axi} indicates, operator $\hat a_ \xi $ is a linear combination of operators $ \hat a(\tp)$ or $ \hat a(\omega)$.
        To derive Eq.~\eqref{eqn:[f,a]}, we first obtain the commutator of operators $\hat a(\tp)$ and $ \hat a_ \xi ^ \dagger $.
        By first noting that $[\hat a(\tp) , \hat a^ \dagger(\tp')]= \delta(\tp- \tp')$, one can write~\cite{blow_pr_1990}
        \begin{equation}
          \begin{split}
            [\hat a(\tp), \hat a_ \xi ^ \dagger]
            & = \int  d\tp' \,  \xi  (\tp')  [\hat a(\tp) , \hat a^ \dagger(\tp')] \\
            &=  \int  d\tp' \,  \xi  (\tp')  \delta(\tp- \tp')\\
            &= \xi (\tp) \, .
          \end{split}
          \label{eqn:[a,axi]}
        \end{equation}
        Using Eq.~\eqref{eqn:[a,axi]} and the commutator relation~$[\hat A, \hat B \hat C]=\hat B[\hat A, \hat C]+[\hat A, \hat B] \hat C$, one can find the commutator of operators $\hat a(\tp)$ and $ \hat a_ \xi ^{\dagger n}$, through the following recursive sequence:	
        \begin{equation}
          \begin{split}        
            [\hat a(\tp), \hat a_ \xi ^ {\dagger  2}] & = \hat a_ \xi ^ {\dagger} [\hat a(\tp), \hat a_ \xi ^ \dagger]+ [\hat a(\tp), \hat a_ \xi ^ \dagger] \hat a_ \xi ^ {\dagger} = 2 \xi(\tp) \hat a_ \xi ^ {\dagger} \, ,\\
            [\hat a(\tp), \hat a_ \xi ^ {\dagger  3}] & = \hat a_ \xi ^ {\dagger} [\hat a(\tp), \hat a_ \xi ^ {\dagger 2}]+ [\hat a(\tp), \hat a_ \xi ^ \dagger] \hat a_ \xi ^ {\dagger 2} = 3 \xi(\tp) \hat a_ \xi ^ {\dagger 2}\, , \\
            \vdots\\
            [\hat a(\tp), \hat a_ \xi ^ {\dagger  n}] & = \hat a_ \xi ^ {\dagger} [\hat a(\tp), \hat a_ \xi ^ {\dagger {n-1}}]+ [\hat a(\tp), \hat a_ \xi ^ \dagger] \hat a_ \xi ^ {\dagger {n-1}} = n \xi(\tp) \hat a_ \xi ^ {\dagger {n-1}}\, .
          \end{split}
          \label{eqn:[at,a^n]}
        \end{equation}
        Considering the series expansion of function $f(\hat a_ \xi ^ \dagger)$ (Eq.~\eqref{eqn:exoff}) and the linearity property of commutators, that is $[\hat A, \hat B + \hat C + \hdots]=[\hat A , \hat B]+[\hat A, \hat C]+\hdots $, the commutator of operators $\hat a(\tp)$ and $f(\hat a_ \xi ^ \dagger)$ is:
        \begin{equation}
          \begin{split}
            [\hat a(\tp),f(\hat a_ \xi ^ \dagger)] &= [ \hat a(\tp) , f(0) +f' (0)  \hat a_ \xi ^ \dagger + \frac {f'' (0)}{2 !}  \hat a_ \xi ^ {\dagger^2}\\
            &\hspace{10em}+ \frac{f'''(0)}{3!}  \hat a_ \xi ^ {\dagger^3} ....]\\
            & = \xi(\tp) (f' (0) + \frac {f'' (0)}{2 !}   2 \hat a_ \xi ^ {\dagger} + \frac{f'''(0)}{3!}  3 \hat a_ \xi ^ {\dagger^2} ....]\\
            & = \xi(\tp) (f' (0) + \frac {f'' (0)}{1 !}  \hat a_ \xi ^ {\dagger} + \frac{f'''(0)}{2!}  \hat a_ \xi ^ {\dagger^2} ....]\\
            & = \xi(\tp) f'(\hat a_ \xi ^ {\dagger}) \, ,
          \end{split}
          \label{eqn:[f,a]-prove}
        \end{equation}
        and its complex conjugate, using the equality $\left(f(\hat a_{\xi}^{
        \dagger})\right)^{\dagger}=f^{\star}(\hat a_{\xi})$, gives
        \begin{equation}
          \begin{split}
            [f^{\star}(\hat a_ \xi),\hat a^{\dagger}(\tp)] = \xi^{\star}(\tp) f'^{\star}(\hat a_ \xi) \, .
          \end{split}
        \end{equation}
        Using Eq.~\eqref{eqn:[f,a]-prove} and the equality of $\hat A \hat B =[\hat A , \hat B] + \hat B \hat A$, one can show 
        \begin{equation}
          \begin{split}
            \hat a(\tp) \lvert \psi \rangle & =  \hat a(\tp) f(\hat a^ \dagger _{\xi}) \lvert 0\rangle\\
            & =  \left([\hat a (\tp), f(\hat a^ \dagger _{\xi}) ] + f(\hat a^ \dagger _{\xi}) \hat a(\tp) \right) \lvert 0\rangle\\
            & =  \xi (\tp)f'(\hat a^ \dagger _{\xi}) \lvert 0\rangle\, ,
          \end{split}
          \label{eqn:apsi}
        \end{equation}
        \noindent
        since $ \hat a(\tp) \lvert 0\rangle=0$.
        If $f' \neq 0 $ (i.e., $f \neq 1$ or $\lvert \psi \rangle \neq \lvert 0 \rangle$), the intensity at time~$\tp$ (see Eq.~\eqref{eqn:I(t)}) is proportional to  the amplitude of the wavepacket at time~$\tp$, that is
        \begin{equation}
          \begin{split}
            I(\tp) = \bar I  \lvert \xi (\tp) \rvert^2 \, ,
          \end{split}
          \label{eqn:Ixiprove}
        \end{equation}
        where $ \bar I = \langle \psi' \vert \psi' \rangle$ corresponds to the mean intensity (the mean photon number) and $\lvert \psi' \rangle = f'(\hat a^ \dagger _{\xi}) \lvert 0\rangle$ corresponds to the state $\lvert \psi \rangle$ when a photon is removed from it (in a practical sense by a photodetector, for example); mathematically speaking $\lvert \psi' \rangle =\hat a_{\xi} \lvert \psi \rangle$.

     \vspace{5mm}        
        \noindent
        {\textbf{Multi Photon-Wavepackets Quantum State}}
        \par
        Let us assume a pure quantum state of light, where its composed photons can occupy M different wavepackets ($\xi_1, \xi_2, \hdots, \xi_{\text{M}}$), which are not necessarily orthogonal.
        Therefore, considering Eq.~\eqref{eqn:axi}, their corresponding field operators may not commute
        \begin{equation}
          \begin{split}
            [\hat a_{\xi_j}, \hat a^ \dagger_{\xi_k}]
            &=\int \int  d\omega'  d\omega \,    \xi^{\star}_j  (\omega') \xi_k  (\omega)  [\hat a (\omega'), \hat a ^\dagger (\omega)]\\
            &=\int \int  d\omega'  d\omega \,    \xi^{\star}_j  (\omega') \xi_k  (\omega) \delta (\omega - \omega')\\
            &=\int  d\omega \,    \xi^{\star}_j  (\omega) \xi_k  (\omega) \\ 
            &= \langle \xi_j \vert \xi_k\rangle\, .
          \end{split}
          \label{eqn:aiadj}
        \end{equation}
        The quantum state vector of light with such M photon-wavepackets can be expressed as
        \begin{equation}
          \begin{split}
            \lvert \psi \rangle &= \sum_{n_1} \sum_{n_2}\hdots \sum_{n_{\text{M}}} c_{n_1 n_2 \hdots n_{\text{M}}} \lvert n_{\xi_1},  n_{\xi_2}, \hdots, n_{\xi_{\text{M}}} \rangle  \\
            &= \sum_{n_1} \sum_{n_2}\hdots \sum_{n_{\text{M}} } c_{n_1 n_2 \hdots n_{\text{M}}}\frac{1}{\sqrt{n_1! n_2!\hdots n_{\text{M}}!}}\\
            & \hspace{11em} \hat a^{\dagger n_1}_{\xi_1} \hat a^{\dagger n_2}_{\xi_2}\hdots \hat a^{\dagger n_\text{M}}_{\xi_{\text{M}}}\lvert 0\rangle \\
            &=f(\hat a^ \dagger _{\xi_1}, \hat a^ \dagger _{\xi_2},\hdots, \hat a^{\dagger}_{\xi_{\text{M}}}) \lvert 0 \rangle \, .
          \end{split}
        \end{equation}
        Considering Taylor polynomials of the corresponding multivariable function $f(z_1,z_2,\hdots,z_{\text{M}})$, one can write
        \begin{equation}
          c_{n_1 n_2 \hdots n_{\text{M}}}= \prod^{\text{M}}_{m=1} \frac{1}{\sqrt{n_{m}!}}  \frac{\partial^{ n_{m}}}{\partial z_m^{n_{m}}} f(z_1,z_2,\hdots,z_{\text{M}}) \Big \lvert_{z_1,z_2,\hdots,z_{\text{M}}\rightarrow 0}
        \end{equation}
        Note that if the wavepackets are orthogonal, that is $\langle \xi_j \vert \xi_k\rangle =\delta_{jk} $, the Fock states $\lvert n_{\xi_1},  n_{\xi_2}, \hdots, n_{\xi_{\text{M}}} \rangle$
        form a complete basis to represent  the light's quantum state.
        It is an overcomplete basis if the wavepackets are not orthogonal. 

        Like Eq.~\eqref{eqn:[a,axi]}-\eqref{eqn:Ixiprove}, to calculate the intensity at time~$\tp$, one needs to know the commutator of operators $\hat a ^{\dagger}(\tp)$ and $f(\hat a^ \dagger _{\xi_1}, \hat a^ \dagger _{\xi_2},\hdots, \hat a^{\dagger}_{\xi_{\text{M}}})$.
        Because of the commutators' linearity property, we first calculate the commutator of operator $ \hat a ^{\dagger}(\tp)$ and various power functions of field creation operators.
        Using Eq.~\eqref{eqn:[at,a^n]}, that is $[\hat a(\tp), \hat a_ \xi ^ {\dagger  n}] =n \xi(\tp) \hat a_ \xi ^ {\dagger {n-1}}$, one can write
        \begin{equation}
          \begin{split}        
            [\hat a(\tp), \hat a_{\xi_1} ^ {\dagger  n_1} \hat a_{\xi_2} ^ {\dagger  n_2}]&=   [\hat a(\tp), \hat a_{\xi_1} ^ {\dagger  n_1} ] \hat a_{\xi_2} ^ {\dagger  n_2}+  \hat a_{\xi_1} ^ {\dagger  n_1} [\hat a(\tp), \hat a_{\xi_2} ^ {\dagger  n_2}]\\
            &=n_1 \xi_1(\tp) \hat a_ {\xi_1} ^ {\dagger {n_1-1}}  \hat a_{\xi_2} ^ {\dagger  n_2} + n_2 \xi_2(\tp) \hat a_{\xi_1} ^ {\dagger {n_1}}  \hat a_{\xi_2} ^ {\dagger  n_2 -1}  \, ,
          \end{split}
        \end{equation}
        and by the recursive process, one can show
        \begin{equation}
          \begin{split}        
            [\hat a(\tp), \hat a_{\xi_1} ^ {\dagger  n_1} \hat a_{\xi_2} ^ {\dagger  n_2}\hdots \hat a_{\xi_{\text{M}}} ^ {\dagger  n_{\text{M}}}]&=\sum_{m=1}^{\text{M}} \Big( \xi_m(\tp)  \\
            & \times n_m\hat a_ {\xi_1} ^ {\dagger {n_1}}  \hat a_{\xi_2} ^ {\dagger  n_2} \hdots\hat a_{\xi_m} ^ {\dagger  n_m-1} \hdots\hat a_{\xi_{\text{M}} } ^ {\dagger  n_{\text{M}} } \Big) \, .
          \end{split}
        \end{equation}
        Considering the power series expansion of multivariable function~$f$,
        one can show the commutator of operators~$\hat a(\tp)$ and $f$ is
        \begin{equation}
          \begin{split}
            [\hat a(\tp),f(\hat a_ {\xi_1} ^ \dagger, \hat a_ {\xi_2} ^ \dagger, \hdots,  \hat a_ {\xi_{\text{M}}} ^ \dagger)]  =\sum_{m=1}^{\text{M}} \xi_m(\tp) \frac{\partial}{\partial \hat a_ {\xi_m} ^ {\dagger}}f(\hat a_ {\xi_1} ^ \dagger, \hat a_ {\xi_2} ^ \dagger, \hdots,  \hat a_ {\xi_{\text{M}}} ^ \dagger)\, .
          \end{split}
          \label{eqn:[a,fmv]}
        \end{equation}                
\section{ \textbf{Spectral Phase Encoding and Decoding Operators}}
\label{sec:pso}

        A simple phase-shifting operator by a phase value of $\theta$  is representable as follows~\cite{leonhardt_book_1997}
        \begin{equation}
          \hat{\text{U}}= e^{-i \theta \hat n} = e^{-i \theta  \hat a^{\dagger} \hat a} \, ,
        \end{equation}
        where $\hat n =  \hat a^{\dagger} \hat a $ is the number operator.        
        In spectrally encoding/decoding QCDMA, since the phase-shifting depends on the frequency, the operator would be
        \begin{equation}
          \begin{split}
            \hat{\text{U}} & = e^{-i  \sum_{\omega} \, \theta(\omega) \hat a ^{\dagger} (\omega) \hat a (\omega) }\\
            & =\prod_{\omega} e^{-i \, \theta(\omega) \hat a ^{\dagger} (\omega) \hat a (\omega) } \\
            & =\prod_{\omega}  \hat{\text{U}} (\omega)  \, ,
          \end{split}
          \label{eqn:U(omeaga)}
        \end{equation}
        where $ \hat{\text{U}} (\omega) =  e^{-i \, \theta(\omega) \hat a ^{\dagger} (\omega) \hat a (\omega) }$ denotes the phase-shifting operator at frequency~$\omega$. 
        Since different frequency field operators commute ($[\hat a (\omega),\hat a ^{\dagger} (\omega')]=\delta(\omega- \omega')$), the transformation of field creation operator $ \hat a ^{\dagger} (\omega)$ by operator $\hat{\text{U}}$ reduces to the transformation by its corresponding operator~$\hat{\text{U}}(\omega)$, that is
        \begin{equation}
          \begin{split}
            \hat{\text{U}} \hat a^\dagger (\omega) \hat{\text{U}}^ \dagger &=  \hat{\text{U}}(\omega) \hat a^\dagger (\omega) \hat{\text{U}}^ \dagger (\omega)\\
            &= e^{-i \, \theta(\omega) \hat a ^{\dagger} (\omega) \hat a (\omega) } \hat a^{\dagger} (\omega) e^{i \, \theta(\omega) \hat a ^{\dagger} (\omega) \hat a (\omega) }\, ,
            \end{split}
          \label{uau->u(omega)}
        \end{equation}
        and now considering Baker-Hausdorff lemma one can expand Eq.~\eqref{uau->u(omega)} as      
        \begin{equation}
          \begin{split}
            \hat{\text{U}} \hat a^\dagger (\omega) \hat{\text{U}}^ \dagger  &= \hat a^{\dagger} (\omega) + (-i \theta (\omega))[\hat a ^{\dagger} (\omega) \hat a (\omega),\hat a ^{\dagger} (\omega)]\\
            &+ \frac{(-i \theta (\omega))^2}{2!}[\hat a ^{\dagger} (\omega) \hat a (\omega),[\hat a ^{\dagger} (\omega) \hat a (\omega), \hat a ^{\dagger} (\omega) ]]\\
            &+ \hdots\\
            &+\frac{ (-i\theta (\omega))^n}{n!}[\hat a ^{\dagger} (\omega) \hat a (\omega),[\hat a ^{\dagger} (\omega) \hat a (\omega),[\hat a ^{\dagger} (\omega) \hat a (\omega),\\
            &\ \ \  \ \ \hdots [\hat a ^{\dagger} (\omega) \hat a (\omega),\hat a ^{\dagger} (\omega) ]]] \hdots]\\
            &+\hdots  \\
            &= \hat a^{\dagger} (\omega) + (-i\theta (\omega))\hat a ^{\dagger} (\omega) + \frac{ (-i\theta (\omega))^2}{2!}\hat a ^{\dagger} (\omega)
            \\ &+ \hdots +\frac{ (-i\theta (\omega))^n}{n!}\hat a ^{\dagger} (\omega) +\hdots  \\
            &=  \hat a ^{\dagger} (\omega) e^{-i \theta (\omega)}\, .
          \end{split}
          \label{eqn:UadUd}
        \end{equation}
        Therefore, operator $\hat{\text{U}}$ phase-shifts each frequency creation operator $ \hat a ^{\dagger} (\omega)$ when acting upon them, and hence the name: spectral phase-shifting operator.                
        Applying Eq.~\eqref{eqn:UadUd} to Eq.~\eqref{eqn:axi} gives
        \begin{equation}
          \begin{split}
            \hat{\text{U}} \hat a^ \dagger _{ \xi} \hat{\text{U}} ^ \dagger
            & =\int  d\omega \,  \xi  (\omega) \hat{\text{U}}   \hat a ^\dagger (\omega) \hat{\text{U}} ^ \dagger\\
            & =\int  d\omega \,  \xi  (\omega)  e^{-i \theta(\omega)}\hat a ^\dagger (\omega)\\
             & =\int  d\omega \,  \xi^{\ep}  (\omega)  \hat a ^\dagger (\omega)\\
            & =  \hat a^ \dagger _{ \xi^{\ep}} \, ,
          \end{split}
          \label{eqn:uadud}
        \end{equation}
        where
        \begin{equation}
          \xi^{\ep}(\omega)= \xi(\omega)e^{-i \theta(\omega)} \, .
          \label{eqn:-theta xi}
        \end{equation}
        Here, superscript $\ep$ indicates the encoded photon-wavepacket.
        As Eq.~\eqref{eqn:ftxi} states, the temporal wavepacket is obtainable by the Fourier transforming of Eq.~\eqref{eqn:-theta xi}, which gives
        \begin{equation}
          \begin{split}
            \xi^{\ep} (\tp) & = \frac{1}{\sqrt{2 \pi}} \int d \omega \,  \xi^{\ep} (\omega) e^{-i \omega \tp} \\
            & = \frac{1}{\sqrt{2 \pi}} \int d \omega \,  \xi (\omega) e^{-i( \omega \tp + \theta(\omega))} \, .
          \end{split}
          \label{eqn:ftk}
        \end{equation}

        \subsection{\textbf{Encoding Operation}}     
        
        Applying the phase-shifting operator onto a general pure state (see Eq.~\eqref{eqn:si=f(adag)app}) gives
        \begin{equation}
          \begin{split}
            \lvert \psi^{\ep} \rangle=\hat{\text{U}} \lvert \psi \rangle & =  \hat{\text{U}} \,  f(\hat a^ \dagger _{\xi}) \lvert 0 \rangle =  \hat{\text{U}} \,  f(\hat a^ \dagger _{\xi})  \hat{\text{U}}^ \dagger \lvert 0 \rangle \, .
          \end{split}
          \label{eqn:psie0}
        \end{equation}
        because for the unitary operator, equality $\hat{\text{U}}^{\dagger} \lvert 0 \rangle = \lvert 0 \rangle$ holds.
        Equation~\eqref{eqn:exoff} shows the expansion of function~$f(\hat a^ \dagger _{\xi})$ in powers of $a^ \dagger _{\xi}$ and since $  \hat{\text{U}} (\hat a^ \dagger _{\xi})^n  \hat{\text{U}}^ \dagger  =  \hat{\text{U}} \hat a^ \dagger _{\xi}  \hat{\text{U}}^ \dagger   \hat{\text{U}} \hat a^ \dagger _{\xi} \hat{\text{U}}^ \dagger  \hdots  \hat{\text{U}} \hat a^ \dagger _{\xi} \hat{\text{U}}^ \dagger =( \hat{\text{U}} \hat a^ \dagger _{\xi} \hat{\text{U}}^ \dagger )^n$, where we used the identity property~$\hat I=\hat{\text{U}}^{\dagger} \hat{\text{U}}$; one can write $\hat{\text{U}} \,  f(\hat a^ \dagger _{\xi})  \hat{\text{U}}^ \dagger= f( \hat{\text{U}} \,  \hat a^ \dagger _{\xi}  \, \hat{\text{U}}^ \dagger)$, for the transformation by unitary operator $ \hat{\text{U}}$.
        Therefore Eq.~\eqref{eqn:psie0} reads
        \begin{equation}
          \begin{split}
             \lvert \psi^{\ep} \rangle =  f( \hat{\text{U}} \,  \hat a^ \dagger _{\xi}  \, \hat{\text{U}}^ \dagger) \lvert 0 \rangle  =  f( \hat a^ \dagger _{\xi^{\ep}}) \lvert 0 \rangle \, ,
           \end{split}
           \label{eqn:psie1}
        \end{equation}
        where Eq.~\eqref{eqn:uadud} is used.
        As stated in Eq.~\eqref{eqn:Ixiprove}, the temporal shape of the intensity is
        \begin{equation}
          \begin{split}
            I(\tp) =\bar I \lvert \xi^{\ep} (\tp) \rvert^2  = \bar I  \frac{1}{2 \pi} \Big \lvert \int d \omega \,  \xi (\omega) e^{-i( \omega \tp + \theta(\omega))} \Big \rvert^2
          \end{split}
          \label{eqn:UIk}
        \end{equation}
        \subsection{\textbf{Decoding Operation}}
   
        To decode the signal, the inverse of operator $\hat{\text{U}} $, specifically $\hat{\text{U}}^ {-1}=\hat{\text{U}}^ \dagger$, should be applied to the quantum state vector.
        Following the same procedure as above, one can show
         \begin{equation}
          \begin{split}
            \hat{\text{U}}^ \dagger \hat a^\dagger (\omega) \hat{\text{U}}  =\hat a ^{\dagger} (\omega)  e^{i \theta} \, ,
          \end{split}
          \label{eqn:UdadU}
        \end{equation}
        and therefore
         \begin{equation}
          \begin{split}
            \hat{\text{U}}^ \dagger \hat a^ \dagger _{ \xi} \hat{\text{U}}  =  \hat a^ \dagger _{ \xi^{d}} \, ,
          \end{split}
          \label{eqn:udadu}
        \end{equation}
        where
        \begin{equation}
          \xi^{d}(\omega)= \xi(\omega)e^{i \theta(\omega)} \, .
          \label{eqn:+theta xi}
        \end{equation}
        and superscript $d$ indicates the decoded photon-wavepacket.
        It implies, decoder operator $\hat{\text{U}}^ \dagger$ transforms quantum state vector $\lvert \psi \rangle$ as follows
        \begin{equation}
          \begin{split}
            \lvert \psi^{d} \rangle&=\hat{\text{U}}^ \dagger \lvert \psi \rangle  =  \hat{\text{U}}^ \dagger f(\hat a^ \dagger _{\xi}) \lvert 0 \rangle =  \hat{\text{U}}^ \dagger f(\hat a^ \dagger _{\xi})  \hat{\text{U}}  \lvert 0 \rangle =  f(\hat{\text{U}}^ \dagger \hat a^ \dagger _{\xi}\hat{\text{U}}) \lvert 0 \rangle\\
            &=  f( \hat a^ \dagger _{\xi^{d}}) \lvert 0 \rangle \, .
          \end{split}
        \end{equation}
        where $ \hat{\text{U}}  \lvert 0 \rangle= \lvert 0 \rangle$ is used.                      
        \vspace{5mm}
        \noindent
        \subsection{\textbf{Barcoding: Binary Spectral Encoding}}
        
        \noindent             
        Let us consider a binary pseudorandom sequence with length $N_c$.
        To apply this code onto the spectrum of the quantum light pulse with the state vector shown in Eq.~\eqref{eqn:si=f(adag)app},
        we divide the spectral range into $N_c$ sequential and non-overlapping spectral chips with boundaries of $\Omega_0, \Omega_1, ...,\Omega_{N_c}$.
        The spectral division is such that the photon-wavepacket's mean absolute square is the same for each spectral chip        
        \begin{equation}
          \int_{\Omega_k}^{\Omega_{k+1}} d \omega \, | \xi (\omega) |^{2}=\frac{1}{N_c}\, .
        \end{equation}
        To encode a binary value onto the spectrum of a quantum light pulse, we phase-shift its corresponding $k$th spectral chip ($\Omega_{k-1}  \leqslant  \omega < \Omega_k$) by value $\theta(\omega)=0$ for code element $+1$ and $\theta(\omega) = \pi$ for code element $-1$.
        Therefore, the corresponding spectral photon-wavepacket amplitude is multiplied by $+1$ and $-1$, respectively, as is shown in Fig.~\ref{fig:coding} .
        We call these sign multiplying factor ``{\it{spectral wavepacket multiplier}}."

    \noindent
    \subsection{\textbf{Multiple-Access Decoding Operator} }

        \noindent
        In QCDMA, assume receiver~$\recv $ decodes the signal sent by the intended transmitter~$\dec$ via the conjugate of the spectral phase-shifter of the $\dec$th code, that is $\hat{\text{U}}^ \dagger_{\dec }$.
        To express the quantum receiver~$\recv$ applies the conjugate phase-shifter of the $\dec$th sender's code, we add a subscript to the decoding phase-shifter as $\hat{\text{U}}^ \dagger_{\{\recv,c_\dec \}}$.
        $c_\dec$ denotes the code associated with the $\dec$th transmitter, and the subscript $\{\recv,c_\dec \}$ indicates that the decoder at output node~$\recv$ decodes based on code~$\dec$ ($\hat{\text{U}}^ \dagger_{\dec}$).
        Then, this decoding phase-shifting operator, according to Eq.~\eqref{eqn:U(omeaga)}, can be expressed as
        \begin{equation}
          \begin{split}
            \hat{\text{U}}^\dagger_{\{\recv,c_{\dec}\}} & = e^{i  \sum_{\omega} \, \theta_{\dec} (\omega) \hat a_{\recv} ^{\dagger} (\omega) \hat a_{\recv} (\omega) } \, . 
          \end{split}
        \end{equation}
        Now, considering all receivers apply their intended decoding phase-shifters, the overall decoding operator $ \hat{\textbf{U}}^ \dagger$ in the extended Hilbert space of M receivers is the tensor product of their decoding operators:
        \begin{equation}
          \begin{split}
             \hat{\textbf{U}}^ \dagger=  \prod^{\text{M}}_{\recv=1} \hat{\text{U}}^ \dagger_{ \{\recv,c_{\dec}\} } \, , 
           \end{split}
           \label{eqn:madec}
        \end{equation}
        where $\dec$ depends on $\recv$, i.e., $\dec = \dec (\recv)$.
        Equivalently, operator $ \hat{\textbf{U}}^ \dagger$ is conceivable as a diagonal M$\times$M matrix where its diagonal elements $ (\hat{\textbf{U}}^ \dagger)_{rr} \, , \ \forall \ \recv \in{1,\ldots,\text{M}},$ are $\hat{\text{U}}^ \dagger_{ {\dec} }$, $\dec \in{1,\ldots,\text{M}}$.

        In a typical QCDMA, we are interested in studying the desired decoder's output when the input encoded signal is not the intended transmitted signal (i.e., is a multiaccess interfering signal).                
        Spectral phase-shifters corresponding to the encoder of code~$j$ ($\hat{\text{U}}_{j}$) and the decoder of code~$k$ ($\hat{\text{U}}^{\dagger}_{k}$) changes the photon-wavepacket $\xi_j(\omega)$ into $\xi_j^{\ep_j \cm d_k} (\omega) =\xi_j(\omega)  e^{-i\theta_{j}(\omega)} e^{i \theta_{k}(\omega)}$, where Eq.~\eqref{eqn:-theta xi} and Eq.~\eqref{eqn:+theta xi} are used.
        The inner product of this photon-wavepacket ($\xi_j^{\ep_j \cm d_k}$) and $\xi_k$ gives
        \begin{equation}
          \begin{split}
            \langle \xi_j^{\ep_j \cm d_k }\vert \xi_k \rangle & = \int d \omega \,  (\xi_j (\omega) e^{-i  \theta_{j}(\omega)} e^{i \theta_{k}(\omega)} )^{\star}\xi_k (\omega)  \\
            & = \int d \omega \,  (\xi_j (\omega) e^{-i  \theta_{j}(\omega)})^{\star} ( \xi_k (\omega)  e^{-i \theta_{k}(\omega)})  \\
            &= \int d \omega \,  \xi_j^{\ep_j \star} (\omega)  \xi_k^{\ep_k} (\omega)  \\
            &= \langle \xi_j^{\ep_j }\vert \xi_k^{\ep_k} \rangle
          \end{split}
          \label{eqn:xiedxi}
        \end{equation}
        It is the inner product of photon-wavepacket $\xi_j$ and $\xi_k$ encoded by code~$j$ and code~$k$, respectively.

        \subsection{\textbf{Special Case: Walsh-Hadamard Orthogonal Codes}}
        Code~$j$ and code~$k$ with code lenght $N_c$ are orthogonal if the inner product of their multiplier sequences is zero ($\langle c_j | c_k \rangle=c_j^{\star}. c_k=0$).
        Walsh-Hadamard sequences are the best-known sequences with such orthogonality property.
        For example, if $N_c$ is 4 and the multiplier sequence of code $j$ is $c_j=(1,-1,1,-1)$ and the multiplier sequence of code $k$ is $c_k=(1,1,-1,-1)$, then $c_j$ and $c_k$ are orthogonal, $\langle c_j | c_k \rangle  =1-1-1+1=0$.
        If two orthogonal codes are encoded onto two quantum light pulses with the same spectral wavepacket $ \xi (\omega)$, their photon-wavepacket's inner product is zero, as is shown in the following
        \begin{equation}
          \begin{split}
            \langle \xi^{\ep_j } \vert \xi^{\ep_k} \rangle & = \int d \omega \,  \xi^{\star} (\omega) e^{i( \theta_{j}(\omega))}  \xi (\omega) e^{-i( \theta_{k}(\omega))} \\
            & = \int d \omega \, | \xi (\omega) |^{2}  e^{-i( \theta_{k}(\omega) - \theta_{j}(\omega))}\\
            & = \frac{1}{N_c} \sum_{l=1}^{N_c} e^{-i( \theta_{k}(l) - \theta_{j}(l))}\\
            & = \frac{1}{N_c} \langle c_j |c_k \rangle\\
            &=0 \, .
          \end{split}
          \label{eqn:ortowp}
        \end{equation}
\section{\textbf{QCDMA Star-Coupler Transformation}}
\label{app:sc}
        \begin{figure}[!t]
          \centering
          \includegraphics[width=\columnwidth]{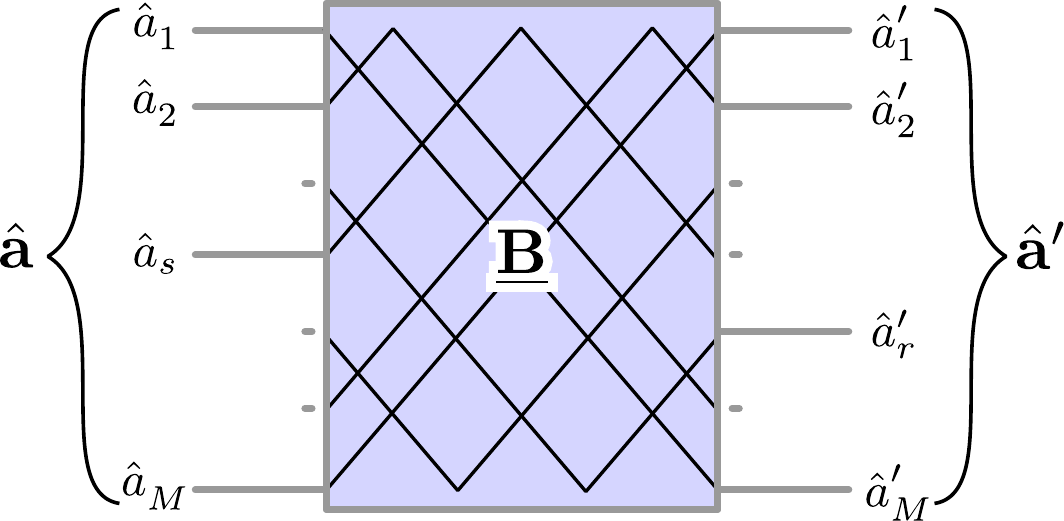}
          \caption{\textbf{Schematic of a star-coupler in the Heisenberg picture for a QCDMA setting.} Each $\hat a $ and each $\hat a'$ operator represents the input and the output field operator of the star-coupler.
            Unitary matrix $\mathbf{\underline B}$ ($\mathbf{\underline B}^{\dagger}$) transforms the input (output) filed operators to the output (input) field operators.
          }\label{fig:starcoupler}
        \end{figure}
        Figure \ref{fig:starcoupler} shows a schematic for an M$\times$M star-coupler, with M input and M output ports.
        The annihilation operator of the input (sender) port~$\tran$ is $\hat a_{\tran}$, and the annihilation operator of the output (receiver) port $\recv$ is shown by $\hat a'_{\recv}$, where both $\tran$ and $\recv$ are elements from $1,\ldots, \text{M}$. 
        Assuming the system is lossless, the transformation of the input filed operators ($\hat{\mathbf{a}}$) to the output field operators ($\hat{\mathbf{a}}'$) can be represented by a unitary matrix ($\mathbf{\underline B}$)
        \begin{equation}
          \hat{\mathbf{a}}'= \mathbf{\underline B} \, \hat{\mathbf{a}}\, ,
          \label{eqn:ap=Ba}
        \end{equation}
        and its corresponding matrix representation is
        \begin{equation}
          \begin{split}
            \begin{pmatrix}
              \hat a'_1\\
              \hat a'_2\\
              \vdots\\
              \hat a'_{\recv}\\
              \vdots \\
              \hat a'_\text{M}
            \end{pmatrix}
            & =            
            \begin{pmatrix}
              B_{11} &B_{12} & \hdots &B_{1 \tran} & \hdots & B_{1 \text{M}}\\
              B_{21} & B_{22} & \hdots &B_{2 \tran}& \hdots & B_{2 \text{M}}\\
              \vdots &\vdots & \ddots & \vdots &\ddots &\vdots\\
              B_{\recv 1} & B_{\recv 2} &\hdots &B_{\recv \tran} & \hdots & B_{\recv  \text{M}}\\
              \vdots &\vdots & \ddots & \vdots &\ddots &\vdots\\
              B_{\text{M} 1} & B_{\text{M} 2} &\hdots &B_{\text{M} \tran} & \hdots & B_{\text{M} \text{M}}
            \end{pmatrix}\,
            \begin{pmatrix}
              \hat a_1 \\
              \hat a_2 \\
              \vdots\\              
              \hat a_{\tran}\\
              \vdots\\
              \hat a_{\text{M}}
            \end{pmatrix} \, .
          \end{split}
          \label{eqn:heispic}
        \end{equation}
        From Eq.~\eqref{eqn:ap=Ba} and Eq.~\eqref{eqn:heispic}, one can write the annihilation operator of a receiver as a linear combination of senders' as follows
        \begin{equation}
          \begin{split}
             \hat a'_{\recv} &= \sum_{\tran =1}^{\text{M}} B_{\recv  \tran} \,  \hat{a}_{\tran}\, .
           \end{split}
           \label{eqn:heispicexpand}
        \end{equation}

        The complex conjugate of Eq.~\eqref{eqn:ap=Ba} gives
        \begin{equation}
          \hat{\mathbf{a}}'^{\star}= \mathbf{\underline B}^{\star} \, \hat{\mathbf{a}}^{\star}\, ,
          \label{eqn:ap*=B*a*}
        \end{equation}
        where $ \hat{\mathbf{a}}^{\star}$ ($ \hat{\mathbf{a}}'^{\star}$) is a column vector with elements of creation operators $\hat a^{\dagger}_{\tran}$ ($ \hat a'^{\dagger}_{\recv}$), and $\mathbf{\underline B}^{\star}$ is an M$\times$M square matrix with elements $B_{\recv \tran}^{\star}$.
        Since $\hat a$  and $\hat a^{\dagger}$ correspond respectively to the electric field's positive and negative frequency parts, they are complex conjugate pairs.
        One may choose to rewrite the star-coupler transformation on the creation operators, Eq.~\eqref{eqn:ap*=B*a*}, as  
        $\hat{\mathbf{a}}'^{\dagger}=   \hat{\mathbf{a}}^{\dagger}\, \mathbf{\underline B}^{\dagger} $,
        which is the transpose of Eq.~\eqref{eqn:ap*=B*a*}, then $\hat{\mathbf{a}}'^{\dagger}$ and $ \hat{\mathbf{a}}^{\dagger}$ are row vectors.
        From Eq.~\eqref{eqn:ap*=B*a*}, we have
        \begin{equation}
          \begin{split}            
            \hat a'^{\dagger}_{\recv}& = \sum_{\tran =1}^{\text{M}} B^{\star}_{\recv  \tran} \,  \hat{a}^{\dagger}_{\tran}\, .
          \end{split}
        \end{equation}
        
        Due to the unitarity of matrix $\mathbf{\underline B}$, $\mathbf{\underline B}^{\dagger} \mathbf{\underline B} = \mathbf{\underline B}^{\top} \mathbf{\underline B}^{\star} = \mathbf{I}$, the inverses of Eq.~\eqref{eqn:ap=Ba} and Eq.~\eqref{eqn:ap*=B*a*} are         
        \begin{equation}
          \begin{split}
            \hat{\mathbf{a}} = \mathbf{\underline B}^{\dagger} \,  \hat{\mathbf{a}}'\, ,
          \end{split}
          \label{eqn:a=Bdap}
        \end{equation}
        and
        \begin{equation}
          \begin{split}
            \hat{\mathbf{a}}^{\star} = \mathbf{\underline B}^{\top} \,  \hat{\mathbf{a}}'^{\star}\, ,
          \end{split}
          \label{eqn:ad=Btapd}
        \end{equation}
        respectively.
        Equivalently, one can write the annihilation operator of a sender ($\hat a_{\tran}$) as a linear combination of receivers' annihilation operators ($ \hat{a}'_{\recv} \, , r=1,2, \hdots, \text{M}$); that is, from Eq.~\eqref{eqn:a=Bdap}, we have        
        \begin{equation}
          \begin{split}
            \hat a_{\tran} =\sum_{\recv =1}^{\text{M}}  B^{\star}_{\recv \tran } \,  \hat{a}'_{\recv} \, ,
          \end{split}
          \label{eqn:at=Bar}
        \end{equation}
        and the relation for the corresponding field creation operators, from Eq.~\eqref{eqn:ad=Btapd}, is
        \begin{equation}
          \begin{split}            
            \hat a_{\tran}^{\dagger} = \sum_{\recv =1}^{\text{M}} B_{\recv   \tran} \,  \hat{a}'^{\dagger}_{\recv}\, .
          \end{split}
          \label{eqn:adt=Badr}
        \end{equation}        
        
        \vspace{5mm}
        \noindent
        \textbf{Approach 1: Quantum Broadcasting Interpretation for Star-Coupler's Transformation}
        
        \begin{figure*}[htp!]
          \includegraphics[width=\textwidth]{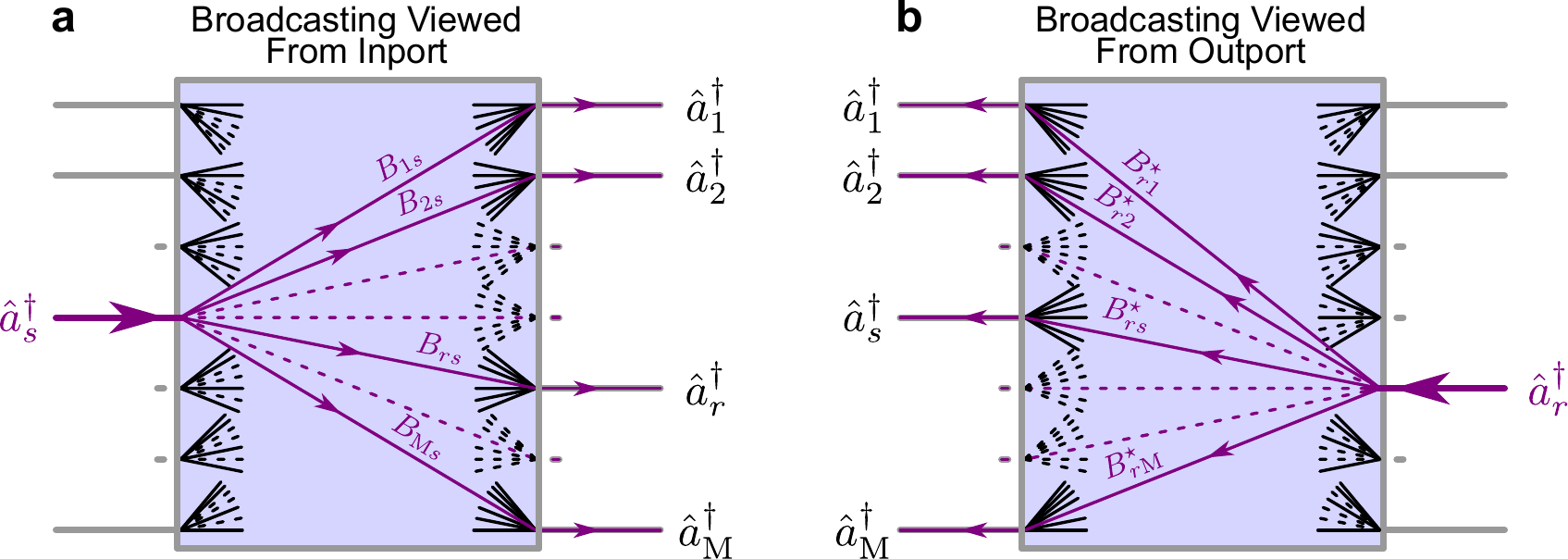}
          \caption{\textbf{Quantum broadcasting interpretation for star-coupler's transformation.}
            \textbf{a} The single-photon creation operator of sender~$\tran$ ($\hat a^{\dagger}_\tran$) conceivable as the linear combination of single-photon creation operators of all receivers ($\hat a^{\dagger}_\recv\, , \ \recv \in{1,\ldots,\text{M}} $) with coefficient $B_{\recv \tran}$.
            \textbf{b}  The single-photon creation operator of receivers $\recv$ ($\hat a^{\dagger}_\recv$) conceivable as the linear combination of single-photon creation operators of all transmitters ($\hat a^{\dagger}_\tran\, , \ \tran \in{1,\ldots,\text{M}} $) with coefficient $B^{\star}_{\recv \tran}$.
          }\label{fig:bc}
        \end{figure*}
        \noindent
        Let us present a quantum broadcasting interpretation to Eq.~\eqref{eqn:adt=Badr} in the context of QCDMA.
        Equation~\eqref{eqn:adt=Badr} indicates that if a photon is created (transmitted) by sender~$\tran$, then the star-coupler would broadcast this photon to a superposition of all receivers' nodes.
        Furthermore, the amplitude of broadcasting from sender~$\tran$ to receiver~$\recv$ is $B_{\recv   \tran}$, which corresponds to the broadcasting weight~$|B_{\recv   \tran}|^2$.
        This broadcasting interpretation is depicted in Fig.~\ref{fig:bc}a and can be expressed as follows
        \begin{equation}
          \begin{split}
            \hat a_{\tran,\, \xi_{\tran}}^{\dagger} \rightarrow  \sum_{\recv =1}^{\text{M}} B_{\recv   \tran} \,  \hat{a}^{\dagger}_{\recv, \, \xi_{\tran}}\, .
          \end{split}
          \label{eqn:at->ar}
        \end{equation}
        In the above equation, we ignored the prime sign used in Eq.~\eqref{eqn:adt=Badr} for the output (receiver) creation operators.         
        Also, we added the transmitted photon's spectral shape ($ \xi_{\tran}$) to the notation.
        Assuming the star-coupler operation is frequency independent, the photon-wavepacket received by user~$\recv$ remains unvaried, with the same wavepacket representation as to transmitted wavepacket~$ \xi_{\tran}$.
        \par
        As discussed before, sender~$\tran$ transmits pure light state $\lvert \psi_{\tran} \rangle= f_{\tran} (\hat a_{\tran, \,\xi_{\tran}}^ \dagger) \lvert 0 \rangle$ (see Eq.~\eqref{eqn:exoff}).
        As  Eq.~\eqref{eqn:at->ar} indicates, the star-coupler broadcasts each photon among all receivers and consequently transforms the transmitted quantum light state $\lvert \psi_{\tran} \rangle$ as follows       
        \begin{equation}
          \begin{split}
            \lvert  \psi_{\tran}\rangle   \rightarrow  f_{\tran}(\sum_{\recv =1}^{\text{M}} B_{\recv \tran} \hat a_{\recv,\xi_{\tran}}^{ \dagger}) \lvert 0 \rangle\, .
          \end{split}
          \label{eqn:psi->phi}
        \end{equation}
        The above transformation shows the broadcasting of a quantum signal with photon-wavepacket~$\xi_{\tran}$.
        Note that, except for exponential $f_{\tran}$, Eq.~\eqref{eqn:psi->phi} contains entanglement among M receivers' quantum signals located at M spatially separated output ports of the star-coupler denoted by $\recv$, where $\recv \in{1,\ldots,\text{M}}$.
        When  $f_{\tran}$ is exponential ($f_{\tran}(x) =  e^{\alpha x}$), indicating the Glauber state, Eq.~\eqref{eqn:psi->phi} transforms into
        $f_{\tran}(\sum_{\recv =1}^{\text{M}} B_{\recv \tran} \hat a_{\recv,\xi_{\tran}}^{ \dagger}) \lvert 0 \rangle \propto e^{\alpha \sum_{\recv =1}^{\text{M}} B_{\recv \tran} \hat a_{\recv,\xi_{\tran}}^{ \dagger}} \lvert 0 \rangle=\prod_{\recv=1}^{\text{M}}e^{\alpha  B_{\recv \tran} \hat a_{\recv,\xi_{\tran}}^{ \dagger}} \lvert 0 \rangle$,
        which is a factorized state denoting that receiver~$\recv$ receives pure quantum state~$e^{\alpha  B_{\recv \tran} \hat a_{\recv,\xi_{\tran}}^{ \dagger}} \lvert 0 \rangle\,$.        
        One may note that only exponential function has the fundamental multiplicative identity $f_{\tran}(x+y) =f_{\tran}(x)f_{\tran}(y)$, resulting in a factorized, non-entangled quantum state among receivers.
        However, for states such as Fock states, one can show the above pure quantum state is not factorizable (see section~\ref{sec:qcdmafock}).
        \par
        Now let us assume, each transmitter sends a pure quantum state, then the input to the star-coupler is the tensor product of all senders' pure quantum states        
        \begin{equation}
          \lvert \Psi \rangle = \prod_{\tran=1}^{\text{M}}  \lvert  \psi_{\tran}\rangle
          \label{eqn:Psi}
        \end{equation}
        and the output quantum state of the star-coupler, employing Eq.~\eqref{eqn:psi->phi}, reads
        \begin{equation}
          \begin{split}
            \lvert \Phi \rangle=\prod_{\tran=1}^{\text{M}} f_{\tran}(\sum_{\recv =1}^{\text{M}} B_{\recv \tran} \hat a_{\recv,\xi_{\tran}}^{ \dagger}) \lvert 0 \rangle\, ,
          \end{split}
          \label{eqn:Phiprove1}
        \end{equation}
        We refer to Eq.~\eqref{eqn:Phiprove1} as the state broadcasting equation.
        In this paper, we use the term broadcasting in a quantum sense.
        However, in classical interpretation, each receiver receives an exact copy of the transmitted signal that is impossible in the quantum domain due to the no-cloning theorem.
        Furthermore, Eq.~\eqref{eqn:Phiprove1} is not, in general, even representable as the tensor product of pure quantum states at the star-coupler's output ports; i.e., not representable as $ \lvert \Phi \rangle= \prod_{\recv=1}^{\text{M}}  \lvert  \phi_{\recv}\rangle$.
         Therefore, star-coupler' output can contain quantum entanglement among the receivers' signals.
         
        \vspace{5mm}
        \noindent
        \textbf{Approach 2: Heisenberg and Schr\"odinger Picture for Quantum Star-Couplers Transformation}

        \noindent
        Equation~\eqref{eqn:ap=Ba}, $\hat{\mathbf{a}}'= \mathbf{\underline B} \, \hat{\mathbf{a}}$, (and also Eq.~\eqref{eqn:ap*=B*a*}, $\hat{\mathbf{a}}'^{\star}= \mathbf{\underline B}^{\star} \, \hat{\mathbf{a}}^{\star}$) is conceivable as the Heisenberg representation of the star-coupler transformation where the unitary matrix~$\mathbf{\underline B}$ transforms the input field operators $ \hat{\mathbf{a}} $ to output field operators $ \hat{\mathbf{a}}'$ (see Fig.~\ref{fig:starcoupler}).
        In the Heisenberg picture, the quantum state of the input beams remains invariant.
        In this manuscript, as Fig.~\ref{fig:cdma} shows, we would take the Schr\"odinger~\cite{leonhardt_book_1997} picture and consider how the light's input quantum state changes to the light's output quantum state.
        In the Schr\"odinger picture, the star-coupler's evolution operator ($\hat{\text{B}}^{\dagger}$) transforms the state vector from $\lvert \Psi \rangle$ to $\lvert \Phi \rangle$, denoted as
        \begin{equation}
          \lvert \Phi \rangle = \hat{\text{B}}^{\dagger} \lvert \Psi \rangle\, .
          \label{eqn:schropic}
        \end{equation}
        \par
        Schr\"odinger and Heisenberg pictures are equivalent. Therefore, choosing any of these pictures, any operator's expectation value at the output ports is the same.        
        In the Heisenberg picture, $ \langle \Psi \rvert  \hat {\mathbf{a}}'\lvert \Psi \rangle$, an M-dimensional vector of expectation values, gives the field annihilation operators' expectation value at the star-couple's output.

        Similarly, the Schr\"odinger representation of this vector at the star-couple's output is $ \langle \Phi \rvert \hat {\mathbf{a}} \lvert \Phi \rangle $.
        The two pictures' equivalency for the field operators' expectation value at the star-couple's output indicates                
        \begin{equation}
          \begin{split}
            \langle \Phi \rvert \hat {\mathbf{a}} \lvert \Phi \rangle  = \langle \Psi \rvert  \hat {\mathbf{a}}'\lvert \Psi \rangle \, .
          \end{split}
        \end{equation}
        This equality for the $\recv$th output element of the star-coupler 
        gives        
        \begin{equation}
          \begin{split}
            \langle \Psi \rvert  \hat{\text{B}}\,  \hat a_{\recv}  \hat{\text{B}}^{\dagger}  \lvert \Psi \rangle = \langle \Psi \lvert  \sum_{\tran =1}^{\text{M}} B_{\recv  \tran} \,  \hat a_{\tran} \lvert \Psi \rangle\\   
          \end{split}
          \label{eqn:H=Sch}
        \end{equation}
        where Eq.~\eqref{eqn:heispicexpand} is used to expand $\hat a'_{\recv}$, and also Eq.~\eqref{eqn:schropic} is used for state~$ \lvert \Phi \rangle$.
        Since Eq.~\eqref{eqn:H=Sch} is valid for any quantum state $\lvert \Psi \rangle$, it gives
        \begin{equation}
          \begin{split}
            \hat{\text{B}}\, \hat a_{\recv}  \hat{\text{B}}^{\dagger} = \sum_{\tran =1}^{\text{M}} B_{\recv  \tran} \,  \hat a_{\tran} \, ;
          \end{split}
          \label{eqn:BarBd}
        \end{equation}
        and its Hermitian adjoint can be shown to be
        \begin{equation}
          \begin{split}
            \hat{\text{B}}\, \hat a^{\dagger}_{\recv}  \hat{\text{B}}^{\dagger} = \sum_{\tran =1}^{\text{M}} B^{\star}_{\recv  \tran} \,  \hat a^{\dagger}_{\tran} \, .
          \end{split}
          \label{eqn:BardBd}
        \end{equation}
        \par
        Expression in Eq.~\eqref{eqn:BardBd} shows the star-coupler's broadcasting transformation viewed from the output port~$\recv$ (see Fig.~\ref{fig:bc}b).
        However, one can obtain a similar expression for the broadcasting from the input ports (see Fig.~\ref{fig:bc}a) if the reverse of the above operation is considered~\cite{furusawa_2015}.
        The inverse of Eq.~\eqref{eqn:schropic}, the Schr\"odinger picture for the quantum state transformation, is
        \begin{equation}
          \lvert \Psi \rangle = \hat{\text{B}}\, \lvert \Phi \rangle\, ,
          \label{eqn:invschropic}
        \end{equation}
        and its corresponding Heisenberg representation is Eq.\eqref{eqn:a=Bdap}, $\hat{\mathbf{a}}= \mathbf{\underline B}^{\dagger} \, \hat{\mathbf{a}}'$, which is the inverse of Eq.~\eqref{eqn:ap=Ba}.
        Again, Schr\"odinger and Heisenberg pictures are equivalent.
        This equivalency for the star-coupler's $\tran$th input port indicates that $\langle \Psi \rvert \hat a'_{\tran} \lvert \Psi \rangle  = \langle \Phi \rvert  \hat a_{\tran}\lvert \Phi \rangle$, and consequently, it gives       
        \begin{equation}
          \begin{split}
            \hat{\text{B}}^{\dagger}  \hat a_{\tran}  \hat{\text{B}} =  \sum_{\recv =1}^{\text{M}}  B^{\star}_{\recv  \tran} \,  \hat a_{\recv} \, ;
          \end{split}
          \label{eqn:sch-heis}
        \end{equation}
        and its Hermitian adjoint is
        \begin{equation}
          \begin{split}
            \hat{\text{B}}^{\dagger}  \hat a^{\dagger}_{\tran}  \hat{\text{B}} =  \sum_{\recv =1}^{\text{M}}  B_{\recv  \tran} \,  \hat a^{\dagger}_{\recv} \, .
          \end{split}
          \label{eqn:sch-heisdag}
        \end{equation}
        To proceed further, assume the star-coupler's transformation is the same for all input quantum signal's frequency components; therefore, their corresponding photon-wavepacket remains unchanged, passing through the star-coupler.
        Furthermore, assume that each of the star-coupler inputs (input $\tran$) carries photons with a distinct wavepacket ($\xi_{\tran}$). Applying these assumptions into Eq.~\eqref{eqn:sch-heisdag}, it reads
        \begin{equation}
          \hat{\text{B}}^{\dagger} \hat a_{\tran,\, \xi_{\tran}}^{\dagger}  \hat{\text{B}}  = \sum_{\recv =1}^{\text{M}}  B_{\recv \tran} \,  \hat{a}^{ \dagger}_{\recv, \,\xi_{\tran}} \, ,
          \label{eqn:bstransform}
        \end{equation}
        which is equivalent to Eq.~\eqref{eqn:at->ar}.
        Therefore, applying the equality~\eqref{eqn:bstransform} and pursuing the Schr\"odinger picture~\eqref{eqn:schropic} show that the quantum star-coupler transforms the input lights' quantum state~\eqref{eqn:Psi} as:
        \begin{equation}
          \begin{split}
            \lvert \Phi \rangle &= \hat{\text{B}}^{\dagger} \lvert \Psi \rangle \\
            &=   \prod_{\tran=1}^{\text{M}} \hat{\text{B}}^{\dagger}f_{\tran}( \hat a_{\tran,\xi_{\tran}}^{ \dagger}) \lvert 0 \rangle\\
            &=   \prod_{\tran=1}^{\text{M}} \hat{\text{B}}^{\dagger}f_{\tran}( \hat a_{\tran,\xi_{\tran}}^{ \dagger}) \hat{\text{B}}\lvert 0 \rangle\\            
            &=  \prod_{\tran=1}^{\text{M}} f_{\tran}(  \hat{\text{B}}^{\dagger} \hat a_{\tran,\xi_{\tran}}^{ \dagger}  \hat{\text{B}} ) \lvert 0 \rangle\\
            &=\prod_{\tran=1}^{\text{M}} f_{\tran}(\sum_{\recv =1}^{\text{M}} B_{\recv \tran} \, \hat a_{\recv,\xi_{\tran}}^{ \dagger}) \lvert 0 \rangle\, ,
          \end{split}
          \label{eqn:phiprove}
        \end{equation}
        which is equivalent to Eq.~\eqref{eqn:Phiprove1}.
        \par
        Let us calculate the intensity's expectation value at output port~$\recv$ of the star-coupler at time~$\tp$.
        For the sake of simplicity, we obtain the expectation value of the intensity by the Heisenberg picture, and that is
        \begin{equation}
          \begin{split}
            I_{\recv}(\tp)&=\langle \Phi \rvert \hat a^{\dagger}_{\recv}(\tp) \hat a_{\recv}(\tp) \rvert \Phi \rangle \\
            &=\langle \Psi \rvert  \hat B \hat a^{\dagger}_{\recv}(\tp) \hat a_{\recv}(\tp) \hat B^{\dagger}\rvert \Psi \rangle \\
            &=\langle \Psi \rvert  \left(\hat B \hat a^{\dagger}_{\recv}(\tp) \hat B^{\dagger} \right)\left( \hat B \hat a_{\recv}(\tp) \hat B^{\dagger}\right)\rvert \Psi \rangle \\           
            &= \sum_{\tran =1}^{\text{M}} \sum_{\tran' =1}^{\text{M}} B^{\star}_{\recv  \tran}   B_{\recv  \tran'} \langle \Psi \rvert  \hat a^{\dagger}_{\tran} (\tp) \hat a_{\tran'} (\tp)  \rvert \Psi \rangle\, ,
          \end{split}
          \label{eqn:igen0}
        \end{equation}
        where Eq.~\eqref{eqn:schropic} and Eq.~\eqref{eqn:BarBd} and their complex conjugates are used.
        Considering the factorized state $\lvert \Psi \rangle $, as stated in Eq.~\eqref{eqn:Psi}, the intensity,  Eq.~\eqref{eqn:igen0}, reduces to 
        \begin{equation}
          \begin{split}
            I_{\recv}(\tp)
            &= \sum_{\tran =1}^{\text{M}} \lvert B_{\recv  \tran}\lvert^2 \langle \psi_{\tran} \rvert \hat a^{\dagger}_{\tran} (\tp) \hat a_{\tran}(\tp)  \lvert \psi_{\tran}\rangle\\
            &+ \sum_{\tran =1}^{\text{M}} \sum_{\tran' \neq \tran}^{\text{M}} B^{\star}_{\recv  \tran}   B_{\recv  \tran'}  \langle \psi_{\tran} \rvert \hat a^{\dagger}_{\tran} (\tp)\lvert \psi_{\tran}\rangle \langle \psi_{\tran'} \rvert  \hat a_{\tran'} (\tp) \rvert \psi_{\tran'} \rangle\\
            &= \sum_{\tran =1}^{\text{M}} \lvert B_{\recv  \tran}\lvert^2 I_{\tran}(\tp)+ \sum_{\tran =1}^{\text{M}} \sum_{\tran' \neq \tran}^{\text{M}} B^{\star}_{\recv  \tran}   B_{\recv  \tran'}   E^{\star}_{\tran} \left( \tp \right)  E_{\tran'} \left(\tp\right) \\            
            &= \sum_{\tran =1}^{\text{M}} \lvert B_{\recv  \tran}\lvert^2 I_{\tran}(\tp)+\sum_{\tran =1}^{\text{M}} \sum_{\tran' \neq \tran}^{\text{M}} Re\left(B^{\star}_{\recv  \tran}   B_{\recv  \tran'}   E^{\star}_{\tran} \left( \tp \right)  E_{\tran'} \left(\tp\right)\right)\, ,        
          \end{split}
        \end{equation}
        where $I_{\tran}(\tp)=\langle \psi_{\tran} \rvert \hat a^{\dagger}_{\tran} (\tp) \hat a_{\tran}(\tp)  \lvert \psi_{\tran}\rangle$ and $E_{\tran}(\tp)= \langle \psi_{\tran'} \rvert  \hat a_{\tran'} (\tp) \rvert \psi_{\tran'} \rangle$  correspond to the intensity and the electric field of input port~$\tran$ at time~$\tp$, respectively.
        Therefore the intensity of output port~$\recv $, $I_{\recv}(\tp)$, of the star-coupler equals the sum of all input ports' intensities, $I_{\tran}(\tp), \tran = 1, 2, \hdots \text{M}$ decreased by the ratio $\lvert B_{\recv  \tran}\lvert^2 $ (for a balanced star-coupler, we have $\lvert B_{\recv  \tran}\lvert^2 =1/\text{M}$) plus an interference term amongst all the input signals.
        The interference term between signals of ports~$\tran$ and~$\tran'$  corresponds to  $ E^{\star}_{\tran} \left( \tp \right)  E_{\tran'} \left(\tp\right)$.
        This inter-signal interference term disappears for some quantum states, such as number states and squeezed coherent states, due to Heisenberg's uncertainty principle.
        From Heisenberg's uncertainty principle, the quantum phase for number state (see appendix~\ref{sec:qcdmafock}) is uniformly distributed between $[0,2 \pi]$, then the electric field expectation value~$E \left(\tp\right)$ becomes time-independent and zero.
        On the other hand, since the field operation on number state~$\lvert n \rangle$, i.e., $\hat a(\tp) \lvert n \rangle$, corresponds to number state~$\lvert n-1 \rangle$; therefore, the expectation value~$ E(\tp)=\langle n \rvert \hat a (\tp) \lvert n \rangle =0$ vanishes.
        Also, field operator~$ \hat a (\tp)$ changes squeezed coherent state~$\lvert \psi \rangle=\sum_{n=0}^{\infty} c_n \lvert 2 n \rangle$, a superposition of all even number states, into a superposition of odd number states, $ \hat a (\tp) \lvert \psi \rangle=\sum_{n=0}^{\infty} d_n \lvert 2 n-1 \rangle$.
        Therefore, again, the expectation value of the field operator is zero, $ E(\tp)=\langle \psi \rvert \hat a (\tp) \lvert \psi \rangle =0$, for squeezed coherent states.

        \vspace{5mm}
        \noindent
        \textbf{Example:  Number States Inputs}
        
        \noindent
        For clarity, we present an example for the quantum star-coupler transformation.
        We assume that the inputs to the star-coupler are Fock number states:
        \begin{equation}
          \begin{split}
            \lvert \Psi \rangle & = \lvert n_1 \rangle \lvert n_2 \rangle \lvert n_3 \rangle \hdots \lvert n_{\text{M}} \rangle \\
            & = \prod_{\tran=1}^{\text{M}} \frac{1}{\sqrt{n_{\tran} !}}(\hat a_{\tran}^{ \dagger})^{n_{\tran}} \lvert 0 \rangle\\
            & = \prod_{\tran=1}^{\text{M}} f_{\tran}(\hat a_{\tran}^{ \dagger}) \lvert 0 \rangle
          \end{split}
          \label{eqn:fockinput}
        \end{equation}
        The above equation denotes that function $f_{\tran}$ is a power function, $f_{\tran}(\hat a_{\tran}^{ \dagger})= \frac{1}{\sqrt{n_{\tran} !}}(\hat a_{\tran}^{ \dagger})^{n_{\tran}}$, and assumes that all the input number states $\lvert n_1\rangle, \lvert n_2\rangle, \hdots, \lvert n_{\text{M}}\rangle$ have identical wavepacket; therefore, the subscript $\xi_{\tran}$ of the creation operators is dropped, i.e., $ \hat a_{\tran, \xi_{\tran} }^{ \dagger}\rightarrow \hat a_{\tran}^{ \dagger}$ .
        Equation~\eqref{eqn:phiprove}, for input Fock number states Eq.~\eqref{eqn:fockinput}, gives the star-coupler's output as     
        \begin{equation}
          \begin{split}
            \lvert \Phi \rangle =\prod_{\tran=1}^{\text{M}}  \frac{1}{\sqrt{n_{\tran} !}} (\sum^{\text{M}}_{\recv=1} B_{\recv  \tran} \hat a_{\recv}^{ \dagger})^{n_{\tran}} \lvert 0 \rangle\, .
          \end{split}
          \label{eqn:fockoutput}
        \end{equation}
        For the single-photon inputs $n_{\tran}=1,\  \forall \ \tran \in{1,\ldots,\text{M}}$, Eq.~\eqref{eqn:fockoutput} reads
        \begin{equation}
          \begin{split}
            \lvert \Phi \rangle =\prod_{\tran=1}^{\text{M}}  \sum^{\text{M}}_{\recv=1} B_{\recv \tran} \hat a_{\recv}^{ \dagger} \lvert 0 \rangle\, ,
          \end{split}
        \end{equation}
        which is studied in more detail in appendix~\ref{sec:qcdmafock}.
        
        In a practical case, we usually want the quantum star-coupler to broadcast the input quantum signals equally to all receiving users.
        Therefore in the following, we briefly study matrix~($\mathbf{\underline{B}}$) for such a balanced quantum star-coupler.
              
        \subsection{\textbf{Balanced Quantum Star-Couplers}}
        \label{sec:qibs}
        
        \noindent
        A balanced quantum star-coupler evenly splits each input quantum optical signals to output ports.
        In this section, we present several possible formalisms for the matrix operation of such a coupler.
        An M$\times$M balanced star-coupler can be modeled as a mesh of 2$\times$2 beamsplitters~\cite{reck_prl_1994, clements_o_2016}.
        Mathematically it can be represented by an M$\times$M matrix 
        \begin{equation}
          \mathbf{\underline{B}} =\frac{1}{\sqrt{\text{M}}}\begin{pmatrix}
            e^{i\phi_{11}} & e^{i\phi_{12}} & e^{i\phi_{13}} & \dots & e^{i\phi_{1\text{M}}} \\
            e^{i\phi_{21}} & e^{i\phi_{22}} & e^{i\phi_{23}} & \dots & e^{i\phi_{2\text{M}}} \\
            e^{i\phi_{31}} & e^{i\phi_{32}} & e^{i\phi_{33}} & \dots & e^{i\phi_{3\text{M}}} \\
            \vdots & \vdots &\vdots & \ddots &\vdots \\
            e^{i\phi_{\text{M}1}} & e^{i\phi_{\text{M}2}} & e^{i\phi_{\text{M}3}} & \dots & e^{i\phi_{\text{M}\text{M}}} 
          \end{pmatrix},
        \end{equation}
        Since $\mathbf{\underline  B}$ is a unitary matrix ($\mathbf{\underline B} . \mathbf{\underline B^{\dagger}}=\mathbf{I}$), therefore
        \begin{equation}
          \sum_{k=1}^{\text{M}} B_{ik} B^{\star}_{jk} =\frac{1}{\text{M}}  \sum_{k=1}^{\text{M}} e^{i( \phi_{ik} - \phi_{jk})}=\delta_{ij}
          \label{eqn:bbdag=I}
        \end{equation}
        There are many solutions to these equations.
        Equation~\eqref{eqn:bbdag=I} shows that if matrix~$\mathbf{\underline  B}$ is a solution for an M$\times$M balanced star-coupler, matrix~$\mathbf{\underline  B}'$, obtained by multiplying each line and column of matrix~$\mathbf{\underline  B}$ by an arbitrary phase factor ($B'_{jk}= e^{i \phi_j} e^{i \phi_k} B_{jk}$), is also  a unitary matrix and hence a feasible solution for a lossless balanced star-coupler transformation matrix.
        Some exciting solutions for the star-coupler are discrete Fourier transform matrices and Hadamard matrices,
        briefly presented in the following.
        \begin{enumerate}[I., \IEEEsetlabelwidth{12)}]
        \item \textbf{DFT Matrix}
          
        An M-point discrete Fourier transform (DFT) matrix can represent a balanced M$\times$M star-coupler, and it is expressed as        
        \begin{equation}
          \mathbf{\underline{B}} =\frac{1}{\sqrt{\text{M}}}\begin{pmatrix}
            1  & 1 & 1 & \dots & 1 \\
            1  & \gamma & \gamma^2  & \dots & \gamma^{\text{M}-1} \\
            1  & \gamma^{2} & \gamma^4  & \dots & \gamma^{2(\text{M}-1)} \\
            \vdots & \vdots &\vdots & \ddots &\vdots \\
            1  & \gamma^{(\text{M}-1)} & \gamma^{2(\text{M}-1)}  & \dots & \gamma^{(\text{M}-1)(\text{M}-1)} \\ 
          \end{pmatrix},
        \end{equation}
        where $\gamma=e^{-\frac{2 \pi i}{\text{M}}}$.
      \item \textbf{Hadamard Matrix}
        
         A Hadamard matrix is a square matrix whose entries are either $+1$ or $-1$, and its rows (columns) are mutually orthogonal. 
         \begin{equation}
           \mathbf{ H.H}^{\top}=\text{M} \, \mathbf I
         \end{equation}
         Therefore, Hadamard matrices can provide solutions for matrix $ \mathbf{\underline{B}}$ of balanced star-coupler with an even number of input and output ports, and that is $ \mathbf{\underline{B}} = \frac{1}{\sqrt{\text{M}}} \, \mathbf{H}$. 
         
         Sylvester's construction of a Hadamard matrix is as follows.
         Let H be a Hadamard matrix of order M$/2$.
         Then the partitioned matrix
         \begin{equation}
           \begin{pmatrix}
             H  & H \\
             H  & -H \\
           \end{pmatrix}
           \label{eqn:Hadamardrec}
         \end{equation}
         is a Hadamard matrix of order M.
         The lowest order of Hadamard matrices is~2, and it is:
         \begin{equation}
           H_2 =
           \begin{pmatrix}
             1  & 1 \\
             1  & -1 \\
           \end{pmatrix}
         \end{equation}
         Recursively, one can find the higher orders of Hadamard matrices using Eq.~\eqref{eqn:Hadamardrec}.
       \end{enumerate}
\section{\textbf{QCDMA}}               
\begin{figure*}[htp!]
         \includegraphics[width=\textwidth]{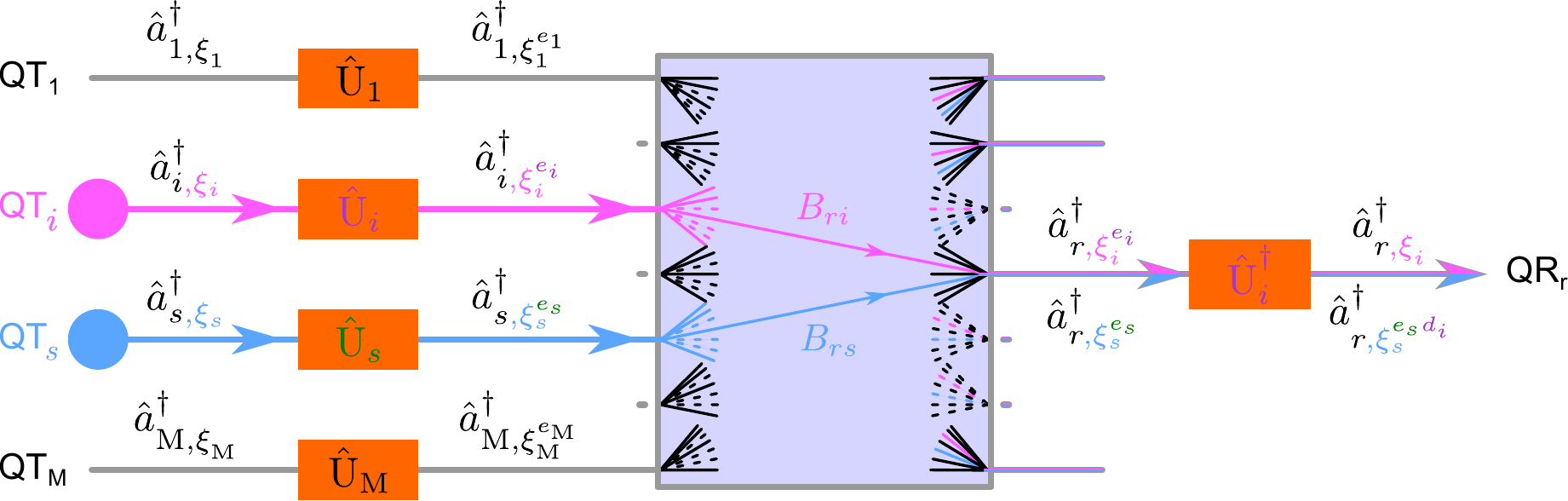}
         \caption{\textbf{ A schematic of a QCDMA.}
           The field operators are shown with two subscripts; the first subscript indicates the path mode (similar to Fig.~\ref{fig:starcoupler}).
           The second subscript indicates the spectral modes that the corresponding photon occupies.
           Spectral encoding changes the spectral mode (photon-wavepacket).
           The superscript added to the photon-wavepacket notation indicates the spectral code applied to the quantum state of light.
           For example, field operator $\hat a^{\dagger}_{m,\xi_{\tran}^{\ep_n}}$  creates a photon in path mode $m$ and spectral modes with amplitude $\xi_{\tran}^{\ep_n}$, where $\tran$ indicates the quantum transmitter (sender) of the photon and $\ep_n$ ($d_n$) indicates the $n$th code is used for the spectral encoding (decoding) via the corresponding operator $\hat{\text{U}}_n$ ($\hat{\text{U}}^{\dagger}_n$).
           The decoding operator used by receiver~$\recv$, i.e., $\hat {\text{U}}^{\dagger}_\dec$, transforms the field related to (intended) transmitter~$\dec$ to its original wavepacket, $\xi_\dec$; however, the field operators' photon-wavepacket related to the other senders ($\tran \ \in{1,\ldots, \dec-1, \dec+1,\ldots, \text{M}}$) remains encoded, and it is shown as $\xi_\tran^{\ep_\tran d_\dec}$.}
         \label{fig:qcdma}
       \end{figure*}

         Figure~\ref{fig:qcdma} shows a simplified QCDMA schematic composed of M transmitters from the point of view of receiver~$\recv$, which decodes the signal sent by the intended ($\dec$th) transmitter.
         Each transmitter sends a pure quantum state composed of photons with a distinct spectrum.
         The spectrum of transmitter~$\tran, \ \forall \  \tran \in{1,\ldots,\text{M}}, $ can be identified by $\xi_{\tran}$.
         Sender~$\tran$ uses code~$\tran$ to encode its quantum signal.
         Receiver~$\recv$ uses the conjugation of the $\dec$th code to decode the intended $\dec$th signal.
         For precise details on two special cases, namely, Glauber states and number states inputs, see sections~\ref{sec:qcdmaglauber} and~\ref{sec:qcdmafock} .

\subsection{\textbf{QCDMA via Continuous Mode Glauber States}}
\label{sec:qcdmaglauber}
        In this section, we assume transmitter~$\tran$ sends its signal via a continuous mode Glauber state
        $\lvert \psi_{\tran} \rangle=\lvert \alpha_{\tran} \xi_{\tran}\rangle=e^{-\frac{\lvert \alpha_{\tran} \rvert^2 }{2}} e^{\alpha_{\tran}  \hat a_{\tran, \xi_{\tran}}^{ \dagger}} \lvert 0 \rangle = f_{\tran}( \hat a_{\tran, \xi_{\tran}}^{ \dagger}) \lvert 0 \rangle$,
        where $|\alpha_{\tran}|^2$ and $\xi_{\tran}$ indicate the total light intensity and the photon-wavepacket of the $\tran$th quantum transmitter, respectively.
        One needs to note that the spectral intensity at angular frequency $\omega$ is $ \lvert \alpha_{\tran} \xi_{\tran} (\omega) \rvert^2$  and photon-wavepacket is normalized ($\int d \omega \, \lvert \xi_{\tran} (\omega) \rvert^2=1 $); therefore, the mean intensity is
        \begin{equation}
          \bar I = \int d \omega \, \lvert \alpha_{\tran} \xi_{\tran} (\omega) \rvert^2=  \lvert \alpha_{\tran} \rvert^2 \int d \omega \, \lvert \xi_{\tran} (\omega) \rvert^2=\lvert \alpha_{\tran} \rvert^2 \, .
        \end{equation}        
        The signal is spectrally encoded via its corresponding binary phase-shifting operator $\hat{\text{U}}_{\tran}$.
        As shown in appendix~\ref{sec:pso} (see Eq.~\eqref{eqn:uadud}), the phase-shifting operator changes the photon-wavepacket to $\xi^{\ep_{\tran}}_{\tran}$.
        And consequently, following the same procedure as Eq.~\eqref{eqn:psie0}-\eqref{eqn:psie1}, the state vector of the encoded $\tran$th sender's signal can be expressed as 
        \begin{equation}
          \begin{split}
            \lvert \psi^\ep_{\tran} \rangle&= \hat{\text{U}}_{\tran} \lvert \alpha_{\tran} \xi_{\tran}\rangle\\
            &=  e^{-\frac{\lvert \alpha_{\tran} \rvert^2 }{2}} e^{\alpha_{\tran}  ( \hat{\text{U}}_{\tran} \hat a_{ \tran, \xi_{\tran}}^{ \dagger} \hat{\text{U}}^\dagger_{\tran})} \lvert 0 \rangle\\
            &=e^{-\frac{\lvert \alpha_{\tran} \rvert^2 }{2}} e^{\alpha_{\tran}  \hat a_{\tran, \xi^{\ep_{\tran}}_{\tran}}^{ \dagger}} \lvert 0 \rangle\\
            &=  f_{\tran}(\hat a_{\tran, \xi^{\ep_{\tran}}_{\tran}}^{ \dagger})\lvert 0 \rangle\\
            &=  \lvert \alpha_{\tran} \xi^{\ep_{\tran}}_{\tran}\rangle\, .
          \end{split}
        \end{equation}
        Let us assume that the network comprises M transmitters, sending their encoded continuous mode Glauber states into the broadcasting star-coupler.
        Therefore, the quantum state at the star-coupler's input is        
        \begin{equation}
          \begin{split}
            \lvert \Psi^{\ep}  \rangle &= \prod_{\tran=1}^{\text{M}} \lvert \psi^\ep_{\tran} \rangle\\
            &=\prod_{\tran=1}^{\text{M}} \lvert \alpha_{\tran} \xi^{\ep_{\tran}}_{\tran}\rangle\\
            &= \prod_{\tran=1}^{\text{M}} e^{-\frac{\lvert \alpha_{\tran} \rvert^2 }{2}} e^{\alpha_{\tran}  \hat a_{\tran,\xi^{\ep_{\tran}}_{\tran}}^{ \dagger}} \lvert 0 \rangle\\
            &=  \prod_{\tran=1}^{\text{M}} f_{\tran}(\hat a_{\tran, \xi^{\ep_{\tran}}_{\tran}}^{ \dagger})\lvert 0 \rangle\, ,
          \end{split}
        \end{equation}
        and the star-coupler's output, using Eq.~\eqref{eqn:phiprove}, gives
        \begin{equation}
          \begin{split}
            \lvert \Phi^{\ep}  \rangle &=\hat{\text{B}}^{\dagger} \lvert \Psi^{\ep}  \rangle\\
            &=\prod_{\tran=1}^{\text{M}} f_{\tran}(\sum_{\recv =1}^{\text{M}} B_{\recv \tran} \, \hat a_{\recv,\xi^{\ep_{\tran}}_{\tran}}^{ \dagger}) \lvert 0 \rangle\\
            &=\prod^{\text{M}}_{\tran=1} e^{-\frac{\lvert \alpha_{\tran} \rvert^2 }{2}} e^{\alpha_{\tran} \sum_{\recv} B_{\recv \tran} \hat a_{\recv,\xi^{\ep_{\tran}}_{\tran}}^{\dagger}} \lvert 0 \rangle \\
            &=\prod^{\text{M}}_{\tran=1} \prod^{\text{M}}_{\recv=1} e^{-\frac{\lvert \alpha_{\tran} \rvert^2 }{2}} e^{\alpha_{\tran} B_{\recv \tran} \hat a_{\recv,\xi^{\ep_{\tran}}_{\tran}}^{\dagger}} \lvert 0 \rangle \\
            &= e^{-\frac{\sum_{\tran} \lvert \alpha_{\tran} \rvert^2 }{2}} \prod^{\text{M}}_{\recv=1} e^{\sum^{\text{M}}_{\tran} \alpha_{\tran} B_{\recv \tran} \hat a_{\recv,\xi^{\ep_{\tran}}_{\tran}}^{ \dagger}} \lvert 0 \rangle \\
            &= \prod^{\text{M}}_{\recv=1} \Big \lvert \sum^{\text{M}}_{\tran=1}  B_{\recv \tran} \alpha_{\tran} \xi^{\ep_{\tran}}_{\tran} \Big \rangle \\
            &= \prod^{\text{M}}_{\recv=1} \lvert \phi^\ep_{\recv} \rangle\, ,
          \end{split}
          \label{eqn:Phiglauber}
        \end{equation}
        where $\lvert \phi^\ep_{\recv} \rangle$ is the encoded quantum state of the signal going towards the $\recv$th quantum decoder and receiver.
        Also, note that this pure quantum state of light at the star-coupler's output is factorized and non-entangled when the inputs are continuous mode Glauber states, as one would expect.
        
        Let us assume receiver~$\recv $ decodes the signal sent by the intended transmitter~$\dec$; specifically, receiver~$\recv $ implements operator $\hat{\text{U}}^ \dagger_{\dec }$, the conjugate of the spectral phase-shifter of the $\dec$th code, see Fig.~\ref{fig:qcdma}.
        Each quantum decoder decodes its corresponding received signal, and therefore, the multiple-access decoding operator, Eq.~\eqref{eqn:madec}, changes the state vector (Eq.~\eqref{eqn:Phiglauber}) as follows
        
        \begin{equation}
          \begin{split}
            \lvert \Phi^{d} \rangle &= \hat{\textbf{U}}^ \dagger \, \ \lvert \Phi^{\ep}  \rangle\ \\
            &= \left( \prod^{\text{M}}_{\recv=1} \hat{\text{U}}^ \dagger_{ \{\recv,c_{\dec}\} } \right)
            \Big ( \prod^{\text{M}}_{r'=1} \lvert \phi^\ep_{r'}\rangle \Big ) \\
            &= \prod^{\text{M}}_{\recv=1} \left( \hat{\text{U}}^ \dagger_{ \{\recv,c_{\dec}\} } \lvert \phi^\ep_{\recv}\rangle \right) \\
            &= \prod^{\text{M}}_{\recv=1} \left( \hat{\text{U}}^ \dagger_{ \{ \recv,c_{\dec} \} } \Big \lvert \sum^{\text{M}}_{\tran=1}  B_{\recv \tran} \alpha_{\tran} \xi^{\ep_{\tran}}_{\tran} \Big \rangle \right)
            \\
            &= \prod^{\text{M}}_{\recv=1}  \Big \lvert \sum^{\text{M}}_{\tran=1}  B_{\recv \tran} \alpha_{\tran} \xi^{\ep_{\tran}  d_{\dec} }_{\tran} \Big \rangle\\
            &= \prod^{\text{M}}_{\recv=1} \Big \lvert B_{\recv \dec} \alpha_{\dec} \xi_{\dec}+\sum^{\text{M}}_{\tran\neq \dec}  B_{\recv \tran} \alpha_{\tran} \xi^{\ep_{\tran}  d_{\dec} }_{\tran}\Big \rangle\, ,
          \end{split}
          \label{eqn:phiDglauber}
        \end{equation}
        where \eqref{eqn:udadu} is used, and  $\xi^{\ep_{\tran}  d_{\dec}}_{\tran} $ is the spectrally phase-shifted photon-wavepacket of photon-wavepacket $\xi^{\ep_{\tran}}_{\tran} $ by the decoder corresponding to transmitter~$\dec$.
        Note that in Eq.~\eqref{eqn:phiDglauber}, $\dec$ is a function of $\recv$, that is, $\dec = \dec(\recv)$.
        \par
        Receivers' decoded state vector $ \lvert \Phi^{d} \rangle$ is a Glauber factorized state, and the Glauber state is an eigenstate of annihilation operator at time~$\tp$, to express mathematically,  for $\recv=\recv_0$:
        \begin{equation}
          \begin{split}
            \hat a _{\recv_0}(\tp) \lvert \Phi^{d} \rangle
            &= \left(B_{\recv_0 \dec} \alpha_{\dec} \xi_{\dec}(\tp)+\sum_{\tran\neq \dec}  B_{\recv_0 \tran} \alpha_{\tran} \xi^{\ep_{\tran}  d_{\dec} }_{\tran}(\tp)  \right)\lvert \Phi^{d} \rangle\, ;
          \end{split}
        \end{equation}
        therefore, it is straightforward to calculate the expectation value of receivers' intensity, which for receiver~$\recv_0$ at time~$\tp$ gives
         \begin{equation}
          \begin{split}
           I_{\recv_0} (\tp)&=\langle \Phi^{d} \rvert \hat a^{\dagger} _{\recv_0}(\tp)\hat a _{\recv_0}(\tp) \lvert \Phi^{d} \rangle\\
            &= \Big \lvert B_{\recv_0 \dec} \alpha_{\dec} \xi_{\dec}(\tp)+\sum^{\text{M}}_{\tran\neq \dec}  B_{\recv_0 \tran} \alpha_{\tran} \xi^{\ep_{\tran}  d_{\dec} }_{\tran}(\tp)\Big \rvert^{2}\, .
          \end{split}
          \label{eqn:Irecv0}
        \end{equation}
        For example, if receiver~$\recv_0=1$ intends to decode the signal of transmitter~$\dec=1$, i.r., $\dec(\recv_0)=1$, its measured intensity would be
        $I_{1} = \lvert B_{1 1} \alpha_{1} \xi_{1}(\tp)+\sum^{\text{M}}_{\tran=2}  B_{1 \tran} \alpha_{\tran} \xi^{\ep_{\tran}  d_{1} }_{\tran}(\tp)\rvert^{2}$.
        It contains the decoded signal of transmitter~1, namely $\alpha_{1} \xi_{1}(\tp)$, and the improperly decoded signals (multiaccess interfering signal) of other transmitters, namely $ \alpha_{\tran} \xi^{\ep_{\tran}  d_{1} }_{\tran}(\tp), \tran =2,3, \hdots, \text{M}\,$.
        
        \vspace{5mm}         
        \noindent
        \textbf{Example: Two User QCDMA with Glauber State Inputs }

        \noindent        
        For simplicity, assume a two user QCDMA, a network composed of two quantum transmitters and two quantum receivers, where receiver~1(2) decodes the signal sent by transmitter~1(2).
        Therefore, the decoded signal's quantum state (Eq.~\eqref{eqn:phiDglauber}) reads 
        \begin{equation}
          \begin{split}
            \lvert \Phi^{d} \rangle &= \Big \lvert B_{11} \alpha_{1} \xi_{1}+ B_{12} \alpha_2 \xi^{\ep_2 d_1 }_{2}\Big \rangle \Big \lvert B_{22} \alpha_{2} \xi_{2}+ B_{21} \alpha_1 \xi^{\ep_1 d _{2} }_{1}\Big \rangle \, .
          \end{split}
        \end{equation}
        Assuming the transform matrix
        \begin{equation}
          \mathbf{\underline B}=
            \begin{pmatrix}
              B_{11} & B_{12}\\
              B_{21} & B_{22} 
            \end{pmatrix}=
            \frac{1}{\sqrt 2 }
            \begin{pmatrix}
              1 & 1\\
              -1 & 1 
            \end{pmatrix}\, ,
          \end{equation}
          for the 2$\times$2 star-coupler, receiver~1 at time~$t$ would measure the following received light intensity (Eq.~\eqref{eqn:Irecv0}):     
        \begin{equation}
          \begin{split}
            I_1(\tp) &=  \Big \lvert B_{11} \alpha_{1} \xi_{1}(\tp)+ B_{12} \alpha_2 \xi^{\ep_2 d_1}_{2}(\tp)\Big \rvert^2\\
             &=\frac{1}{2}\Big \lvert \alpha_{1} \xi_{1}(\tp)+ \alpha_2 \xi^{\ep_2 d_1}_{2}(\tp)\Big \rvert^2\\
            &= \frac{1}{2} | \alpha_{1} \xi_{1}(\tp) |^2+ \frac{1}{2}| \alpha_2 \xi^{\ep_2 d_1}_{2}(\tp) |^2 \\
            & \ \ +\text{Re}( \alpha_{1} \alpha^{\star}_{2}  \xi_{1}(\tp)  \xi^{\ep_2 d_1 \star}_{2}(\tp) )\\
          \end{split}
          \label{eqn:glauberio1}
        \end{equation}
        Equation~\eqref{eqn:glauberio1} is the sum of intensities of the desired decoded signal $|\xi_{1}(\tp) |^2$ and the multiaccess undesired improperly decoded signal $| \xi^{\ep_2 d_1}_{2}(\tp) |^2$, and inter-signal interference $\text{Re} (| \xi_{1}(\tp)  \xi^{\ep_2 d_1 \star}_{2}(\tp) |)$.
        The second receiver's intensity $I_2(\tp)$ is similar; however, the intra-signal interference appears with a minus sign:
        \begin{equation}
          \begin{split}
            I_2(\tp)&=\frac{1}{2}\lvert \alpha_2\xi_{2}(\tp)|^2+ \frac{1}{2}\lvert \alpha_1 \xi^{\ep_1 d _{2} }_{1} (\tp) \rvert^2\\
            & \ \ -\text{Re}(\alpha_2 \alpha_1^{\star}\xi_{2}(\tp) \xi^{\ep_1 d _{2} \star}_{1} (\tp))\, .
          \end{split}
        \end{equation}        

        Typically, undesired users' transmitted signals are asynchronous, meaning that each undesired transmitter can transmit its signal in all manner of time.
        Therefore, the temporal centers of photon-wavepackets of the undesired users ($\tp_0$ in Eq.~\eqref{eqn:xiGomega} and~\eqref{eqn:xit}) are random variable with a uniform distribution within a one-bit period.
        The worst-case scenario from the strength of multiaccess interfering signals would be when transmitters transmit their signals simultaneously, which is equivalent to synchronous QCDMA.        
        Let us consider this scenario in Eq.~\eqref{eqn:glauberio1}.
        Assuming  $| \xi_1(\tp_0)|=1$,  for $\alpha_1=1$ and $\alpha_2 =0$ then $I_1(\tp_0)=\nicefrac{1}{2}$ and signal is decoded without error, however for $\alpha_1=1$ and $\alpha_2=1$  (both signals send binary one) then $I(\tp_0)= \frac{1}{2}|1+\xi^{\ep_2 d_1 }_{2}(\tp_0)|^2$.
        For large code length $N_c$ and by invoking central limit theorem, one can show that ${\xi^{\ep_2 d_1 }_{2}(\tp_0)}$ is a Gaussian random variable with mean zero and variance equal to $1/N_c$. 
        Therefore the average value of the output intensity at port~1 is approximately $\mathbb E \{I_1(\tp_0)\}  \approx \frac{1}{2}(1+\frac{1}{N_c})$.
        In this case and for a large $N_c$~value, one can claim that $I_1$ is decoded correctly with high probability.
        A similar argument can be applied for the second receiver's multiaccess signal with quantum state vector        
         $ \lvert B_{22} \alpha_{2} \xi_{2}+ B_{21} \alpha_1 \xi^{\ep_1 d _{2} }_{1} \rangle = \frac{1}{\sqrt{2}}\lvert \alpha_{2} \xi_{2}- \alpha_1 \xi^{\ep_1 d _{2} }_{1} \rangle$ .                  
\subsection{ \textbf{QCDMA via Continuous Mode Fock States}}
\label{sec:qcdmafock}

\noindent
        This section assumes transmitters send their signals via a continuous mode Fock number states; specifically, sender~$\tran$ sends $\lvert \psi_{\tran} \rangle=\lvert n_{\xi_{\tran}} \rangle=  \frac{1}{\sqrt{n_{\tran}!}}(\hat a^{\dagger}_{\tran,\, \xi_{\tran}} )^{n_{\tran}} \lvert 0\rangle$.
        Afterward, encoding the signal via a binary spectral phase-shifting operator (see appendix~\ref{sec:pso} and Eq.~\eqref{eqn:uadud}) changes the photon-wavepacket from $\xi_{\tran}$ to $\xi^{\ep_{\tran}}_{\tran}$; therefore, $\tran$th sender's  encoded  quantum state reads
        \begin{equation}
          \begin{split}
            \lvert \psi^{\ep} _{\tran} \rangle&=\Big\lvert n_{\xi^{\ep_{\tran}}_{\tran}} \Big\rangle=  \frac{1}{\sqrt{n_{\tran}!}}(\hat a^{\dagger}_{\tran,\, \xi^{\ep_{\tran}}_{\tran}} )^{n_{\tran}} \lvert 0\rangle\, .
          \end{split}
          \label{eqn:fnin}
        \end{equation}              
        Each of M transmitters sends its desired quantum signal via an encoded continuous mode number state; therefore, the input into the quantum broadcasting star-coupler is         
        \begin{equation}
          \begin{split}
            \lvert \Psi^{\ep}  \rangle&=\Big \lvert n_{\xi^{\ep_1}_{1}} \Big \rangle \Big \lvert n_{\xi^{\ep_2}_{2}} \Big \rangle  \dots  \Big \lvert n_{\xi^{\ep_{\text{M}}}_{\text{M}}} \Big \rangle \\
            &=\prod_{\tran=1}^{\text{M}} \Big \lvert n_{\xi^{\ep_{\tran}}_{\tran}} \Big \rangle \\
            &=\prod_{\tran=1}^{\text{M}} \frac{1}{\sqrt{n_{\tran}!}}(\hat a^{\dagger}_{\tran,\, \xi^{\ep_{\tran}}_{\tran}} )^{n_{\tran}} \lvert 0\rangle \, .
          \end{split}
          \label{eqn:fnsin}
        \end{equation}
        Equation~\eqref{eqn:phiprove} gives the output of the star-coupler, where $f$ is a power function for number state input-signals (Eq.~\eqref{eqn:fnin});
        therefore, the output reads
        \begin{equation}
          \begin{split}
            \lvert \Phi^{\ep}  \rangle &=\hat{\text{B}}^{\dagger} \lvert \Psi^{\ep}  \rangle\\
            &= \hat{\text{B}}^{\dagger} \prod_{\tran=1}^{\text{M}}  \frac{1}{\sqrt{n_{\tran}!}}(\hat a^{\dagger}_{\tran,\, \xi^{\ep_{\tran}}_{\tran}} )^{n_{\tran}}  \lvert 0\rangle\\
            &=  \prod_{\tran=1}^{\text{M}}  \frac{1}{\sqrt{n_{\tran}!}}( \hat{\text{B}}^{\dagger} \hat a^{\dagger}_{\tran,\, \xi^{\ep_{\tran}}_{\tran}} \hat{\text{B}})^{n_{\tran}}  \lvert 0\rangle\\
            &= \prod_{\tran=1}^{\text{M}}  \frac{1}{\sqrt{n_{\tran}!}} \left(\sum^{\text{M}}_{\recv=1} B_{\recv \tran} \hat a_{\recv,\xi^{\ep_{\tran}}_{\tran}}^{\dagger} \right )^{n_{\tran}}  \lvert 0\rangle \, .
          \end{split}
          \label{eqn:Phifock}
        \end{equation}
        Note that since the above pure quantum state is not factorizable, it contains entanglement among the star-coupler's output ports; as appose to Eq.~\eqref{eqn:Phiglauber}, the corresponding state for Glauber state input-signals, which is factorized and non-entangled.
            
        Assume each receiver decodes its intended transmitter's signal by applying the appropriate decoding operator, that is, for all $ \recv \in{1,\ldots,\text{M}}$ exists $\dec = \dec(\recv)$  so that receiver $\recv$ applies $\hat{\text{U}}^ \dagger_{\dec }$.
        Therefore, the multiple-access decoding operator, Eq.~\eqref{eqn:madec}, transforms the state of Eq.~\eqref{eqn:Phifock} to the following state
        \begin{equation}
          \begin{split}
            \lvert \Phi^{d} \rangle &= \hat{\textbf{U}}^ \dagger \, \ \lvert \Phi^{\ep}  \rangle\ \\
            &=\hat{\textbf{U}}^ \dagger \, 
            \prod_{\tran=1}^{\text{M}}  \frac{1}{\sqrt{n_{\tran}!}} \left(\sum^{\text{M}}_{\recv=1} B_{\recv \tran} \hat a_{\recv,\xi^{\ep_{\tran}}_{\tran}}^{\dagger} \right )^{n_{\tran}}  \lvert 0\rangle  \\
             &= \prod_{\tran=1}^{\text{M}}  \frac{1}{\sqrt{n_{\tran}!}} \left(\sum^{\text{M}}_{\recv=1} B_{\recv \tran}  \hat{\textbf{U}}^ \dagger \hat a_{\recv,\xi^{\ep_{\tran}}_{\tran}}^{\dagger} \hat{\textbf{U}} \right )^{n_{\tran}}  \lvert 0\rangle  \\
              &=  \prod_{\tran=1}^{\text{M}}  \frac{1}{\sqrt{n_{\tran}!}} \left(\sum^{\text{M}}_{\recv=1} B_{\recv \tran}  \hat{\text{U}}^ \dagger_{ \{\recv,c_{\dec}\} }  \hat a_{\recv,\xi^{\ep_{\tran}}_{\tran}}^{\dagger} \hat{\text{U}}_{ \{\recv,c_{\dec}\} } \right )^{n_{\tran}}  \lvert 0\rangle \\
               &= 
            \prod_{\tran=1}^{\text{M}}  \frac{1}{\sqrt{n_{\tran}!}} \left(\sum^{\text{M}}_{\recv=1} B_{\recv \tran} \hat a_{\recv,\xi^{\ep_{\tran} d_{\dec}}_{\tran}}^{\dagger} \right )^{n_{\tran}}  \lvert 0\rangle
          \end{split}
          \label{eqn:Phidfockgen}
        \end{equation}
        As stated above,  $\dec$ is a function of $\recv$, let us assume $\dec=\recv$, that is, receiver $1,2,\hdots,\text{M}$ decode the signal of transmitter  $1,2,\hdots,\text{M}$, respectively.
        This assumption ($\hat{\text{U}}^{\dagger}_{ \{\recv,c_{\dec}\} } =\hat{\text{U}}^{\dagger}_{ \{\recv,c_{\recv}\} } $) reduces Eq.~\eqref{eqn:Phidfockgen} to  
        
        \begin{equation}
          \begin{split}
            \lvert \Phi^{d} \rangle &=   \prod_{\tran=1}^{\text{M}}  \frac{1}{\sqrt{n_{\tran}!}} \left(\sum^{\text{M}}_{\recv=1} B_{\recv \tran} \hat a_{\recv,\xi^{\ep_{\tran} d_{\recv}}_{\tran}}^{\dagger} \right )^{n_{\tran}}  \lvert 0\rangle \\
             &=   \prod_{\tran=1}^{\text{M}}  \frac{1}{\sqrt{n_{\tran}!}} \left(  B_{\tran \tran} \hat a_{\tran ,\xi_{\tran}}^{\dagger} + \sum^{\text{M}}_{\recv \neq \tran} B_{\recv \tran} \hat a_{\recv,\xi^{\ep_{\tran} d_{\recv}}_{\tran}}^{\dagger} \right )^{n_{\tran}}  \lvert 0\rangle 
           \end{split}
           \label{eqn:PhidFock}
        \end{equation}        
        To calculate the expectation value of the intensity measured by receiver~$\recv_0$ at time~$\tp$,
        similar to Eq.~\eqref{eqn:apsi}-\eqref{eqn:Ixiprove}, we first calculate the effect of the corresponding annihilation operator at time~$\tp$, $\hat a_{\recv_0}(\tp)$, on the decoded state~$\lvert \Phi^{d} \rangle$. That gives
        \begin{equation}
          \begin{split}
            \hat a_{\recv_0}(\tp)\lvert \Phi^{d} \rangle &=  \hat a_{\recv_0}(\tp)  \prod_{\tran=1}^{\text{M}}  \frac{1}{\sqrt{n_{\tran}!}} \left(\sum^{\text{M}}_{\recv=1} B_{\recv \tran} \hat a_{\recv,\xi^{\ep_{\tran} d_{\recv}}_{\tran}}^{\dagger} \right )^{n_{\tran}}  \lvert 0\rangle \\
            &=\Big[\hat a_{\recv_0}(\tp),  \prod_{\tran=1}^{\text{M}}  \frac{1}{\sqrt{n_{\tran}!}} \left(\sum^{\text{M}}_{\recv=1} B_{\recv \tran} \hat a_{\recv,\xi^{\ep_{\tran} d_{\recv}}_{\tran}}^{\dagger} \right )^{n_{\tran}} \Big] \lvert 0\rangle \\
            \\
            &=\sum_{\tran=1}^{\text{M}} \xi^{\ep_{\tran} d_{{\recv}_0}}_{\tran} (\tp) \prod_{\tran'\neq \tran }^{\text{M}}  \frac{1}{\sqrt{n_{\tran'}!}} \left(\sum^{\text{M}}_{\recv=1} B_{\recv \tran'} \hat a_{\recv,\xi^{\ep_{\tran'} d_{\recv}}_{\tran'}}^{\dagger} \right )^{n_{\tran'}}  \\
             & \hspace{1em}\times \frac{1}{\sqrt{n_{\tran}!}}
            \frac{\partial}{\partial \hat a^ {\dagger}_{ \recv_0, \xi^{\ep_{\tran} d_{{\recv}_0}}_{\tran}} }
             \left(\sum^{\text{M}}_{\recv=1} B_{\recv \tran} \hat a_{\recv,\xi^{\ep_{\tran} d_{\recv}}_{\tran}}^{\dagger} \right )^{n_{\tran}}\lvert 0 \rangle\\
       \\
            &  =\sum_{\tran=1}^{\text{M}} \xi^{\ep_{\tran} d_{{\recv}_0}}_{\tran} (\tp) \prod_{\tran'\neq \tran}^{\text{M}}  \frac{1}{\sqrt{n_{\tran'}!}} \left( \hat b^{\dagger}_{\tran'}\right )^{n_{\tran'}} \\
            & \hspace{1em}\times
            \frac{n_s B_{r_0 s}}{\sqrt{n_{\tran}!}} \left(\hat b^{\dagger}_{\tran} \right )^{n_{\tran}-1}
             \lvert 0\rangle\, ,
           \end{split}
           \label{eqn:atPhidfock}
        \end{equation}        
        where Eq.~\eqref{eqn:[a,fmv]} is used, and operators $\hat b^{\dagger}_{\tran}$ are defined to be      
        \begin{equation}
          \begin{split}         
            \hat b^{\dagger}_{\tran}&=\sum^{\text{M}}_{\recv=1} B_{\recv \tran} \hat a_{\recv,\xi^{\ep_{\tran} d_{\recv}}_{\tran}}^{\dagger}  \, , \qquad \tran \in{1,\ldots,\text{M}} \, .
          \end{split}
          \label{eqn:bdag}
        \end{equation}
        and they correspond to M~orthogonal modes~$\tran =1, 2, \hdots, \text{M}$.
        Therefore, between them there is the canonical commutation relation~$ [ \hat b_{\tran}  , \hat b^{\dagger}_{\tran'} ]=\delta_{\tran \tran'}$.
        The commutation relation can be proved as follows
        \begin{equation}
          \begin{split}            
             [ \hat b_{\tran}  , \hat b^{\dagger}_{\tran'} ] &=\left[
            \sum^{\text{M}}_{\recv=1} B^{\star}_{\recv \tran} \hat a_{\recv,\xi^{\ep_{\tran} d_{\recv}}_{\tran}} , \sum^{\text{M}}_{\recv'=1} B_{\recv' \tran'} \hat a_{\recv',\xi^{\ep_{\tran'} d_{\recv'}}_{\tran'}}^{\dagger} 
          \right]\\
          &=\sum^{\text{M}}_{\recv=1} \sum^{\text{M}}_{\recv'=1} B^{\star}_{\recv \tran} B_{\recv' \tran'} \left[
             \hat a_{\recv,\xi^{\ep_{\tran} d_{\recv}}_{\tran}} ,  \hat a_{\recv',\xi^{\ep_{\tran'} d_{\recv'}}_{\tran'}}^{\dagger} 
           \right]\\
            &=\sum^{\text{M}}_{\recv=1} \sum^{\text{M}}_{\recv'=1} B^{\star}_{\recv \tran} B_{\recv \tran'} \left[
             \hat a_{\recv,\xi^{\ep_{\tran} d_{\recv}}_{\tran}} ,  \hat a_{\recv,\xi^{\ep_{\tran'} d_{\recv}}_{\tran'}}^{\dagger} 
            \right] \delta_{\recv, \recv'} \\
            &= \sum^{\text{M}}_{\recv=1} B^{\star}_{\recv \tran} B_{\recv \tran'} \left[
             \hat a_{\recv,\xi^{\ep_{\tran} d_{\recv}}_{\tran}} ,  \hat a_{\recv,\xi^{\ep_{\tran'} d_{\recv}}_{\tran'}}^{\dagger} 
           \right]  \, .
         \end{split}
         \label{eqn:[bs,bds']pre}
       \end{equation}
       Eq.~\eqref{eqn:aiadj} indicates that for the commutation relation, equality $[\hat a_{\recv,\xi^{\ep_{\tran} d_{\recv}}_{\tran}} ,  \hat a_{\recv,\xi^{\ep_{\tran'} d_{\recv}}_{\tran'}}^{\dagger}] =  \langle \xi^{\ep_{\tran} d_{\recv}}_{\tran} \vert \xi^{\ep_{\tran'} d_{\recv}}_{\tran'}\rangle$ holds and, using Eq.~\eqref{eqn:xiedxi}, the commutation relation becomes $ [\hat a_{\recv,\xi^{\ep_{\tran} d_{\recv}}_{\tran}} ,  \hat a_{\recv,\xi^{\ep_{\tran'} d_{\recv}}_{\tran'}}^{\dagger}] =\langle \xi^{\ep_{\tran}}_{\tran} \vert \xi^{\ep_{\tran'} d_{\recv}  e_{\recv}}_{\tran'}\rangle =  \langle \xi^{\ep_{\tran}}_{\tran} \vert \xi^{\ep_{\tran'}}_{\tran'}\rangle$, which is independent of parameter~$\recv$.
       Therefore, Eq.~\eqref{eqn:[bs,bds']pre} reduces to             
       \begin{equation}
         \begin{split}            
             [ \hat b_{\tran}  , \hat b^{\dagger}_{\tran'} ]  &= \langle \xi^{\ep_{\tran} }_{\tran} \vert \xi^{\ep_{\tran'}}_{\tran'}\rangle \sum^{\text{M}}_{\recv=1} B^{\star}_{\recv \tran} B_{\recv \tran'}   \\           
           &=  \langle \xi^{\ep_{\tran} }_{\tran} \vert \xi^{\ep_{\tran}}_{\tran}\rangle \delta_{\tran, \tran'}\\
           &= \delta_{\tran, \tran'}\, ,
          \end{split}          
        \end{equation}
         where the unitary property of matrix~$\underline{\mathbf{B}}$, i.e., $ \left( \underline{\mathbf{B}}^{\dagger} \underline{\mathbf{B}}\right)_{\tran \tran'}=\textbf{I}_{\tran,\tran'}=\delta_{\tran,\tran'}$, is used.
        One can express Eq.~\eqref{eqn:atPhidfock} as an expansion of the corresponding number states of field creation operators $\hat b^{\dagger}_{\tran}$, more precisely as: $ \lvert n_{\tran}\rangle =\frac{1} {\sqrt{n_{\tran}!}} \left( \hat b^{\dagger}_{\tran}\right )^{n_{\tran}} \lvert 0 \rangle$. 
        Therefore, Eq.~\eqref{eqn:atPhidfock} gives
        \begin{equation}
          \begin{split}
            \hat a_{\recv_0}(\tp)\lvert \Phi^{d} \rangle  & =\sum_{\tran=1}^{\text{M}} \sqrt{n_{\tran}} B_{r_0 s}  \xi^{\ep_{\tran} d_{{\recv}_0}}_{\tran}(\tp)
            \lvert n_1\rangle \lvert n_2\rangle \hdots  \lvert n_{\tran}-1\rangle \hdots \lvert n_{\text{M}}\rangle\, , \\
            \end{split}
          \end{equation}
          where M modes number state~$ \lvert n_1\rangle \lvert n_2\rangle \hdots  \lvert n_{\tran}-1\rangle \hdots \lvert n_{\text{M}}\rangle$ denotes $ \frac{1}{\sqrt{(n_{\tran}-1)!}} (\hat b^{\dagger}_{\tran} )^{n_{\tran}-1} \prod_{\tran'\neq \tran}^{\text{M}}  \frac{1}{\sqrt{n_{\tran'}!}}( \hat b^{\dagger}_{\tran'} )^{n_{\tran'}} \lvert 0\rangle$.
           Since field operators $\hat b^{\dagger}_{\tran}$ correspond to M~orthogonal modes, one can easily calculate the intensity at time $\tp$, which gives
          \begin{equation}
          \begin{split}
            I_{\recv_0}(\tp)&=\langle \Phi^{d}\rvert  \hat a^{\dagger}_{\recv_0}(\tp)  \hat a_{\recv_0}(\tp)\lvert \Phi^{d} \rangle\\
            & =\sum_{\tran=1}^{\text{M}}  n_{\tran}  \lvert B_{r_0 s} \rvert^2 \lvert\xi^{\ep_{\tran} d_{{\recv}_0}}_{\tran} (\tp)\rvert^2\, .
            \end{split}
          \end{equation}
          Let us assume the star-coupler is balanced, $\lvert B_{r_0 s} \rvert^2=1/\text{M}$; therefore, the measured intensity at receiver~$\recv_0$ at time~$\tp$ would be $ I_{\recv_0}(\tp) =1/\text{M}\sum_{\tran=1}^{\text{M}}  n_{\tran}\lvert\xi^{\ep_{\tran} d_{{\recv}_0}}_{\tran} (\tp)\rvert^2$.
          For example, the intensity of receiver~1 is
          \begin{equation}
            I_{1}(\tp) =\frac{1}{\text{M}}\left(n_1 \lvert\xi_{1}(\tp)\rvert^2 + \sum_{\tran=2}^{\text{M}}  n_{\tran} \lvert\xi^{\ep_{\tran} d_{1}}_{\tran} (\tp)\rvert^2 \right)\, ,
          \end{equation}
          which is the sum of the decoded signal of transmitter~1, $n_1 \lvert\xi_{1}(\tp)\rvert^2$, and the multiaccess interfering signal, $ \sum_{\tran=2}^{\text{M}}  n_{\tran} \lvert\xi^{\ep_{\tran} d_{1}}_{\tran} (\tp)\rvert^2$.
          One should note that there is no inter-signal interference because of the complete phase uncertainty of the input number state signals.
        \vspace{5mm}         
        \noindent
        \textbf{Example: Two User QCDMA with Single-Photon Inputs }

        \noindent
        Let us give an example for the number states QCDMA.
        Assume that the transmitters send quantum signals via an encoded continuous mode single-photon. 
        Therefore $n_{\tran}$ is 1, if user~$\tran$ sends a quantum signal for binary one and  vacuum state $\lvert 0 \rangle $, i.e.,  $n_{\tran}=0$, for binary zero.
        Consider a challenging case where two quantum transmitters simultaneously send their corresponding encoded single-photon pulse into the star-coupler.
        Then the state vector of the star-coupler input (Eq.~\eqref{eqn:fnsin}) is as follows
        \begin{equation}
          \begin{split}
            \lvert \Psi^{\ep} \rangle&=\lvert 1_{\xi^{\ep_1}_1} \rangle \lvert 1_{\xi^{\ep_2}_2} \rangle \\
            &=\hat a^{\dagger}_{1,\, \xi^{\ep_1}_1} \hat a^{\dagger}_{2,\, \xi^{\ep_2}_2}\lvert 0\rangle \, .
          \end{split}
          \label{eqn:1in}
        \end{equation}
        And the output (Eq.~\eqref{eqn:Phifock}) gives
        \begin{equation}
          \begin{split}
            \lvert \Phi^{\ep} \rangle &= \left( B_{11} \hat a_{1,\xi^{\ep_1}_{1}}^{\dagger}+B_{21} \hat a_{2,\xi^{\ep_1}_{1}}^{\dagger} \right) \left( B_{12} \hat a_{1,\xi^{\ep_2}_{2}}^{\dagger}+B_{22} \hat a_{2,\xi^{\ep_2}_{2}}^{\dagger} \right) \lvert 0 \rangle\\
            &= \Big( B_{11}  B_{12} \hat a_{1,\xi^{\ep_1}_{1}}^{\dagger} \hat a_{1,\xi^{\ep_2}_{2}}^{\dagger}
            +B_{11} B_{22} \hat a_{1,\xi^{\ep_1}_{1}}^{\dagger}   \hat a_{2,\xi^{\ep_2}_{2}}^{\dagger}\\
            & \ \  +B_{21}  B_{12} \hat a_{2,\xi^{\ep_1}_{1}}^{\dagger} \hat a_{1,\xi^{\ep_2}_{2}}^{\dagger}
            +B_{21} B_{22}  \hat a_{2,\xi^{\ep_1}_{1}}^{\dagger} \hat a_{2,\xi^{\ep_2}_{2}}^{\dagger} \Big) \lvert 0 \rangle\\
            &= B_{11}  B_{12} \lvert( 1_{\xi^{\ep_1}_{1}},1_{\xi^{\ep_2}_{2}})\rangle \lvert 0\rangle
            +B_{11} B_{22} \lvert 1_{\xi^{\ep_1}_{1}}\rangle\lvert 1_{\xi^{\ep_2}_{2}}\rangle \\
            & \ \  +B_{21}  B_{12} \lvert  1_{\xi^{\ep_2}_{2}} \rangle \lvert 1_{\xi^{\ep_1}_{1}}\rangle
            +B_{21} B_{22}  \lvert 0 \rangle \lvert (1_{\xi^{\ep_1}_{1}}, 1_{\xi^{\ep_2}_{2}}) \rangle\, ,
          \end{split}
          \label{eqn:Phi1}
        \end{equation}
        where $ \lvert( 1_{\xi^{\ep_1}_{1}},1_{\xi^{\ep_2}_{2}})\rangle=  \hat a_{1,\xi^{\ep_1}_{1}}^{\dagger} \hat a_{1,\xi^{\ep_2}_{2}}^{\dagger} \lvert 0 \rangle$
        indicates a two-photon state where one photon is in wavepacket~${\xi^{\ep_1}_{1}}$ and the other in wavepacket~${\xi^{\ep_2}_{2}}$ and its normalization factor~$N_f$ is
        \begin{equation}
          \begin{split}
            N_f&=\sqrt{\langle( 1_{\xi^{\ep_1}_{1}},1_{\xi^{\ep_2}_{2}}) \vert( 1_{\xi^{\ep_1}_{1}},1_{\xi^{\ep_2}_{2}})\rangle}\\
            &=\sqrt{\langle 0 \rvert \hat a_{1,\xi^{\ep_2}_{2}} \hat a_{1,\xi^{\ep_1}_{1}} \hat a_{1,\xi^{\ep_1}_{1}}^{\dagger} \hat a_{1,\xi^{\ep_2}_{2}}^{\dagger} \lvert 0 \rangle}\\
            &=\sqrt{\langle 0 \rvert \hat a_{1,\xi^{\ep_2}_{2}} \hat a_{1,\xi^{\ep_2}_{2}}^{\dagger} \lvert 0 \rangle + \langle 0 \rvert \hat a_{1,\xi^{\ep_2}_{2}}  \hat a_{1,\xi^{\ep_1}_{1}}^{\dagger}  \hat a_{1,\xi^{\ep_1}_{1}}\hat a_{1,\xi^{\ep_2}_{2}}^{\dagger} \lvert 0 \rangle}\\
            &=\sqrt{1 + \langle \xi^{\ep_1}_{1} \vert \xi^{\ep_2}_{2} \rangle  \langle 0 \rvert \hat a_{1,\xi^{\ep_2}_{2}}  \hat a_{1,\xi^{\ep_1}_{1}}^{\dagger}  \lvert 0 \rangle}\\
             &=\sqrt{1 + \Big|\langle \xi^{\ep_1}_{1} \vert \xi^{\ep_2}_{2} \rangle \Big|^2}\, ,
           \end{split}
           \label{eqn:Nf}
          \end{equation}
          where Eq.~\eqref{eqn:aiadj} is used.
        Again, similar to section~\ref{sec:qcdmaglauber}, we assume receiver~1(2) decodes the signal sent by transmitter~1(2) via the conjugate of spectral phase-shifting operator  $\hat{\text{U}}^ \dagger_{1}$ ($\hat{\text{U}}^ \dagger_{2}$).
        We rewrite the receiver decoding operator as $\hat{\text{U}}^ \dagger_{ \{1,c_1\} }$ ($\hat{\text{U}}^ \dagger_{ \{2,c_2\} }$) to make the receiver which applies the phase-shifter more clear.
        As Eq.~\eqref{eqn:Phidfockgen}-\eqref{eqn:PhidFock} states, the multiple-access decoding operator $\hat{\textbf{U}}^ \dagger= \hat{\text{U}}^ \dagger_{ \{1,c_1\} }\hat{\text{U}}^ \dagger_{ \{2,c_2\} }$ changes the state vector (Eq.~\eqref{eqn:Phi1}) to
        \begin{equation}
          \begin{split}
            \lvert \Phi^{d} \rangle &= \hat{\text{U}}^ \dagger_{ \{2,c_2\} }\hat{\text{U}}^ \dagger_{ \{1,c_1\} }  \lvert \Phi^{\ep} \rangle \\
            &=   \prod_{\tran=1}^{2}  \left(  B_{\tran \tran} \hat a_{\tran ,\xi_{\tran}}^{\dagger} + \sum^{\text{M}}_{\recv \neq \tran} B_{\recv \tran} \hat a_{\recv,\xi^{\ep_{\tran} d_{\recv}}_{\tran}}^{\dagger} \right ) \lvert 0\rangle \\
            &=  \left(  B_{1 1} \hat a_{1 ,\xi_{1}}^{\dagger} +  B_{2 1} \hat a_{2,\xi^{\ep_{1} d_{2}}_{1}}^{\dagger} \right )\\
            & \qquad \times \left(   B_{1 2} \hat a_{1,\xi^{\ep_{2} d_{1}}_{2}}^{\dagger} +  B_{2 2} \hat a_{2,\xi_{2}}^{\dagger}  \right )  \lvert 0\rangle \\
            &= \Big( B_{11}  B_{12} \hat a_{1,\xi_{1}}^{\dagger} \hat a_{1,\xi^{\ep_2 d_1}_{2}}^{\dagger}
            +B_{11} B_{22} \hat a_{1,\xi_{1}}^{\dagger}   \hat a_{2,\xi_{2}}^{\dagger}\\
            & \ \  +B_{21}  B_{12} \hat a_{2,\xi^{\ep_1 \cm d_2}_{1}}^{\dagger} \hat a_{1,\xi^{\ep_2 d_1}_{2}}^{\dagger}
            +B_{21} B_{22}  \hat a_{2,\xi^{\ep_1 \cm d_2}_{1}}^{\dagger} \hat a_{2,\xi_{2}}^{\dagger} \Big) \lvert 0 \rangle\\  
            &= B_{11}  B_{12} \lvert (1_{\xi_{1}},1_{\xi^{\ep_2 d_1}_{2}})\rangle \lvert 0\rangle
            +B_{11} B_{22} \lvert 1_{\xi_{1}}\rangle\lvert 1_{\xi_{2}}\rangle \\
            & \ \  +B_{21}  B_{12} \lvert  1_{\xi^{\ep_2 d_1}_{2}} \rangle \lvert 1_{\xi^{\ep_1 \cm d_2}_{1}}\rangle
            +B_{21} B_{22}  \lvert 0 \rangle \lvert (1_{\xi^{\ep_1 \cm d_2}_{1}}, 1_{\xi_{2}}) \rangle\, ,            
          \end{split}
          \label{eqn:PhiD1}
        \end{equation}
        where $\lvert (1_{\xi_{1}},1_{\xi^{\ep_2 d_1}_{2}})\rangle$ and $\lvert (1_{\xi^{\ep_1 \cm d_2}_{1}}, 1_{\xi_{2}}) \rangle$ are two-photon states and their state normalization factor, as stated in Eq.~\eqref{eqn:Nf}, is:
        \begin{equation}
          \begin{split}
          N_f &=
          \sqrt{\langle (1_{\xi_{1}},1_{\xi^{\ep_2 d_1}_{2}}) \vert (1_{\xi_{1}},1_{\xi^{\ep_2 d_1}_{2}})\rangle}\\
          &= \sqrt{\langle (1_{\xi^{\ep_1 \cm d_2}_{1}}, 1_{\xi_{2}}) \vert (1_{\xi^{\ep_1 \cm d_2}_{1}}, 1_{\xi_{2}}) \rangle}\\
             &=\sqrt{1 + \Big|\langle \xi^{\ep_1}_{1} \vert \xi^{\ep_2}_{2} \rangle \Big|^2}\, ,          
          \end{split}
        \end{equation}
        where Eq.~\eqref{eqn:xiedxi} is used.
        Now, we calculate the expectation value of the light intensity at time~$\tp$.
        First, annihilation (detection) of a photon at time~$\tp$ by the quantum receiver~1 projects the state (Eq.~\eqref{eqn:PhiD1}) to the following state
        
        \begin{equation}
          \begin{split}
            \hat a_1(\tp) \lvert \Phi^{d} \rangle &=\Big( B_{11}  B_{12}  \xi_{1}(\tp) \hat a_{1,\xi^{\ep_2 d_1}_{2}}^{\dagger} + B_{11}  B_{12} \hat a_{1,\xi_{1}}^{\dagger} \hat a_1(\tp) \hat a_{1,\xi^{\ep_2 d_1}_{2}}^{\dagger}\\
            & \ \ +B_{11} B_{22} \xi_{1}(\tp)  \hat a_{2,\xi_{2}}^{\dagger}  +B_{21}  B_{12}  \xi^{\ep_2 d_1}_{2}(\tp) \hat a_{2,\xi^{\ep_1 \cm d_2}_{1}}^{\dagger} \Big) \lvert 0 \rangle\\           
            &=\Big( B_{11}  B_{12} \hat \xi_{1}(\tp) \hat a_{1,\xi^{\ep_2 d_1}_{2}}^{\dagger} + B_{11}  B_{12}  \xi^{\ep_2 d_1}_{2}(\tp) \hat a_{1,\xi_{1}}^{\dagger}\\
            & \ \ +B_{11} B_{22} \xi_{1}(\tp)  \hat a_{2,\xi_{2}}^{\dagger}  +B_{21}  B_{12}  \xi^{\ep_2 d_1}_{2}(\tp) \hat a_{2,\xi^{\ep_1 \cm d_2}_{1}}^{\dagger} \Big) \lvert 0 \rangle\, ,
          \end{split}
        \end{equation}
        where Eqs.~\eqref{eqn:[a,axi]} and~\eqref{eqn:apsi} are used. 
        Therefore, intensity measurement by receiver~1 gives:
        \begin{equation}
          \begin{split}
            I_1(\tp)&= \langle \Phi^{d} \lvert \hat a^{\dagger}_1(\tp) \hat a_1(\tp) \lvert \Phi^{d} \rangle\\
            &= |B_{11}  B_{12} \xi_{1}(\tp)|^2 + |B_{11}  B_{12}  \xi^{\ep_2 d_1}_{2}(\tp) |^2\\
            & \ \ +2\text{Re}\Big( B^{\star}_{1 1}   B^{\star}_{1 2} B_{11}  B_{12} \xi^{\star}_{1}(\tp)  \xi^{\ep_2 d_1}_{2}(\tp) \langle \xi^{\ep_2 d_1 }_{2} \vert \xi_{1}\rangle \Big)\\  
            & \ \ +|B_{11} B_{22} \xi_{1}(\tp) |^2 +| B_{21}  B_{12}  \xi^{\ep_2 d_1}_{2}(\tp) |^2\\
            & \ \ +2\text{Re}\Big( B^{\star}_{1 1} B^{\star}_{2 2}   B_{21}  B_{12}  \xi^{\star}_{1}(\tp) \xi^{\ep_2 d_1}_{2}(\tp)  \langle \xi_{2} \vert \xi^{\ep_1 \cm d_2}_{1}\rangle \Big)\, .
          \end{split}
          \label{eqn:I1presps}
        \end{equation}
        As Eq.~\eqref{eqn:xiedxi} shows, $\langle \xi_{2} \vert \xi^{\ep_1 \cm d_2}_{1}\rangle = \langle \xi^{\ep_2 d_1}_{2} \vert  \xi_{1}\rangle =\langle \xi^{\ep_2 }_{2} \vert  \xi^{\ep_1}_{1}\rangle$.
        Furthermore, $\mathbf{\underline  B}$ is a unitary matrix, $B_{21} B^{\star}_{11}=- B_{22} B^{\star}_{12} $ (see Eq.~\eqref{eqn:bbdag=I}).
       Therefore Eq.~\eqref{eqn:I1presps} gives:
        \begin{equation}
          \begin{split}
            I_1(\tp) &= |B_{11} |^2\left( | B_{12}|^2  + |B_{22}|^2 \right) |\xi_{1}(\tp)|^2\\  
            &+ | B_{12}|^2 \left( |B_{11} |^2+| B_{21}|^2 \right) | \xi^{\ep_2 d_1}_{2}(\tp) |^2  \, .
          \end{split}
        \end{equation}
        For balanced star-couplers (see appendix~\ref{sec:qibs}), $|B_{ij}|^2=\nicefrac{1}{\text{M}}$, where M is 2 in our example; therefore, the intensity reads 
        \begin{equation}
          \begin{split}
            I_1(\tp) &= \frac{1}{2} \left(|\xi_{1}(\tp)|^2+ | \xi^{\ep_2 d_1}_{2}(\tp) |^2 \right)           \, ,
          \end{split}
          \label{eqn:I1sps}
        \end{equation}
         which is composed of decoded signal intensity $| \xi_{1}(\tp) |^2$ and multiaccess signal intensity $|\xi^{\ep_2 d_1}_{2}(\tp) |^2$.
         Interestingly, the intensity in Eq.~\eqref{eqn:I1sps} does not contain inter-signal interference,  $\text{Re} (|\xi_{1}(\tp)  \xi^{\ep_2 d_1 \star}_{2}(\tp) |)$, which plays a part in the QCDMA via Glauber states, as Eq.~\eqref{eqn:glauberio1} shows.
         In other words, because a single-photon quantum state is an eigenstate of the number operator, the particle-like single-photon phase is totally random and can not produce interference.
         To put it another way, a single-photon state with sub-Poissonian photon statistics and zero photon-number uncertainty is subject to Heisenberg’s uncertainty principle and has a complete phase uncertainty.    

\section*{Acknowledgment}
    M. Rezai acknowledges the funding and the support from the Iran National Elite Foundation.

\ifCLASSOPTIONcaptionsoff
  \newpage
\fi
\bibliographystyle{IEEEtran}

\begin{thebibliography}{10}
\providecommand{\url}[1]{#1}
\csname url@samestyle\endcsname
\providecommand{\newblock}{\relax}
\providecommand{\bibinfo}[2]{#2}
\providecommand{\BIBentrySTDinterwordspacing}{\spaceskip=0pt\relax}
\providecommand{\BIBentryALTinterwordstretchfactor}{4}
\providecommand{\BIBentryALTinterwordspacing}{\spaceskip=\fontdimen2\font plus
\BIBentryALTinterwordstretchfactor\fontdimen3\font minus
  \fontdimen4\font\relax}
\providecommand{\BIBforeignlanguage}[2]{{%
\expandafter\ifx\csname l@#1\endcsname\relax
\typeout{** WARNING: IEEEtran.bst: No hyphenation pattern has been}%
\typeout{** loaded for the language `#1'. Using the pattern for}%
\typeout{** the default language instead.}%
\else
\language=\csname l@#1\endcsname
\fi
#2}}
\providecommand{\BIBdecl}{\relax}
\BIBdecl

\bibitem{helstrom_1976}
\BIBentryALTinterwordspacing
C.~W. Helstrom, \emph{{Quantum detection and estimation theory}}, ser. Math.
  Sci. Eng.\hskip 1em plus 0.5em minus 0.4em\relax New York, NY: Academic
  Press, 1976. [Online]. Available: \url{http://cds.cern.ch/record/110988}
\BIBentrySTDinterwordspacing

\bibitem{cariolaro_2015}
G.~Cariolaro, \emph{Quantum Communications}.\hskip 1em plus 0.5em minus
  0.4em\relax Springer Publishing Company, Incorporated, 2015.

\bibitem{wilde_2013}
M.~M. Wilde, \emph{Quantum Information Theory}.\hskip 1em plus 0.5em minus
  0.4em\relax Cambridge University Press, 2013.

\bibitem{razavi_2018}
\BIBentryALTinterwordspacing
M.~Razavi, \emph{An Introduction to Quantum Communications Networks}, ser.
  2053-2571.\hskip 1em plus 0.5em minus 0.4em\relax Morgan \& Claypool
  Publishers, 2018. [Online]. Available:
  \url{http://dx.doi.org/10.1088/978-1-6817-4653-1}
\BIBentrySTDinterwordspacing

\bibitem{kimble_n_2008}
\BIBentryALTinterwordspacing
H.~J. Kimble, ``The quantum internet,'' \emph{Nature}, vol. 453, no. 7198, pp.
  1023--1030, Jun. 2008. [Online]. Available:
  \url{http://dx.doi.org/10.1038/nature07127}
\BIBentrySTDinterwordspacing

\bibitem{wehner_s_2018}
\BIBentryALTinterwordspacing
S.~Wehner, D.~Elkouss, and R.~Hanson, ``Quantum internet: A vision for the road
  ahead,'' \emph{Science}, vol. 362, no. 6412, 2018. [Online]. Available:
  \url{https://science.sciencemag.org/content/362/6412/eaam9288}
\BIBentrySTDinterwordspacing

\bibitem{yard_tit_2011}
J.~{Yard}, P.~{Hayden}, and I.~{Devetak}, ``Quantum broadcast channels,''
  \emph{IEEE Transactions on Information Theory}, vol.~57, no.~10, pp.
  7147--7162, 2011.

\bibitem{zhang_sr_2013}
\BIBentryALTinterwordspacing
J.~Zhang, Y.-x. Liu, S.~K. {\"O}zdemir, R.-B. Wu, F.~Gao, X.-B. Wang, L.~Yang,
  and F.~Nori, ``Quantum internet using code division multiple access,''
  \emph{Scientific Reports}, vol.~3, no.~1, p. 2211, 2013. [Online]. Available:
  \url{https://doi.org/10.1038/srep02211}
\BIBentrySTDinterwordspacing

\bibitem{escartin_2015_jstonqe}
J.~C. {Garcia-Escartin} and P.~{Chamorro-Posada}, ``Quantum spread spectrum
  multiple access,'' \emph{IEEE Journal of Selected Topics in Quantum
  Electronics}, vol.~21, no.~3, pp. 30--36, 2015.

\bibitem{sharma_oqe_2020}
\BIBentryALTinterwordspacing
V.~Sharma and S.~Banerjee, ``Quantum communication using code division multiple
  access network,'' \emph{Optical and Quantum Electronics}, vol.~52, no.~8, pp.
  381--, 2020. [Online]. Available:
  \url{https://doi.org/10.1007/s11082-020-02494-3}
\BIBentrySTDinterwordspacing

\bibitem{giovannetti_prl_2004}
\BIBentryALTinterwordspacing
V.~Giovannetti, S.~Guha, S.~Lloyd, L.~Maccone, J.~H. Shapiro, and H.~P. Yuen,
  ``Classical capacity of the lossy bosonic channel: The exact solution,''
  \emph{Phys. Rev. Lett.}, vol.~92, p. 027902, Jan 2004. [Online]. Available:
  \url{https://link.aps.org/doi/10.1103/PhysRevLett.92.027902}
\BIBentrySTDinterwordspacing

\bibitem{brent_pra_2005}
\BIBentryALTinterwordspacing
B.~J. Yen and J.~H. Shapiro, ``Multiple-access bosonic communications,''
  \emph{Phys. Rev. A}, vol.~72, p. 062312, Dec 2005. [Online]. Available:
  \url{https://link.aps.org/doi/10.1103/PhysRevA.72.062312}
\BIBentrySTDinterwordspacing

\bibitem{guha_pra_2007}
\BIBentryALTinterwordspacing
S.~Guha, J.~H. Shapiro, and B.~I. Erkmen, ``Classical capacity of bosonic
  broadcast communication and a minimum output entropy conjecture,''
  \emph{Phys. Rev. A}, vol.~76, p. 032303, Sep 2007. [Online]. Available:
  \url{https://link.aps.org/doi/10.1103/PhysRevA.76.032303}
\BIBentrySTDinterwordspacing

\bibitem{shapiro_jstqe_2009}
J.~H. {Shapiro}, ``The quantum theory of optical communications,'' \emph{IEEE
  Journal of Selected Topics in Quantum Electronics}, vol.~15, no.~6, pp.
  1547--1569, 2009.

\bibitem{guha_itp_2011}
S.~{Guha}, Z.~{Dutton}, and J.~H. {Shapiro}, ``On quantum limit of optical
  communications: Concatenated codes and joint-detection receivers,'' in
  \emph{2011 IEEE International Symposium on Information Theory Proceedings},
  2011, pp. 274--278.

\bibitem{wilde_ita_2012}
M.~M. {Wilde} and S.~{Guha}, ``Explicit receivers for pure-interference bosonic
  multiple access channels,'' in \emph{2012 International Symposium on
  Information Theory and its Applications}, 2012, pp. 303--307.

\bibitem{wilde_prl_2012}
\BIBentryALTinterwordspacing
M.~M. Wilde, P.~Hayden, and S.~Guha, ``Information trade-offs for optical
  quantum communication,'' \emph{Phys. Rev. Lett.}, vol. 108, p. 140501, Apr
  2012. [Online]. Available:
  \url{https://link.aps.org/doi/10.1103/PhysRevLett.108.140501}
\BIBentrySTDinterwordspacing

\bibitem{xu_qip_2013}
\BIBentryALTinterwordspacing
S.~C. Xu and M.~M. Wilde, ``Sequential, successive, and simultaneous decoders
  for entanglement-assisted classical communication,'' \emph{Quantum
  Information Processing}, vol.~12, no.~1, pp. 641--683, 2013. [Online].
  Available: \url{https://doi.org/10.1007/s11128-012-0410-y}
\BIBentrySTDinterwordspacing

\bibitem{papen_blahut_2019}
G.~C. Papen and R.~E. Blahut, \emph{Lightwave Communications}.\hskip 1em plus
  0.5em minus 0.4em\relax Cambridge University Press, 2019.

\bibitem{green_b_2005}
P.~E. Green, \emph{Fiber to the Home: The New Empowerment (Wiley Survival
  Guides in Engineering and Science)}.\hskip 1em plus 0.5em minus 0.4em\relax
  USA: Wiley-Interscience, 2005.

\bibitem{aburgheff_2007}
M.~A. Abu-Rgheff, \emph{Introduction to CDMA Wireless Communications}.\hskip
  1em plus 0.5em minus 0.4em\relax USA: Academic Press, Inc., 2007.

\bibitem{viterbi_1995}
A.~J. Viterbi, \emph{CDMA: Principles of Spread Spectrum Communication}.\hskip
  1em plus 0.5em minus 0.4em\relax USA: Addison Wesley Longman Publishing Co.,
  Inc., 1995.

\bibitem{salehi_tran_1989}
J.~A. Salehi, ``Code division multiple-access techniques in optical fiber
  networks---part {I}: Fundamental principles,'' \emph{IEEE Transactions on
  Communications}, vol.~37, pp. 824 -- 833, 09 1989.

\bibitem{golomb_b_1967}
S.~Golomb, \emph{Shift Register Sequences}, ser. Holden-Day Series in
  Information Systems.\hskip 1em plus 0.5em minus 0.4em\relax Holden-Day, 1967.

\bibitem{salehi_jlwt_1990}
J.~A. {Salehi}, A.~M. {Weiner}, and J.~P. {Heritage}, ``Coherent ultrashort
  light pulse code-division multiple access communication systems,''
  \emph{Journal of Lightwave Technology}, vol.~8, no.~3, pp. 478--491, March
  1990.

\bibitem{boozer_prl_2007}
\BIBentryALTinterwordspacing
A.~D. Boozer, A.~Boca, R.~Miller, T.~E. Northup, and H.~J. Kimble, ``Reversible
  state transfer between light and a single trapped atom,'' \emph{Phys. Rev.
  Lett.}, vol.~98, p. 193601, May 2007. [Online]. Available:
  \url{https://link.aps.org/doi/10.1103/PhysRevLett.98.193601}
\BIBentrySTDinterwordspacing

\bibitem{yang_np_2016}
\BIBentryALTinterwordspacing
S.~Yang, Y.~Wang, D.~D.~B. Rao, T.~Hien~Tran, A.~S. Momenzadeh, M.~Markham,
  D.~J. Twitchen, P.~Wang, W.~Yang, R.~St{\"o}hr, P.~Neumann, H.~Kosaka, and
  J.~Wrachtrup, ``High-fidelity transfer and storage of photon states in a
  single nuclear spin,'' \emph{Nature Photonics}, vol.~10, no.~8, pp. 507--511,
  2016. [Online]. Available: \url{https://doi.org/10.1038/nphoton.2016.103}
\BIBentrySTDinterwordspacing

\bibitem{okamoto_b_2000}
\BIBentryALTinterwordspacing
K.~Okamoto, \emph{Fundamentals of Optical Waveguides}, ser. Optics and
  Photonics.\hskip 1em plus 0.5em minus 0.4em\relax Academic Press, 2000.
  [Online]. Available: \url{https://books.google.com/books?id=Igfx0KJc7ZoC}
\BIBentrySTDinterwordspacing

\bibitem{weiner_ol_1988}
\BIBentryALTinterwordspacing
A.~M. Weiner, J.~P. Heritage, and J.~A. Salehi, ``Encoding and decoding of
  femtosecond pulses,'' \emph{Opt. Lett.}, vol.~13, no.~4, pp. 300--302, Apr
  1988. [Online]. Available:
  \url{http://ol.osa.org/abstract.cfm?URI=ol-13-4-300}
\BIBentrySTDinterwordspacing

\bibitem{kues_np_2019}
\BIBentryALTinterwordspacing
M.~Kues, C.~Reimer, J.~M. Lukens, W.~J. Munro, A.~M. Weiner, D.~J. Moss, and
  R.~Morandotti, ``Quantum optical microcombs,'' \emph{Nature Photonics},
  vol.~13, no.~3, pp. 170--179, 2019. [Online]. Available:
  \url{https://doi.org/10.1038/s41566-019-0363-0}
\BIBentrySTDinterwordspacing

\bibitem{loudon_2000}
\BIBentryALTinterwordspacing
R.~Loudon, \emph{The Quantum Theory of Light (Third Edition)}, ser. Oxford
  science publications.\hskip 1em plus 0.5em minus 0.4em\relax OUP Oxford,
  2000. [Online]. Available:
  \url{https://books.google.de/books?id=BpnYmAEACAAJ}
\BIBentrySTDinterwordspacing

\bibitem{louisell_1990}
W.~H. {Louisell}, \emph{{Quantum Statistical Properties of Radiation}}.\hskip
  1em plus 0.5em minus 0.4em\relax Wiley, 1990.

\bibitem{rezai_prx_2018}
\BIBentryALTinterwordspacing
M.~Rezai, J.~Wrachtrup, and I.~Gerhardt, ``Coherence properties of molecular
  single photons for quantum networks,'' \emph{Phys. Rev. X}, vol.~8, p.
  031026, Jul 2018. [Online]. Available:
  \url{https://link.aps.org/doi/10.1103/PhysRevX.8.031026}
\BIBentrySTDinterwordspacing

\bibitem{torrieri_b_2018}
D.~Torrieri, \emph{Principles of Spread-Spectrum Communication Systems},
  4th~ed.\hskip 1em plus 0.5em minus 0.4em\relax Springer International
  Publishing, 2018.

\bibitem{reck_prl_1994}
\BIBentryALTinterwordspacing
M.~Reck, A.~Zeilinger, H.~J. Bernstein, and P.~Bertani, ``Experimental
  realization of any discrete unitary operator,'' \emph{Phys. Rev. Lett.},
  vol.~73, pp. 58--61, Jul 1994. [Online]. Available:
  \url{https://link.aps.org/doi/10.1103/PhysRevLett.73.58}
\BIBentrySTDinterwordspacing

\bibitem{clements_o_2016}
\BIBentryALTinterwordspacing
W.~R. Clements, P.~C. Humphreys, B.~J. Metcalf, W.~S. Kolthammer, and I.~A.
  Walmsley, ``Optimal design for universal multiport interferometers,''
  \emph{Optica}, vol.~3, no.~12, pp. 1460--1465, Dec 2016. [Online]. Available:
  \url{http://www.osapublishing.org/optica/abstract.cfm?URI=optica-3-12-1460}
\BIBentrySTDinterwordspacing

\bibitem{knill_n_2001}
\BIBentryALTinterwordspacing
E.~Knill, R.~Laflamme, and G.~J. Milburn, ``A scheme for efficient quantum
  computation with linear optics,'' \emph{Nature}, vol. 409, no. 6816, pp.
  46--52, Jan. 2001. [Online]. Available:
  \url{http://dx.doi.org/10.1038/35051009}
\BIBentrySTDinterwordspacing

\bibitem{rezai_qst_2019}
\BIBentryALTinterwordspacing
M.~Rezai, J.~Sperling, and I.~Gerhardt, ``What can single photons do what
  lasers cannot do?'' \emph{Quantum Science and Technology}, vol.~4, no.~4, p.
  045008, sep 2019. [Online]. Available:
  \url{https://doi.org/10.1088%2F2058-9565%2Fab3d56}
\BIBentrySTDinterwordspacing

\bibitem{leonhardt_rpp_2003}
\BIBentryALTinterwordspacing
U.~Leonhardt, ``Quantum physics of simple optical instruments,'' \emph{Reports
  on Progress in Physics}, vol.~66, no.~7, pp. 1207--1249, jun 2003. [Online].
  Available: \url{https://doi.org/10.1088%2F0034-4885%2F66%2F7%2F203}
\BIBentrySTDinterwordspacing

\bibitem{legero_aamo_2006}
\BIBentryALTinterwordspacing
T.~Legero, T.~Wilk, A.~Kuhn, and G.~Rempe, ``Characterization of single photons
  using two-photon interference,'' \emph{Advances In Atomic, Molecular, and
  Optical Physics}, vol.~53, pp. 253 -- 289, 2006. [Online]. Available:
  \url{http://www.sciencedirect.com/science/article/pii/S1049250X06530095}
\BIBentrySTDinterwordspacing

\bibitem{leonhardt_book_1997}
U.~Leonhardt, \emph{Measuring the Quantum State of Light}.\hskip 1em plus 0.5em
  minus 0.4em\relax Cambrdge University Press, 1997.

\bibitem{furusawa_2015}
\BIBentryALTinterwordspacing
A.~Furusawa, \emph{Quantum States of Light}, ser. Oxford science
  publications.\hskip 1em plus 0.5em minus 0.4em\relax Springer Japan, 2015.
  [Online]. Available: \url{https://www.springer.com/gp/book/9784431559580}
\BIBentrySTDinterwordspacing

\bibitem{shih_prl_1988}
\BIBentryALTinterwordspacing
Y.~H. Shih and C.~O. Alley, ``New type of einstein-podolsky-rosen-bohm
  experiment using pairs of light quanta produced by optical parametric down
  conversion,'' \emph{Phys. Rev. Lett.}, vol.~61, pp. 2921--2924, Dec 1988.
  [Online]. Available:
  \url{http://link.aps.org/doi/10.1103/PhysRevLett.61.2921}
\BIBentrySTDinterwordspacing

\bibitem{rezai_o_2019}
\BIBentryALTinterwordspacing
M.~Rezai, J.~Wrachtrup, and I.~Gerhardt, ``Polarization-entangled photon pairs
  from a single molecule,'' \emph{Optica}, vol.~6, no.~1, pp. 34--40, Jan 2019.
  [Online]. Available:
  \url{http://www.osapublishing.org/optica/abstract.cfm?URI=optica-6-1-34}
\BIBentrySTDinterwordspacing

\bibitem{hong_prl_1987}
\BIBentryALTinterwordspacing
C.~K. Hong, Z.~Y. Ou, and L.~Mandel, ``Measurement of subpicosecond time
  intervals between two photons by interference,'' \emph{Phys. Rev. Lett.},
  vol.~59, pp. 2044--2046, Nov 1987. [Online]. Available:
  \url{http://link.aps.org/doi/10.1103/PhysRevLett.59.2044}
\BIBentrySTDinterwordspacing

\bibitem{guo_sr_2014}
\BIBentryALTinterwordspacing
Y.~Guo and S.~Wu, ``Quantum correlation exists in any non-product state,''
  \emph{Scientific Reports}, vol.~4, no.~1, pp. 7179--, 2014. [Online].
  Available: \url{https://doi.org/10.1038/srep07179}
\BIBentrySTDinterwordspacing

\bibitem{ritter_n_2012}
\BIBentryALTinterwordspacing
S.~Ritter, C.~Nolleke, C.~Hahn, A.~Reiserer, A.~Neuzner, M.~Uphoff, M.~Mucke,
  E.~Figueroa, J.~Bochmann, and G.~Rempe, ``An elementary quantum network of
  single atoms in optical cavities,'' \emph{Nature}, vol. 484, no. 7393, pp.
  195--200, Apr. 2012. [Online]. Available:
  \url{http://dx.doi.org/10.1038/nature11023}
\BIBentrySTDinterwordspacing

\bibitem{rezus_prl_2012}
\BIBentryALTinterwordspacing
Y.~L.~A. Rezus, S.~G. Walt, R.~Lettow, A.~Renn, G.~Zumofen, S.~G\"otzinger, and
  V.~Sandoghdar, ``Single-photon spectroscopy of a single molecule,''
  \emph{Phys. Rev. Lett.}, vol. 108, p. 093601, Feb 2012. [Online]. Available:
  \url{https://link.aps.org/doi/10.1103/PhysRevLett.108.093601}
\BIBentrySTDinterwordspacing

\bibitem{gagliardi_book_1995}
\BIBentryALTinterwordspacing
R.~Gagliardi and S.~Karp, \emph{Optical Communications}, ser. Wiley Series in
  Telecommunications and Signal Processing.\hskip 1em plus 0.5em minus
  0.4em\relax Wiley, 1995. [Online]. Available:
  \url{https://books.google.com/books?id=ySAfAQAAIAAJ}
\BIBentrySTDinterwordspacing

\bibitem{blow_pr_1990}
\BIBentryALTinterwordspacing
K.~J. Blow, R.~Loudon, S.~J.~D. Phoenix, and T.~J. Shepherd, ``Continuum fields
  in quantum optics,'' \emph{Phys. Rev. A}, vol.~42, pp. 4102--4114, Oct 1990.
  [Online]. Available: \url{https://link.aps.org/doi/10.1103/PhysRevA.42.4102}
\BIBentrySTDinterwordspacing

\end{thebibliography}

\begin{IEEEbiography}[{\includegraphics[width=1in,height=1.25in,clip,keepaspectratio]{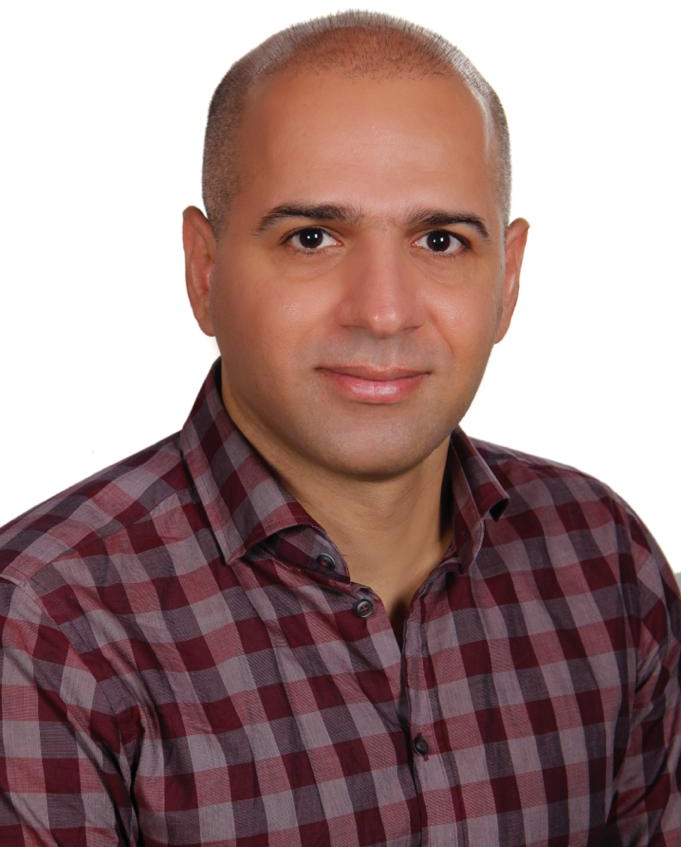}}]{Mohammad Rezai}
  was born in Firoozabad, Iran in 1983.
  He received the B.S. degree from the University of Sistan and Baluchestan in 2006, the M.S. degree in physics from the Sharif University of Technology in 2009, and the Ph.D. degree in physics from the University of Stuttgart, Germany, in 2018. 
    From 2010 to 2013, he was a member of the International Max Planck Research School for Advanced Materials and a member of research staff in the field of condensed matter physics in the Institute for Theoretical Physics III, University of Stuttgart, Germany.
  In 2013 he joined the 3rd Physikalisches Institut, University of Stuttgart, Germany, where he engaged in optical quantum information processing experiments.
  \par
  Since 2019, he has been a postdoctoral researcher with Sharif Quantum Center and Electrical Engineering Department, Sharif University of Technology, Tehran, Iran.
  His current research interests include quantum holography, quantum Fourier optics, quantum multiple access communication systems and quantum coherence in photosynthetic systems.
  \par
  Dr. Rezai was elected to the Iran National Elite Foundation in 2019 and a recipient of the Max Planck scholarship in 2010. 

\end{IEEEbiography}
\begin{IEEEbiography}[{\includegraphics[width=1in,height=1.25in,clip,keepaspectratio]{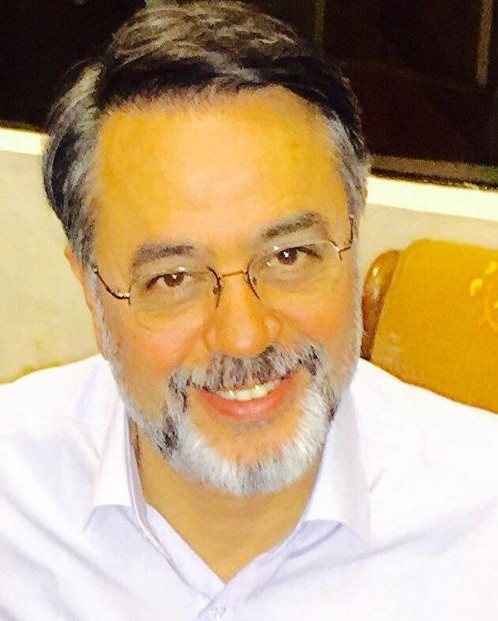}}]{Jawad A. Salehi}
Jawad A. Salehi (M’84–SM’07–F’11) was born in Kazemain, Iraq, in 1956. He received the B.Sc.degree from the University of California at Irvine in 1979, and the M.Sc. and Ph.D. degrees in electrical engineering from the University of Southern California (USC), in 1980 and 1984, respectively.
From 1984 to 1993, he was a Member of the Technical Staff of the Applied Research Area, Bell Communications Research (Bellcore), Morristown, New Jersey.
In 1990, he was with the Laboratory of Information and Decision Systems, Massachusetts Institute of Technology, as a visiting research scientist conducting research on optical multiple-access networks.
He was an Associate Professor from 1997 to 2003 and currently he is a Distinguished Professor with the Department of Electrical Engineering (EE), Sharif University of Technology (SUT), Tehran, Iran.
\par
From 2003 to 2006, he was the Director of the National Center of Excellence in Communications Science at the EE department of SUT.
In 2003, he founded and directed the Optical Networks Research Laboratory 
for advanced theoretical and experimental research in futuristic all-optical networks. 
Currently he is the Head 
of Sharif Quantum Center emphasizing in advancing quantum communication systems, quantum optical signal processing and quantum information science.
\par
His current research interests include quantum optics, quantum communications signals and systems, quantum CDMA, quantum Fourier optics, and optical wireless communication (indoors and underwater).
He is the holder of 12 U.S. patents on optical CDMA. 
\par
Dr. Salehi was named as among the 250 preeminent and most influential researchers worldwide by the Institute for Scientific Information Highly Cited in the Computer-Science Category, 2003.
He is a recipient of the Bellcore’s Award of Excellence, the Outstanding Research Award of the EE Department of SUT in 2002 and 2003, 
the Nationwide Outstanding Research Award 
2003, and the Nation’s Highly Cited Researcher Award 2004.
\par
From 2001 to 2012, he was an Associate Editor of the Optical CDMA of the IEEE TRANSACTIONS ON COMMUNICATIONS.
Professor Salehi is a member of the Iran Academy of Science and a Fellow of the Islamic World Academy of Science, Amman, Jordan.
\end{IEEEbiography}

\end{document}